\documentclass[fleqn,usenatbib]{mnras}

\usepackage{newtxtext}
\usepackage[varvw]{newtxmath}

\usepackage[T1]{fontenc}

\DeclareRobustCommand{\VAN}[3]{#2}
\let\VANthebibliography\thebibliography
\def\thebibliography{\DeclareRobustCommand{\VAN}[3]{##3}\VANthebibliography}

\usepackage{graphicx}	
\usepackage{amsmath}	
\usepackage{multirow}
\usepackage{verbatim}
\usepackage{threeparttable}

\title[An unexplored regime of shock breakout]{An unexplored regime of shock breakout with a distinct spectral signature}

\author[Irwin \& Hotokezaka]{
Christopher M. Irwin,$^{1}$\thanks{E-mail: irwincm@g.ecc.u-tokyo.ac.jp (CMI)}
Kenta Hotokezaka$^{1}$
\\
$^{1}$Research Center for the Early Universe, Graduate School of Science, The University of Tokyo, Bunkyo, Tokyo 113-0033, Japan
}

\date{Accepted XXX. Received YYY; in original form ZZZ}

\pubyear{\the\year{}}

\begin{document}
\label{firstpage}
\pagerange{\pageref{firstpage}--\pageref{lastpage}}
\maketitle

\begin{abstract}
The first light that escapes from a supernova explosion is the shock breakout emission, which produces a bright flash of UV or X-ray radiation. Standard theory predicts that the shock breakout spectrum will be a blackbody if the gas and radiation are in thermal equilibrium, or a Comptonized free-free spectrum if not. Using recent results for the post-breakout evolution which suggest that lower-temperature ejecta are probed earlier than previously thought, we show that another scenario is possible in which the gas and radiation are initially out of equilibrium, but the time when thermalized ejecta are revealed is short compared to the light-crossing time of the system. In this case, the observed spectrum differs significantly from the standard expectation, as the non-negligible light travel time acts to smear the spectrum into a complex multi-temperature blend of blackbody and free-free components. For typical parameters, a bright multi-wavelength transient is produced, with the free-free emission being spread over a wide frequency range from optical to hard X-rays, and the blackbody component peaking in soft X-rays.  We explore the necessary conditions to obtain this type of unusual spectrum, finding that it may be relevant for bare blue supergiant progenitors, or for shocks with a velocity of $v_{\rm bo} \sim 0.1\,c$ breaking out from an extended medium of radius $R_{\rm env}$ with a sufficiently high density $\rho_{\rm bo} \ga 4\times 10^{-12}\,\text{g}\,\text{cm}^{-3} (R_{\rm env}/10^{14}\,\text{cm})^{-15/16}$. An application to low-luminosity gamma-ray bursts is considered in a companion paper. 
\end{abstract}




\section{Introduction}
\label{sec:introduction}

The first emission observed from a supernova (SN) is a burst of emission escaping the radiation-mediated shock known as shock breakout.  Characterizing the shock breakout emission has long been of theoretical interest, as observing the breakout pulse can provide crucial information about the progenitor star and its immediate environment. Building on the pioneering work of the 1970s \citep{colgate,weaver,falk,klein}, the subject has seen a resurgence since the late 2000s, as observational capabilities have evolved to probe ever shorter time-scales, finally bringing the detection of shock breakout within reach.  Significant theoretical effort \citep[see][for a review]{waxmankatz} has been devoted to understanding the breakout signal in the classical case of a non-relativistic spherical explosion breaking out from a bare progenitor star \citep{imshennik,ensman,mm,katz,ns,piro,sapir1,katz2,sapir2,fs}, as well as extending these results to cover breakout in extended media \citep{waxman,ofek,balberg,ci11,moriya1,chatzopoulos,ginzburg1,moriya2,svirski1,ginzburg2,svirski2,svirski3,moriya3}, relativistic shocks \citep{budnik,ns2,granot,fs2} and deviations from spherical symmetry \citep{couch,matzner,suzuki,afsariardchi,ohtani,il,goldberg}.

By now, several noteworthy shock breakout candidates have emerged; here we highlight only a few salient examples \citep[for a more extensive discussion of candidates, see, e.g.,][]{waxmankatz}. An early milestone was the \textit{Swift} discovery of prompt X-ray emission associated with the Type Ib SN 2008D \citep{mazzali2,soderberg2,malesani,modjaz}.  Assuming this event was powered by shock breakout \citep[see, however,][]{li08,xu}, the $600\,$s duration of the X-ray event significantly exceeded the expectation for a compact Type Ib progenitor, prompting the suggestion that the breakout took place in a stellar wind \citep[e.g.,][]{cf,svirski3}.  At around the same time, a new class of low-luminosity gamma-ray bursts ($ll$GRBs) with faint and soft prompt emission was recognized \citep[e.g.,][and references therein]{liang,virgili}.  Some of these events, like GRB 060218 \citep{campana,mazzali,pian,soderberg,sollerman,amati,kaneko} and GRB 100316D \citep{chornock,cano,fan2,starling,margutti1} have smooth, long-lived light curves distinct from the typical GRB population.  Several authors have proposed explanations for these $ll$GRBs involving shock breakout \citep[e.g.,][]{campana,li,wang,waxman,ns2,nakar}.  Curiously, these events are also associated with broad-lined Type Ic SNe with a peculiar early optical peak \citep[e.g.,][]{campana}, similar to the early peak seen in Type IIb events \citep[see][for a recent review of double-peaked SNe]{modjazarcavi}.  This has been taken to indicate a non-standard progenitor with an extended, low-mass envelope \citep[e.g.,][]{np,nakar,ic16}.  More recently, exquisite early data obtained from two nearby supernovae, SN 2023ixf \citep{hiramatsu,hosseinzadeh,jacobson1,smith,teja,vasylyev,yamanaka,bostroem,li24,shrestha1,singh,zimmerman} and SN 2024ggi \citep{shrestha2,jacobson2,chen1,chen2,pessi,zhang}, have shed new light on the early emission in Type II events.  The long-lived bolometric peak, spectropolarimetric evolution, and presence of narrow emission features in these objects have painted the picture of a progenitor with a complex circumstellar environment, consistent with the pre-explosion imaging, which in both cases has revealed a dust-enshrouded red supergiant \citep{kilpatrick,jencson,xiang}.  These and other events have underscored the essential role played by the circumstellar environment in shaping the shock breakout emission.

However, along with these revelations, there were also surprises.  Of particular interest in the present work is the unusual spectrum of $ll$GRBs: some events, most notably GRB 060218 and GRB 100316D, exhibit a peculiar spectrum with both a soft, $\sim0.1\,$keV thermal component, and a hard, $\sim10\,$keV power-law or Band function component \citep[e.g.,][]{campana,kaneko,toma,starling}.  This complicated spectrum has led some authors to question whether a shock breakout origin is indeed possible for these events \citep[e.g.,][]{ghisellini1,ghisellini2,ic16}.

The aim of this paper is, first of all, to understand whether such a spectrum can ever be produced by shock breakout, and if so, to understand the requisite conditions and the overall spectral evolution.  The structure of the paper is as follows.  First, we introduce the basic idea of our model in Section~\ref{sec:basicidea}. We start by reviewing the physics of non-equilibrium shock breakout in Section~\ref{sec:nonequilibrium}.  After discussing the temperature evolution in Section~\ref{sec:thermalization}, we show in Section~\ref{sec:blendedspectrum} that a peculiar type of breakout can occur when the temperature evolution is sufficiently fast. Next, we investigate the necessary conditions for this non-standard shock breakout scenario in Section~\ref{sec:conditions}, considering the general theory in Section~\ref{sec:parameterspace}, and specifying to the case of shock breakout in extended dense environments in Section~\ref{sec:envelopebreakout}.  A particularly interesting limiting case of our model and its connection to past work are discussed in Section~\ref{sec:n0} and Section~\ref{sec:connections}.  We then develop a model for the shock breakout spectrum in this new regime in Section~\ref{sec:model}. We describe the model parameters and assumptions in Section~\ref{sec:modelparameters}, the basic method for computing the spectrum in Section~\ref{sec:sourceframe}, and the effects of light travel time and Comptonization in Sections~\ref{sec:observedspectrum} and~\ref{sec:comptonization} respectively.  We summarize our key results in Section~\ref{sec:conclusions}, and conclude by discussing the outlook for observations of this unusual type of shock breakout.  We also provide a glossary in Appendix~\ref{sec:appendixa} to help the reader keep track of the symbols defined throughout the paper. 

This is the first in a series of three papers related to shock breakout spectra.  In the second paper (Irwin \& Hotokezaka 2024a, submitted, hereafter Paper II), we consider a simple means of classifying the spectra produced by planar shock breakout and characterizing their time evolution.  The third paper (Irwin \& Hotokezaka 2024b, submitted, hereafter Paper III) considers an application of our results to a well-studied $ll$GRB, GRB 060218, which has long eluded a satisfactory explanation.

\section{A non-standard shock breakout scenario}
\label{sec:basicidea}
Motivated by the observations of $ll$GRBs discussed in Section~\ref{sec:introduction}, our primary goal in this section is to answer a simple question: is it ever possible for a shock breakout to produce thermal and non-thermal emission simultaneously?  To address this, we will first review the standard theoretical expectations for the initial breakout properties and the subsequent evolution.  We then consider a thought experiment that leads to the answer we seek.

\subsection{Non-equilibrium breakout: basic idea and initial properties}
\label{sec:nonequilibrium}

In the general shock breakout problem, we consider a radiation mediated shock propagating in a dense region of size $R$.  The density profile near the edge of the region is assumed to steepen as $r\rightarrow R$, with $\rho(r) \propto (R-r)^n$, causing the shock to accelerate leading up to the breakout.  The shock's acceleration is described by the self-similar solution of \citet{sakurai}, which shows that the velocity scales with density as $v \propto \rho^{-\mu}$, with $\mu\approx 0.19$.  Shock breakout occurs when the diffusion time from the shock to the surface becomes equal to the time it takes for the shock to cross that same distance; this happens at an optical depth of $c/v_{\rm{bo}}$, where $v_{\rm{bo}}$ is the shock velocity at the moment of breakout and $c$ is the speed of light.  We treat here the case where the breakout occurs close to the edge of the medium, so that the shock radius at breakout, $R_{\rm bo}$, satisfies $R_{\rm bo} \approx R$ and $R_{\rm bo} - R \ll R$.  This is typical of breakout from the edge of a star, but also applies to breakout from an extended medium, if the medium has a sharp edge and is sufficiently optically thick \citep[e.g.,][see also Section~\ref{sec:conditions} below]{ci11,hp,khatami}.

To aid the discussion, it will be helpful to introduce some terminology.  Following \citet[][hereafter NS10]{ns}, we define three special locations within the breakout ejecta: the `breakout shell,' the `luminosity shell,' and the `colour shell.'  As in \citet[][hereafter FS19]{fs}, we will label the properties of these shells using the subscripts `bo,' `ls,' and `cs' respectively.  The breakout shell is the shell from which the first breakout radiation escapes, and its properties determine the initial conditions in the problem.  The luminosity shell is the shell from which the observed photons originate at a given time, i.e. it is the shell which satisfies the condition $\tau \approx c/v$ (up to a logarithmic correction, as discussed below).  Initially, the luminosity shell coincides with the breakout shell, but as time goes on and progressively more of the ejecta are exposed, it recedes inwards.  Thus, after the initial breakout flash, the luminosity shell is always interior to the breakout shell.  Finally, the colour shell is the shell out to which the gas can modify the spectrum of the radiation, i.e. it is the shell where the observed temperature is determined.  Its location depends on whether the luminosity shell is in thermal equilibrium or not (NS10).  If it is, then the colour shell is located outside the luminosity shell, at the outermost location where thermal equilibrium can be maintained (i.e., the thermalization depth).  Otherwise, the colour shell is simply the luminosity shell.

The breakout shell has the special property that its dynamical time, $t_{\rm bo},$ equals its diffusion time.  The value of $t_{\rm{bo}}$ is approximately given by (see Section~\ref{sec:envelopebreakout} for a derivation)
\begin{equation}
    \label{tbo}
    t_{\rm{bo}} \approx \dfrac{c(n+1)}{\kappa \rho_{\rm{bo}} v_{\rm{bo}}^2} \approx 1700\,\text{s} \,(n+1)\rho_{\rm{bo,}-11}^{-1}v_{\rm{bo,}-1}^{-2},
\end{equation}
where $\kappa$ is the opacity, and $\rho_{\rm{bo}}$ is the upstream density at the breakout shell's location.  We have assumed an opacity of $\kappa = 0.2\,\rm{cm}^2\rm{g}^{-1}$ for ionized hydrogen-free gas (as appropriate for $ll$GRB progenitors) and scaled the density and velocity according to ${\rho_{\rm{bo,}-11} = \rho_{\rm{bo}}/(10^{-11}\,\text{g}\,\text{cm}^{-3})}$ and ${v_{\rm{bo,}-1} = v_{\rm{bo}}/(0.1\,c)}$, as expected for shocks in the extended envelope in $ll$GRBs \citep[e.g.,][see also Paper III]{nakar,ic16}.  In order to satisfy our assumption that the breakout occurs at the edge of the medium, we require $v_{\rm bo} t_{\rm bo} < R_{\rm bo}$.

The thermal equilibrium conditions at breakout can be gauged by utilizing the thermal coupling coefficient $\eta$ introduced by NS10.  Effectively, $\eta$ is the ratio of the photon number density required to achieve thermal equilibrium in a given shell, to the number density of photons with energy $\approx 3kT_{\rm{BB}}$ generated by free-free processes in that shell.\footnote{The analysis of NS10 assumes that the gas and radiation are mainly coupled by free-free, not bound-free, processes.  For the low-metallicities expected in $ll$GRBs, this should be appropriate.}  If $\eta<1$, then enough photons are generated to reach thermal equilibrium, and the radiation temperature is equal to the blackbody temperature of the shell, $T_{\rm{BB}}$.  On the other hand, if $\eta > 1$, then thermal equilibrium cannot be attained and the radiation temperature is $> T_{\rm{BB}}$.

The thermal coupling coefficient of the breakout shell, $\eta_{\rm{bo}}$, is an important quantity because it determines whether or not the observed radiation is initially in thermal equilibrium with the gas, which strongly influences the observed spectrum.  If $\eta_{\rm{bo}} <1$, the gas and radiation are in thermal equilibrium from the beginning, and the observed spectrum is a blackbody.  However, if $\eta_{\rm{bo}} > 1$, the gas and radiation are initially out of equilibrium, and the observed spectrum is a self-absorbed free-free spectrum, which may also be modified by Comptonization (NS10).  In this case, blackbody emission is not observed until later, when slower and denser layers of the ejecta which can maintain equilibrium have been revealed to the observer.

The value of $\eta_{\rm{bo}}$ can be estimated from equation 10 of NS10:
\begin{equation}
\label{etabo}
    \eta_{\rm{bo}} \approx 22 q (n+1)^{-1} g_{\rm B}^{-1} \rho_{\rm{bo,}-11}^{-1/8} v_{\rm{bo,}-1}^{15/4},
\end{equation}
where we have introduced an order-unity dimensionless constant $q$ which will be discussed below, and have included the dependence on $g_{\rm B}$, the frequency average of the velocity-averaged Gaunt factor, which was dropped by NS10.\footnote{Although NS10 considered a hydrogen composition, for the same assumed mass density, the value of $\eta_{\rm{bo}}$ is insensitive to whether the shocked medium is composed of hydrogen or helium.  This is because $\eta_{\rm{bo}} \propto (\dot{n}_{\rm{ff}} t_{\rm{bo}})^{-1}$ (NS10), and $t_{\rm{bo}} \propto \kappa^{-1}$ (equation~\ref{tbo}), while the free-free emissivity goes as ${\dot{n}_{\rm{ff}} \propto Z^2 n_{\rm{e}} n_{\rm{i}}}$, where $Z$ is the atomic number and $n_{\rm{e}}$ and $n_{\rm{i}}$ are the electron and ion number densities respectively.  For helium, $Z$ is increased by a factor of 2, and the values of $n_{\rm{e}}$, $n_{\rm{i}}$, and $\kappa$ are respectively reduced to $\frac{1}{2}$, $\frac{1}{4}$, and $\approx \frac{1}{2}$ of the values in the pure hydrogen case.  In the calculation of $\eta_{\rm{bo}}$, these factors cancel and $\eta_{\rm{bo}}$ is left unchanged.}  The factor of $(n+1)^{-1}$, which was neglected by NS10, comes from the fact that $\eta_{\rm{bo}} \propto t_{\rm{bo}}^{-1}$ (see Section~\ref{sec:envelopebreakout}).  As expected, we find that breakout shell is far from thermal equilibrium for standard $ll$GRB parameters, but this depends strongly on the breakout velocity. In calculating $\eta_{\rm{bo}}$, following NS10, we approximated the blackbody temperature of the breakout shell as 
\begin{equation}
    \label{TBB}
    kT_{\rm{BB,bo}} \approx k\left(\dfrac{\rho_{\rm{bo}} v_{\rm{bo}}^2}{a}\right)^{1/4} \approx 28\,\text{eV}\, \rho_{\rm{bo,}-11}^{1/4} v_{\rm{bo,}-1}^{1/2},
\end{equation}
where $k$ and $a$ are the Bolztmann constant and the radiation energy density constant. 

The parameter $q$ in equation~\ref{etabo} is included to account for possible uncertainties in the value of $\eta_{\rm{bo}}$.  As we will see, some quantities in our model are quite sensitive to the value of $\eta_{\rm{bo}}$, so it is important to keep track of how order-unity variations in $\eta_{\rm{bo}}$ affect our results.  Equation~\ref{etabo} agrees with equation 10 in NS10 when $q=1$ and the factor of  $(n+1)^{-1} g_{\rm B}^{-1}$ is dropped.  However, as pointed out by NS10, the value obtained by \citet{katz} is effectively twice as large, corresponding to $q\approx 2$.  This difference possibly stems from the fact that \citet{katz} obtained the same scaling for the blackbody temperature, but with a different numerical prefactor: $T_{\rm{BB,bo}} \approx (2.6 \rho_{\rm{bo}} v_{\rm{bo}}^2/a)^{1/4}$ according to their equation 5. Although this difference only slightly affects $T_{\rm{BB,bo}}$, since $\eta_{\rm{bo}} \propto T_{\rm{BB,bo}}^{7/2}$ via equation 9 of NS10,  the value of $\eta_{\rm{bo}}$ is about doubled.  On the other hand, if bound-free or other photon generation processes contribute non-negligible numbers of photons, the value of $\eta_{\rm{bo}}$ will be reduced compared to the value of NS10 which only includes free-free photon production.  A steep density profile also leads to a  reduction in $\eta_{\rm{bo}}$ via the $(1+n)^{-1}$ prefactor (for the same $v_{\rm{bo}}$ and $\rho_{\rm{bo}}$).  Later, we will argue that a value of $q(n+1)^{-1} \sim 0.5$ leads to good agreement with Monte-Carlo simulations of radiative shocks \citep[e.g.,][see Section~\ref{sec:comptonization} for further discussion]{ito,ito1,ito2}. 

Now, let us consider some directly observable properties of the breakout flash. The most basic of these is the observed duration of the breakout.  The breakout signal will be smeared over the light-crossing time of the system or the diffusion time of the breakout layer, whichever is longer, and therefore we have 
\begin{equation}
    \label{tobs}
    t_{\rm{obs,bo}} = \max(t_{\rm{bo}},t_{\rm{lc}}),
\end{equation}
where
\begin{equation}
    \label{tlc}
    t_{\rm{lc}} = R_{\rm bo}/c \approx 3300\,\text{s}\, R_{\rm bo,14}
\end{equation}  
is the light-crossing time, and ${R_{\rm bo,14} = R_{\rm bo}/(10^{14}\,\text{cm})}$.  The observed luminosity is also straightforward to estimate via
\begin{equation}
    \label{Lobs}
    L_{\rm{obs,bo}} \approx L_{\rm{bo}} \left(\dfrac{t_{\rm{obs,bo}}}{t_{\rm{bo}}}\right)^{-1},
\end{equation}
where according to NS10,
\begin{equation}
\label{Lbo}
L_{\rm{bo}} \approx 4 \uppi R_{\rm bo}^2 \tau_{\rm{bo}}^{-1} a c T_{\rm{BB,bo}}^4 \approx 3.4 \times 10^{46}\,\text{erg\,s}^{-1}\,\rho_{\rm{bo,}-11} v_{\rm{bo,}-1}^3 R_{\rm bo,14}^2,
\end{equation}
and we have used $\tau_{\rm{bo}} \approx c/v_{\rm{bo}}$.  The quantity $L_{\rm{bo}}$ is the luminosity \textit{produced} by the breakout shell.  It is equal to the observed luminosity only if $t_{\rm{bo}} > t_{\rm{lc}}$ so that the light travel time is negligible.  If on the other hand $t_{\rm{lc}} > t_{\rm{bo}}$, the energy emitted by the breakout shell over its dynamical time, $L_{\rm{bo}} t_{\rm{bo}}$, is spread over an arrival time $\approx t_{\rm{lc}}$, so that the observed luminosity is $L_{\rm{obs,bo}} \approx L_{\rm{bo}} t_{\rm{bo}}/t_{\rm{lc}}$ (see also Section~\ref{sec:observedspectrum}).  Equations~\ref{tobs} and~\ref{Lobs} together cover both of these possibilities.

As for the initial observed temperature of the breakout, $T_{\rm{obs,bo}}$, if the gas and radiation in the breakout shell are in thermal equilibrium (i.e., if $\eta_{\rm{bo}} < 1$), then $T_{\rm{obs,bo}}$ is simply $T_{\rm{BB,bo}}$ as given by equation~\ref{TBB}.  However, in the non-equilibrium case, determining the observed temperature is more nuanced.  As shown by NS10, for $\eta_{\rm{bo}} > 1$, $T_{\rm{obs,bo}}$ is given by the implicit equation $T_{\rm{obs,bo}} \approx T_{\rm{BB,bo}} \eta_{\rm{bo}}^2/\xi_{\rm{bo}}(T_{\rm{obs,bo}})^2$, where $\xi_{\rm{bo}}(T_{\rm obs,bo})$ is a Comptonization parameter describing the breakout shell, defined as the ratio of the total number of photons at energy $3kT_{\rm obs,bo}$ (including those upscattered to that energy by Compton processes) to the number of photons generated with energy $3kT_{\rm obs,bo}$ by free-free processes (see NS10, FS19, and Section~\ref{sec:comptonization} below).  When Comptonization is important, $\xi_{\rm{bo}}>1$, and when it is not, $\xi_{\rm{bo}}=1$.  Therefore, $\xi_{\rm{bo}}(T_{\rm{obs,bo}}) \ge 1$ and the inequality 
\begin{equation}
\label{Tinequality}
kT_{\rm{obs,bo}} \le kT_{\rm{obs,bo}} \xi_{\rm{bo}}^2 \approx kT_{\rm{BB,bo}}\eta_{\rm{bo}}^2 \approx 13\,\text{keV}\, q^2 (n+1)^{-2} g_{\rm B}^{-2} v_{\rm{bo,}-1}^{8}
\end{equation}  
holds for non-equilibrium breakouts, where $T_{\rm{BB,bo}}$ and $\eta_{\rm{bo}}$ are given by equations~\ref{etabo} and~\ref{TBB}, and $k$ is Boltzmann's constant.  In some parts of parameter space, it is possible to eliminate $\xi_{\rm{bo}}$ from equation~\ref{Tinequality} to obtain an implicit equation for $T_{\rm{obs,bo}}$ in terms of $v_{\rm{bo}}$ and ${q/(n+1)}$ only (see Section~\ref{sec:comptonization}). Note that equation~\ref{Tinequality} is only applicable when the observed temperature is below the energy threshold for pair creation.  We will assume that this is the case throughout this paper.  Once pairs become important ($T_{\rm{obs,bo}} \ga 50\,$keV), the temperature rises much more slowly with $v_{\rm{bo}}$ and it eventually saturates around $\approx 200\,$keV \citep[see][for further discussion]{weaver,katz,ns2,fs2}.  Although the pair creation condition is marginal for GRB 060218, which had a peak energy of $\sim 30\,$keV \citep{campana,toma,kaneko}, a recent analysis by \citet{fs2} concluded that relativistic effects are most likely unimportant for this event. 

If we assume that $ll$GRBs are powered by shock breakout, then for an observed peak energy of $3kT_{\rm{obs,bo}} \sim 30 \,\rm{keV}$, as is typical in these events, equation~\ref{Tinequality} implies $v_{\rm{bo}} \ga 0.1 c$.  Due to the extreme sensitivity of $T_{\rm obs,bo}$ on $v_{\rm{bo}}$, the breakout velocity cannot be much larger than this value unless the Comptonization parameter $\xi_{\rm{bo}}$ is very large.  Even if $\xi_{\rm{bo}} = 100$, the implied breakout velocity is only $v_{\rm{bo}} \approx 0.3\,c$ for the same observed temperature.  Moreover, observations seem to disfavour a large Comptonization, as this would lead to a prominent Wien peak in the spectrum, which so far has not been observed in any $ll$GRB.  We therefore conclude that if $ll$GRBs are powered by shock breakout, the shock velocity must have been $\approx 0.1\,c$, consistent with the discussion in Paper III.  We will return to the topic of Comptonization when discussing our spectral model in Section~\ref{sec:model}. 

The free-free spectrum produced during a non-equilibrium breakout is expected to be self-absorbed below a frequency $\nu_{\rm{a}}$.  For a given shell with density $\rho$ and photon temperature $T_{\rm{obs}}$ in which photons have been trapped for a time $t$, $\nu_{\rm{a}}$ can be found by equating the blackbody photon energy density $(4\uppi/c)B_\nu(T_{\rm{obs}})$ in the Rayleigh-Jeans limit to the energy density of photons produced by free-free processes, $\epsilon_{\nu\rm{,ff}}(\rho,T) t$, where $B_\nu$ is the Planck function and $\epsilon_{\nu\rm{,ff}}$ is the free-free energy generation rate per unit volume per unit frequency. This leads to the result (see also, e.g., FS19\footnote{We note that there is an error in equation 47 of FS19.  The self-absorption frequency should scale as the number density $n$, not as $n^2$.})
\begin{equation}
    \label{nuadef}
    \nu_{\rm{a}} = \left(\dfrac{c^3 \epsilon_{\nu\rm{,ff}} t}{8 \uppi k T_{\rm{obs}}}\right)^{1/2}.
\end{equation}
Using the fact that photons in the breakout shell escape at the time $t_{\rm{bo}}$, given by equation~\ref{tbo}, we find from equation~\ref{nuadef} that the self-absorption frequency of the breakout shell is\footnote{Equation~\ref{nuabo} assumes ionized hydrogen, but as with equation~\ref{etabo}, it is unaffected when considering a helium composition instead.  The decrease in $\epsilon_{\nu\rm{,ff}}$ for a helium composition is compensated by an increase in $t_{\rm{bo}}$ due to the decreased opacity, so that $\nu_{\rm{a,bo}} \propto (\epsilon_{\nu\rm{,ff}} t_{\rm{bo}})^{1/2}$ remains unchanged.}
\begin{align}
    \label{nuabo}
    h\nu_{\rm{a,bo}} & \approx 0.11\,\text{eV}\,g_{\rm{ff}}^{1/2} (n+1)^{1/2} \rho_{\rm{bo,}-11}^{1/2} v_{\rm bo,-1}^{-1} \left(\dfrac{kT_{\rm{obs,bo}}}{10\,\text{keV}}\right)^{-3/4}  \\
    &\approx 0.092\,\text{eV}\,g_{\rm{ff}}^{1/2} g_{\rm B}^{3/2} q^{-3/2} (n+1)^2 \xi_{\rm{bo}}^{3/2}\rho_{\rm{bo,}-11}^{1/2} v_{\rm{bo,}-1}^{-7}  \nonumber ,
\end{align}
where $g_{\rm{ff}}$ is the free-free Gaunt factor, and as in equation~\ref{etabo} we have taken account that the postshock gas is compressed by a factor of 7 for an adiabatic index of 4/3.  In the second equality of equation~\ref{nuabo}, we have estimated $T_{\rm{obs,bo}}$ via equation~\ref{Tinequality}.
 
Interestingly, we see that the breakout shell's self-absorption frequency is extremely sensitive to the breakout velocity, as $\nu_{\rm{a,bo}} \propto v_{\rm{bo}}^{-7}$.  Since the initial observed temperature also depends very strongly on the breakout velocity, $T_{\rm{obs,bo}} \propto v_{\rm{bo}}^8$ according to equation~\ref{Tinequality}, systems with fast shocks that produce out-of-equilibrium breakouts will inevitably have a very large separation between $h\nu_{\rm{a,bo}}$ and $kT_{\rm{obs,bo}}$, and therefore will emit over a very wide range in frequencies.  As seen from equation~\ref{nuabo}, in low-density environments $h\nu_{\rm{a,bo}}$ can easily be below the optical band, while $kT_{\rm{bo}}$ can be tens of keV via equation~\ref{Tinequality}, resulting in a transient that is bright from optical up to hard X-rays.  A similar conclusion about enhanced optical/UV emission was reached by \citet{ito2}, who considered radiation-mediated shocks in stellar winds. 

The fact that $h\nu_{\rm{a,bo}}/kT_{\rm{obs,bo}} \propto \rho_{\rm{bo}}^{1/2} v_{\rm{bo}}^{-15} \propto \eta_{\rm{bo}}^{-4}$ (as can be seen from equations~\ref{etabo}, \ref{Tinequality}, and~\ref{nuabo}) hints at a possible connection between $\eta$, $\nu_{\rm{a}}$, and $kT_{\rm{obs}}$.  Indeed, the defining equations for these quantities (equations ~\ref{Tinequality} and ~\ref{nuadef}, along with equation 9 of NS10) can be reworked into the form
\begin{align}
    \label{etanuaTobs}
    \eta & = \left(\dfrac{\uppi^4 q g_{\rm{ff}}}{45 g_{\rm{B}}}\right) \dfrac{(kT_{\rm{BB}})^{7/2}}{(kT_{\rm{obs}})^{3/2}(h\nu_{\rm{a}})^2} \nonumber\\
    & \approx \left(\dfrac{\uppi^4 q g_{\rm{ff}}}{45 g_{\rm{B}}}\right)^{1/8} [\xi(T_{\rm{obs}})]^{7/8} \left(\dfrac{kT_{\rm{obs}}}{h\nu_{\rm{a}}}\right)^{1/4},
\end{align}
where in the second equality we have used the relation ${\eta \approx \xi(T_{\rm{obs}}}) (T_{\rm{obs}}/T_{\rm{BB}})^{1/2}$ derived by NS10 to eliminate $T_{\rm{BB}}$.  Eliminating $\eta$ instead also generates a useful result, ${kT_{\rm{BB}} \sim [\xi(T_{\rm{obs}})]^{1/4} (h\nu_{\rm{a}} kT_{\rm{obs}})^{1/2}}$. Equation~\ref{etanuaTobs} is very general and applies to any shell, not just the breakout shell.  Inspecting the equation, we see that as thermal equilibrium is approached (i.e., as $\eta \rightarrow 1$ and $\xi(T_{\rm{obs}})\rightarrow 1$), $h\nu_{\rm{a}} \rightarrow kT_{\rm{obs}}$ up to an order-unity factor, and $T_{\rm{obs}} \rightarrow T_{\rm{BB}}$ as well.  Thus, as expected, once equilibrium is attained the spectrum evolves into a blackbody with temperature $T_{\rm{BB}}$.  Determining the time it takes for this to happen will be the topic of the next subsection.

\subsection{Non-equilibrium breakout: temperature evolution and rapid thermalization}
\label{sec:thermalization}

Early works on shock breakout (e.g., NS10) assumed that the luminosity shell is always found at the location where the optical depth is $\tau \approx c/v$.   Because the breakout shell satisfies this condition at $t=t_{\rm{bo}}$, and because the optical depth is approximately constant during the early planar phase of evolution, this assumption implies that the luminosity shell is equivalent to the breakout shell during the entire planar phase, i.e. $m_{\rm{ls}} \approx m_{\rm{bo}}$ for $t<t_{\rm s}$, where $m_{\rm{ls}}$ and $m_{\rm bo}$ are respectively the mass coordinates of the luminosity and breakout shells, and
\begin{equation}
    \label{ts}
    t_{\rm s} \approx R_{\rm bo}/v_{\rm{bo}} \approx 3.3 \times 10^4\,\text{s}\,R_{\rm bo,14} v_{\rm{bo,}-1}^{-1}
\end{equation} is the time it takes for the breakout shell's radius to double.  However, recent work by FS19 has shown that this assumption is not quite correct.  By obtaining a self-similar solution describing the specific energy in the outer layers of the ejecta, they demonstrated that $m_{\rm{ls}}$ actually grows logarithmically with time during the planar phase as ${m_{\rm{ls}} = m_{\rm{bo}}[1+\ln(t/t_{\rm{bo}})]^{(n+1)/(n+1-\mu n)}}$.  With this correction, the luminosity shell is located slightly deeper in the ejecta, at an optical depth of ${\tau_{\rm{ls}} \approx (c/v_{\rm{ls}})[1+\ln(t/t_{\rm{bo}})]}$ rather than at ${\tau_{\rm ls} \approx c/v_{\rm ls}}$, where $v_{\rm ls}$ is the luminosity shell's velocity.

This logarithmic correction has important implications for the observed behaviour at early times. Although the effect on the luminosity is minor (to a good approximation, $L_{\rm{ls}} \propto t^{-4/3}$ throughout the planar phase with an extremely weak dependence on logarithmic terms, see FS19), the effect on the temperature can be drastic, particularly if the breakout is initially out of thermal equilibrium, with $\eta_{\rm{bo}} > 1$.  As the luminosity shell moves into the ejecta, it encounters denser and slower ejecta, which are more readily thermalized.  Because the thermal coupling coefficient depends strongly on location in the outer parts of the flow, even a small change in the luminosity shell's position can significantly influence the spectrum (recall that when $\eta > 1$, the luminosity shell is equivalent to the colour shell, i.e. the properties of the luminosity shell fully determine the observed spectrum). Therefore, compared to previous models which neglect the logarithmic dependence of $m_{\rm{ls}}$, thermal equilibrium of the luminosity shell can be achieved much sooner in the model of FS19.  This can be seen by examining the evolution of the luminosity shell's thermal coupling coefficient, which in the planar phase is given by (FS19)
\begin{equation}
\label{etals}
\eta_{\rm{ls}} = \eta_{\rm{bo}} \left(t/t_{\rm{bo}}\right)^{-1/6} \left[1+\ln(t/t_{\rm{bo}})\right]^{-p_\eta(n,\mu)},
\end{equation}
where 
\begin{equation}
    \label{peta}
    p_\eta(n,\mu) = \dfrac{20+27n+62\mu n}{24(1+n-\mu n)} \approx \dfrac{5(1+1.94n)}{6(1+0.81n)},
\end{equation}
and the second equality in equation~\ref{peta} assumes $\mu \approx 0.19$.
The logarithmic factor in brackets in equation~\ref{etals} is a correction due to FS19, which results in a significantly faster evolution than the $\eta_{\rm{ls}} \propto t^{-1/6}$ behaviour predicted by NS10.  For example, at $t=10 t_{\rm{bo}}$, neglecting the logarithmic prediction gives $\eta_{\rm{ls}} \approx 0.7 \eta_{\rm{bo}}$, while including it results in $\eta_{\rm{ls}} \approx 0.1 \eta_{\rm{bo}}$, for $n=3$.

Thermal equilibrium of the luminosity shell is achieved once $\eta_{\rm{ls}}$ becomes equal to unity.  Let us define the time when this occurs as $t_{\rm{eq,ls}}$. Since the luminosity shell determines the observed spectrum during the non-equilibrium phase (as discussed above), the time $t_{\rm{eq,ls}}$ is also the time when the observer first starts to see blackbody emission.  For a typical shock breakout occurring at the edge of a standard blue supergiant or Wolf-Rayet star progenitor, the breakout is initially out of equilibrium (i.e., $\eta_{\rm{bo}} >1$), and it remains that way throughout the planar phase (NS10,FS19).  In other words, $t_{\rm{eq,ls}} > t_{\rm s}$ in the usual case.  However, we point out that in principle, $t_{\rm{eq,ls}} < t_{\rm s}$ may also be possible if the progenitor's properties differ somewhat from the typically assumed values.  In fact, as we will see below, in some parts of parameter space $t_{\rm{eq,ls}}$ can even be shorter than the light-crossing time of the system, $t_{\rm{lc}}$ (see Section~\ref{sec:parameterspace}).

In the case where $t_{\rm{eq,ls}} < t_{\rm s}$, equation~\ref{etals} applies up until equilibrium is achieved, so we can use it to estimate $t_{\rm{eq,ls}}$. Since by definition $\eta_{\rm{ls}} = 1$ at $t=t_{\rm{eq,ls}}$, we find via equation~\ref{etals} that $t_{\rm{eq,ls}}$ satisfies
\begin{equation}
    \label{teq}
    (t_{\rm{eq,ls}}/t_{\rm{bo}})[1+\ln(t_{\rm{eq,ls}}/t_{\rm{bo}})]^{6p_\eta} = \eta_{\rm{bo}}^6
\end{equation}
for $t_{\rm{eq,ls}} < t_{\rm s}$.  For a given $\rho_{\rm{bo}}$, $v_{\rm{bo}}$, and $n$, $t_{\rm{bo}}$ and $\eta_{\rm{bo}}$ can be obtained from equations~\ref{tbo} and~\ref{etabo}, and then equation~\ref{teq} can be solved to yield $t_{\rm{eq,ls}}$.  Note that equation~\ref{teq} is only valid when $\eta_{\rm{bo}} > 1$, in which case $t_{\rm{eq,ls}} > t_{\rm{bo}}$.  As $\eta_{\rm{bo}} \rightarrow 1$, $t_{\rm{eq,ls}} \rightarrow t_{\rm{bo}}$.  For $\eta_{\rm{bo}} \le 1$, $t_{\rm{eq,ls}}$ has no meaning since the breakout is already in thermal equilibrium from the beginning.  In this case we simply define $t_{\rm{eq,ls}} = t_{\rm{bo}}$ for convenience.

The rapid decrease of the thermalization parameter of the luminosity shell causes the observed temperature of the breakout emission to quickly decline as (FS19)
\begin{equation}
    \label{Tobsls}
    T_{\rm{obs,ls}} = T_{\rm{obs,bo}} \left[\dfrac{\xi_{\rm ls}(T_{\rm obs,ls})}{\xi_{\rm{bo}}}\right]^{-2} (t/t_{\rm{bo}})^{-2/3} [1+\ln(t/t_{\rm{bo}})]^{-\frac{4+6n+16\mu n}{3(1+n-\mu n)}},
\end{equation}
where $\xi_{\rm ls}(T_{\rm obs,ls})$ is the Comptonization parameter of luminosity shell. As the gas and radiation approach equilibrium,  $\xi_{\rm{ls}}(T_{\rm{obs,ls}})$ continuously decreases towards unity (see Section~\ref{sec:comptonization} for further information).  If we approximate the evolution of $T_{\rm{obs}}$ as a power-law between, e.g., $t_{\rm{bo}}$ and $10 t_{\rm{bo}}$, then we find that for $n=3$, the decrease could be as steep as $T_{\rm{obs}} \propto t^{-2.2}$ if $\xi_{\rm{ls}}$ depends weakly on time (the evolution is shallower for smaller $n$, or if $\xi_{\rm{ls}}$ has a strong time dependence).

The decrease in observed temperature is accompanied by a rapid increase in the self-absorption frequency, according to 
\begin{equation}
\label{nuals}
\nu_{\rm{a,ls}} = h\nu_{\rm{a,bo}} [\xi_{\rm{ls}}(T_{\rm{obs,ls}})/\xi_{\rm{bo}}]^{3/2}[1+\ln(t/t_{\rm{bo}})]^{\frac{4+5n+10\mu n}{2(1+n-\mu n)}},
\end{equation}
as can be derived from equation~\ref{nuadef} with $\epsilon_{\nu\rm{,ff}} \propto \rho_{\rm{ls}}^2 T_{\rm{obs,ls}}^{-1/2}$, ${\rho_{\rm{ls}} \propto t^{-1} [1+\ln(t/t_{\rm{bo}})]^{(1+n+\mu n)/(1+n-\mu n)}}$ (FS19), and $T_{\rm{obs,ls}}$ given by equation~\ref{Tobsls}.  Thus, the enhanced rate of thermalization affects not only the high-frequency emission, but the low-frequency emission as well. 

Meanwhile, the condition $T_{\rm{obs,ls}} \xi_{\rm{ls}}^2 = T_{\rm BB,ls} \eta_{\rm{ls}}^2$ implies, from equations~\ref{etals} and~\ref{Tobsls}, that
\begin{equation}
\label{TBBls}
T_{\rm BB,ls} = T_{\rm{BB,bo}}(t/t_{\rm{bo}})^{-1/3} [1+\ln(t/t_{\rm{bo}})]^{\frac{4+3n-2\mu n}{12(1+n-\mu n)}}.
\end{equation}
Although $T_{\rm BB,ls}$ starts out with a lower value than $T_{\rm{obs,bo}}$ when ${\eta_{\rm{bo}} > 1}$, it declines much more slowly. The two temperatures eventually become equal at $t_{\rm{eq,ls}}$.  It is  convenient to define the equilibrium temperature $T_{\rm{eq,ls}} = T_{\rm{BB}}(t_{\rm{eq,ls}})  = T_{\rm{obs}}(t_{\rm{eq,ls}})$, which provided that equilibrium is achieved during the planar phase, is given by
\begin{align}
    \label{Teqls}
    T_{\rm{eq,ls}} &= T_{\rm{BB,bo}} \left(\dfrac{t_{\rm{eq,ls}}}{t_{\rm{bo}}}\right)^{-1/3} [1+\ln(t_{\rm{eq,ls}}/t_{\rm{bo}})]^{\frac{4+3n-2\mu n}{12(1+n-\mu n)}} \nonumber \\
    & = T_{\rm{obs,bo}} \xi_{\rm{bo}}^2 \left(\dfrac{t_{\rm{eq,ls}}}{t_{\rm{bo}}}\right)^{-2/3} [1+\ln(t_{\rm{eq,ls}}/t_{\rm{bo}})]^{-\frac{4+6n+16\mu n}{3(1+n-\mu n)}},
\end{align}
where in the second equality we used the fact that $\xi_{\rm{ls}} \rightarrow 1$ as $t\rightarrow t_{\rm{eq,ls}}$.  The equivalence of these two expressions can be verified by directly substituting equation~\ref{teq} and applying the condition $\xi_{\rm{bo}}^2 T_{\rm{obs,bo}} = \eta_{\rm{bo}}^2 T_{\rm{BB,bo}}$.  As discussed above, $h\nu_{\rm{a}}/k$ also approaches $T_{\rm{eq,ls}}$ at $t_{\rm{eq,ls}}$, up to an order-unity factor.

\begin{figure*}
\centering
\includegraphics[width=\textwidth]{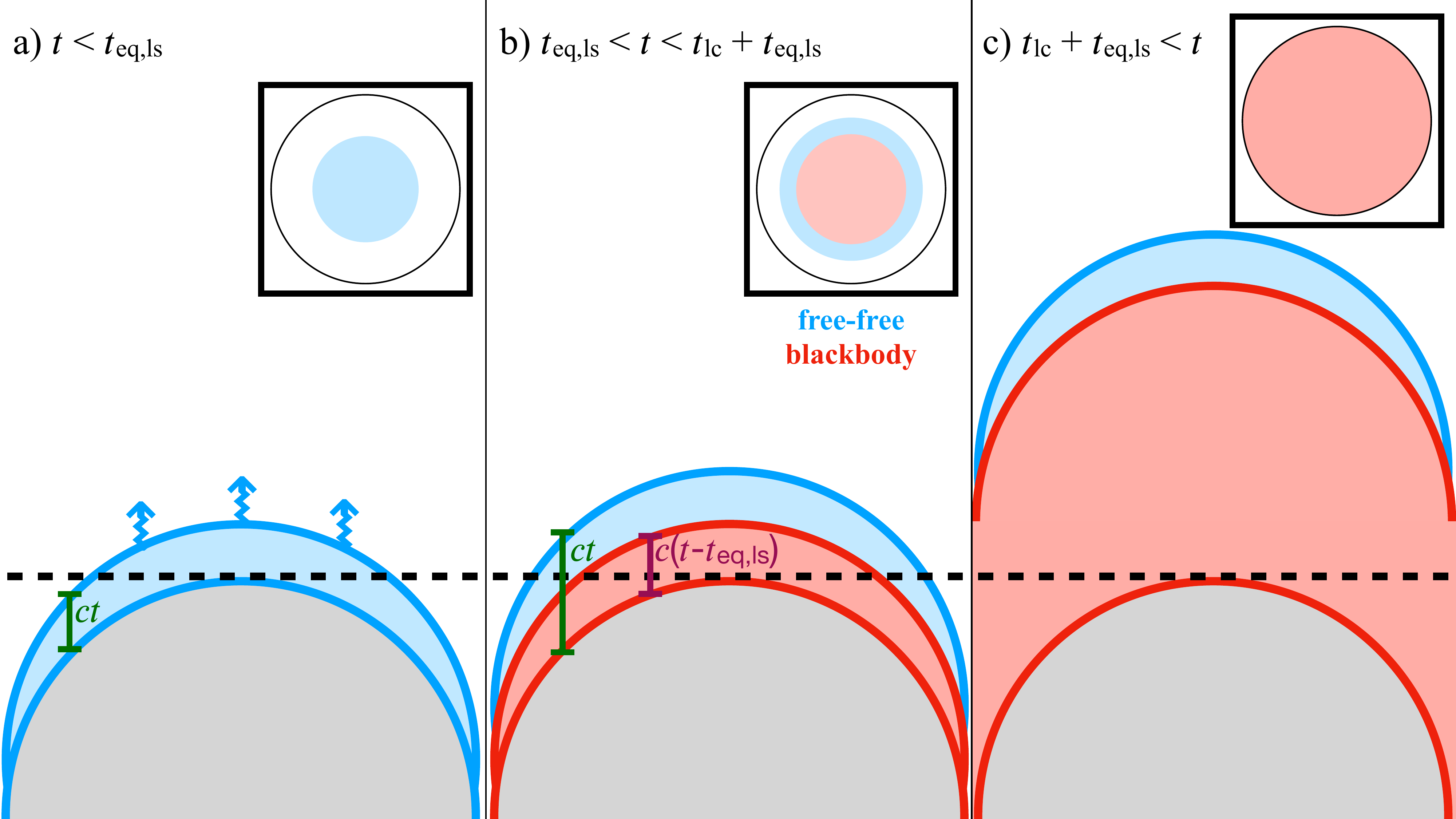}
    \caption{An illustration of the system's geometry when $t_{\rm bo} < t_{\rm eq,ls} < t_{\rm lc}$.  A source of size $R$ produces a flash of free-free emission lasting for a time $t_{\rm eq,ls}$ (shown in blue), and subsequently radiates as a blackbody (shown in red). Each panel shows where the radiation has reached at a different time.  The black dashed line indicates the slice probed by the observer at time $t$.  The inset shows the emission regions that an observer who could resolve the source would see. \textit{Panel a):} For times $t < t_{\rm eq,ls}$, the source produces free-free emission.  The leading edge of the radiation has travelled a distance $ct$, as indicated by the dark green bar.  Only a small part of the surface, which is within a distance of $ct$ from the dashed line (corresponding to angles within $\theta \la (2t/t_{\rm lc})^{1/2}$ of the line of sight), can be observed. \textit{Panel b):} For times $t>t_{\rm eq,ls}$, the surface is radiating as a blackbody.  The blackbody emission has travelled a distance of $c(t-t_{\rm eq,ls})$, as shown by the purple bar.  Only the blackbody radiation emitted within a distance of $c(t-t_{\rm eq,ls})$ from the dashed line (or within an angle of $\theta \la [2(t-t_{\rm eq,ls})/t_{\rm lc}]^{1/2}$) has had time to reach the observer.  For $t<t_{\rm lc}$, free-free emission is observed from a ring, with $[2(t-t_{\rm eq,ls})/t_{\rm lc}]^{1/2} \la \theta \la (2t/t_{\rm lc})^{1/2}$, while emission from near the equator, which has travelled a distance of $ct < R$, still has not arrived.  The free-free emission produced at the equator finally arrives at $t=t_{\rm lc}$ and it lasts for a time $t_{\rm eq,ls}$ after that.  \textit{Panel c):} At $t=t_{\rm lc}+t_{\rm eq,ls}$, the blackbody emission from the equator finally arrives.  From this point on, the entire source is visible and only blackbody radiation is observed. }  
    \label{fig:cartoon}
\end{figure*}

As a final remark, it is often desirable to replace the complicated logarithmic dependence in equations~\ref{etals} and~\ref{Tobsls}--\ref{Teqls} with a power-law approximation which is easier to work with.  Focusing on the directly observable quantities $\nu_{\rm{a}}$ and $T_{\rm{obs}}$, the approximate power-law behaviour between $t_{\rm{bo}}$ and $t_{\rm{eq,ls}}$ can be expressed as 
\begin{align}
\label{powerlaws}
T_{\rm{obs}} &\propto t^{-\alpha} \nonumber \\
\nu_{\rm{a}} &\propto t^\beta,
\end{align}
where
\begin{equation} 
\label{Tpowerlaw}
 \alpha = -\dfrac{\ln\left(T_{\rm{eq,ls}}/T_{\rm{obs,bo}}\right)}{\ln\left(t_{\rm{eq,ls}}/t_{\rm{bo}}\right)}
\end{equation}
and 
\begin{equation}
\label{nuapowerlaw}
\beta = \dfrac{\ln\left[(\uppi^4 q g_{\rm{ff}}/45 g_{\rm{B}})^{1/2} (T_{\rm{eq,ls}}/h \nu_{\rm{a,bo}})\right]}{\ln\left(t_{\rm{eq,ls}}/t_{\rm{bo}}\right)}.
\end{equation}
When $\xi_{\rm{bo}}\approx 1$ and the logarithmic correction is ignored in equation~\ref{Teqls}, we obtain $\alpha=2/3$ from equation~\ref{Tpowerlaw}, implying $T_{\rm{obs}}\propto t^{-2/3}$ as in NS10.  Including Comptonization reduces $\alpha$, while including the logarithmic correction increases it.  By manipulating equations~\ref{Tpowerlaw} and~\ref{nuapowerlaw} using equations~\ref{etanuaTobs} and~\ref{Tobsls}--\ref{Teqls}, we find that $\alpha$ and $\beta$ are related by
\begin{equation}
\label{betavsalpha}
    \beta = \lambda(\alpha-2/3) + (2\lambda -3/2) \dfrac{\ln \xi_{\rm{bo}}}{\ln(t_{\rm{eq,ls}}/t_{\rm{bo}})},
\end{equation}
where
\begin{equation}
\label{lambda}
    \lambda = \dfrac{3(4+5n+10\mu n)}{2(4+6n+16\mu n)} \approx \dfrac{3(1+1.73n)}{2(1+2.26n)}.
\end{equation}
The value of $\lambda$ ranges from 1.5 at $n=0$, to 1.14 as $n\rightarrow \infty$; for the commonly used values $n=1.5$ and $n=3$, $\lambda \approx 1.2$.  Furthermore, we find numerically that the logarithmic term in equation~\ref{betavsalpha} is not very important, being at most about 0.3 over a wide range of parameters. The fundamental reason for this is that $t_{\rm eq,ls}/t_{\rm bo}$ and $\xi_{\rm bo}$ both depend mainly on $v_{\rm bo}$ (we discuss how to calculate $\xi_{\rm{bo}}$ in Section~\ref{sec:comptonization}), but $t_{\rm eq,ls}/t_{\rm bo}$ grows faster than $\xi_{\rm bo}$ as $v_{\rm bo}$ is increased, which keeps the correction term small.  Thus, to a reasonable approximation, 
\begin{equation}
    \label{betaapprox}
    \beta \approx 1.2 (\alpha-2/3)
\end{equation} 
for typical parameters.  This rough closure relation could be used to check the consistency of the non-equilibrium breakout scenario, if the evolution of both $T_{\rm{obs}}$ and $\nu_{\rm{a}}$ can be measured in a future shock breakout event.

\subsection{An interesting new regime}
\label{sec:blendedspectrum}
With the results of Sections~\ref{sec:nonequilibrium} and~\ref{sec:thermalization}, we are now equipped to address the question posed at the beginning of this section.  Since we have seen that shock breakout either produces blackbody or free-free emission, we can rephrase the question as: do we ever expect to see both of these at the same time?  Let us consider what would be required for this to happen. 

First of all, as we saw in Section~\ref{sec:thermalization},  in order to observe free-free emission, we require that $t_{\rm{eq,ls}} > t_{\rm{bo}}$ so that the gas and radiation are initially out of equilibrium.  We also saw that in this case, blackbody emission is only produced at times $t>t_{\rm{eq,ls}}$.  Meanwhile, as discussed in Section~\ref{sec:nonequilibrium}, the observed duration of the breakout flash is ${t_{\rm{obs}} = \max(t_{\rm{bo}},t_{\rm{lc}})}$.  Therefore, in order to see blackbody emission while the breakout is ongoing, we require $
t_{\rm{eq,ls}} < \max(t_{\rm{bo}},t_{\rm{lc}}).$
Clearly, this condition contradicts the one above if the right hand side equals $t_{\rm{bo}}$, so to avoid this we must take $t_{\rm{lc}} > t_{\rm{bo}}$.  The condition for seeing blackbody emission during the breakout then becomes $t_{\rm{eq,ls}} < t_{\rm{lc}}$.  Finally, combining the conditions for seeing blackbody and free-free emission, we have 
\begin{equation}
\label{twocomponent}
    t_{\rm{bo}} < t_{\rm{eq,ls}} < t_{\rm{lc}}.
\end{equation}
Note that our assumption that the breakout occurs near the edge of the medium (i.e., the breakout layer's thickness is $v_{\rm bo} t_{\rm bo} < R_{\rm bo}$) is required in order for equation~\ref{twocomponent} to hold.  Otherwise, the diffusion time is $t_{\rm bo} \sim R_{\rm bo}/v_{\rm bo} > t_{\rm lc}$ \citep[e.g.,][]{ci11,khatami}.  The physical conditions necessary to satisfy equation~\ref{twocomponent} will be discussed further in Section~\ref{sec:conditions}. 

Assuming that equation~\ref{twocomponent} holds, how would the observed emission be affected?  Let us first clarify how the spectrum is impacted by light travel time.  As shown in equations~\ref{etals}--\ref{TBBls}, the quantities shaping the spectrum evolve on a characteristic time-scale of $t_{\rm bo}$.  If the light-crossing time were negligible (i.e., if $t_{\rm lc} \ll t_{\rm bo})$, then the observed spectrum would be a blackbody or free-free spectrum described by a single characteristic temperature, $T_{\rm obs,ls}(t)$.  The effect of a non-negligible light-crossing time (i.e., $t_{\rm lc} \ga t_{\rm bo}$) is to smear this spectrum over a range of temperatures (for further details, see Section~\ref{sec:observedspectrum}).  This occurs because, at a given observer time, the emission arriving from close to the line of sight originates from deeper layers of the ejecta compared to the emission from higher latitudes, which arrives with a delay.  Since the deeper layers are also cooler, we expect to simultaneously see cooler emission coming from near the line of sight, and hotter emission coming from higher latitudes. 

Now, suppose we have a breakout which satisfies equation~\ref{twocomponent}.  An illustration of this scenario at several representative times is given in Fig.~\ref{fig:cartoon}. The emission located at an angle $\theta$ from the line of sight arrives with a time delay of 
\begin{equation}
    \label{tdel}
    t_{\rm{del}}(\theta) = t_{\rm{lc}}(1-\cos\theta) \approx t_{\rm{lc}}\theta^2/2,
\end{equation}
so at a given time $t<t_{\rm{lc}}$, only emission from within $\theta \la \sqrt{2t/t_{\rm{lc}}}$ has had enough time to reach the observer. Initially, at times $t<t_{\rm{eq,ls}}$, we would see emission only from the outermost layers of the ejecta, where the gas and radiation are not in equilibrium (as in the left panel of Fig.~\ref{fig:cartoon}).  The expected spectrum would be a Comptonized free-free spectrum, although smeared by light-crossing effects as discussed above (see also Section~\ref{sec:model}).  However, at time $t=t_{\rm{eq,ls}}$, ejecta which are in thermal equilibrium would start to be revealed along the line of sight, and a blackbody component would emerge in the spectrum.  Then, for times $t_{\rm{eq,ls}} < t < t_{\rm{lc}}$, the emitting material within $\theta \la \sqrt{2(t-t_{\rm{eq,ls}})/t_{\rm{lc}}}$ of the line of sight would have already reached thermal equilibrium, while the emitting material beyond that would still be out of equilibrium, resulting in an observed spectrum with both blackbody and free-free components, as in the central panel of Fig.~\ref{fig:cartoon}.  Finally, at times $t>t_{\rm{lc}} + t_{\rm{eq,ls}}$, all of the observed material would be in equilibrium, as in the right panel of Fig.~\ref{fig:cartoon}.  The free-free component would fade, and the observed spectrum would be a blackbody.  

Thus, the answer to our question is yes, it is possible to see both blackbody and free-free emission at the same time in shock breakout, for times $t_{\rm{eq,ls}} \la t \la t_{\rm{lc}} + t_{\rm{eq,ls}}$, provided that equation~\ref{twocomponent} can be satisfied.  For convenience, we will refer to this type of non-standard shock breakout as `initially non-equilibrium rapid thermalization breakout,' or INERT breakout for short.  The INERT case is distinguished from the usual non-equilibrium case by a thermalization time-scale which is short compared to the light-crossing time.  To highlight this distinction, we will refer to the standard non-equilibrium breakout scenario, which produces only free-free emission during the breakout flash, as `non-equilibrium slow thermalization breakout,' or NEST breakout.  As for the classical case where equilibrium holds from the beginning, we simply call this `thermal breakout.'   

\section{The shock breakout parameter space}
\label{sec:conditions}

The results of  Section~\ref{sec:basicidea} suggest that the observed spectrum exhibits interesting behaviour when equation~\ref{twocomponent} is satisfied.  We now wish to consider more carefully the physical conditions under which this occurs.  To do this, it will be convenient to introduce two dimensionless parameters,
\begin{equation}
    \label{chidef}
    \chi = t_{\rm{lc}}/t_{\rm{bo}} \approx 2.0 \,(n+1)^{-1} R_{\rm bo,14} \rho_{\rm{bo,}11} v_{\rm{bo,}-1}^2
\end{equation}
and
\begin{equation}
    \label{zetadef}
    \zeta = t_{\rm{eq,ls}}/\max{(t_{\rm{lc}},t_{\rm{bo}})}.
\end{equation}
The parameter $\chi$ encodes whether the breakout duration is set by the light-crossing time ($\chi>1$) or by the diffusion time of the breakout shell ($\chi <1$), while the parameter $\zeta$ expresses the thermal equilibrium time-scale as a fraction of the duration of the breakout.  With these definitions, in order to obtain an INERT breakout spectrum we require $\zeta < 1$ and $\chi> 1$ while $\eta_{\rm{bo}} >1$. 

Our aim in this section is to determine how the dimensionless parameters $\eta_{\rm{bo}}$, $\chi$, and $\zeta$ vary throughout the shock breakout parameter space.  Throughout this study, we fix the values of $\mu =0.19$ and $\kappa = 0.2\,\rm{cm}^2\,\rm{g}^{-1}$.\footnote{In stars which retain a significant hydrogen envelope, $\kappa = 0.34\,\rm{cm}^2\,\rm{g}^{-1}$ is more suitable, but this difference does not significantly affect our conclusions.  However, if there are significant opacity sources other than electron scattering, the analysis presented here is not valid.}  For simplicity, we also take $g_{\rm ff} \approx g_{\rm B} \approx 1$ for the remainder of the paper.  In this case, the behaviour is determined by four independent parameters.  We will first consider the general case in Section~\ref{sec:parameterspace}, taking $v_{\rm{bo}}$, $\rho_{\rm{bo}}$, $R_{\rm bo}$, and $n$ as the four parameters.  The motivation for choosing $\rho_{\rm{bo}}$ and $v_{\rm{bo}}$ is that in principle, if sufficiently early spectra were available, they could be constrained directly from observations.  Additionally, this parametrization is convenient because $\eta_{\rm{bo}}$ effectively depends on only one parameter, $v_{\rm{bo}}$, with only a very weak dependence on $\rho_{\rm{bo}}$ and no dependence on $R_{\rm bo}$ or $n$. 
After studying the general case, we will specify to the case of breakout from an extended envelope in Section~\ref{sec:envelopebreakout}, and consider how the breakout depends on the envelope's mass and radius, and the deposited energy.  

\subsection{The general case}
\label{sec:parameterspace}
\begin{figure}
\includegraphics[width=\columnwidth]{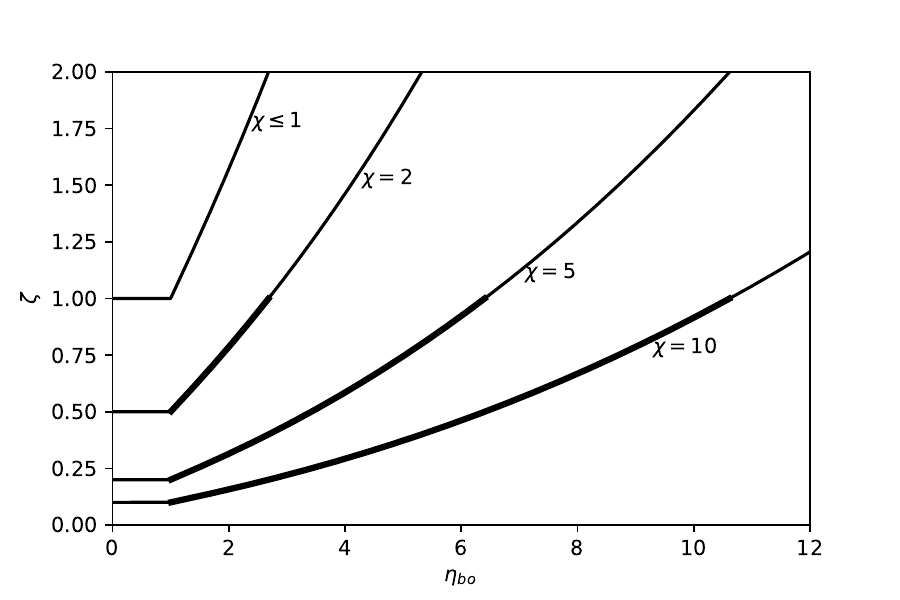}
\caption{$\zeta$ versus $\eta_{\rm{bo}}$ for $\chi = (1,2,5,10)$. When $\chi \le 1$, $\zeta$ is independent of $\chi$.  The region where the conditions $\eta_{\rm{bo}} > 1$ and $\zeta < 1$ are satisfied is shown by a heavy line. }
\label{fig:zeta}
\end{figure}

The dependence of $\eta_{\rm{bo}}$ and $\chi$ on the breakout parameters is already given in equations~\ref{etabo} and~\ref{chidef}.  Let us now obtain an explicit expression relating $\zeta$ to $\chi$ and $\eta_{\rm{bo}}$.  Applying equation~\ref{chidef} and~\ref{zetadef} to equation~\ref{teq}, we obtain the relations
\begin{align}
    \label{zetachieta}
    \zeta\max(1,\chi)[1+\ln(\zeta\max(1,\chi))]^{6p_\eta} = \eta_{\rm{bo}}^6 &, \eta_{\rm{bo}} >1 \nonumber\\
    \zeta \max(1,\chi) = 1&, \eta_{\rm{bo}} <1 
\end{align}
where in the second equation we used the definition introduced in Section~\ref{sec:thermalization}, $t_{\rm{eq,ls}} = t_{\rm{bo}}$ for $\eta_{\rm{bo}} <1$.  By making use of the Lambert $W$-function defined by $W(xe^x)=x$, we can write an explicit expression for $\zeta$ which is valid for any $\chi$ and $\eta_{\rm{bo}}$:
\begin{equation}
    \label{zeta}
    \zeta = \min(1,\chi^{-1})\max\left\{1,\exp\left[6p_\eta W\left(\frac{(e \eta_{\rm{bo}}^6)^{1/6p_\eta}}{6p_\eta}\right)-1\right]\right\}.
\end{equation}
Note that the exponential term is 1 when $\eta_{\rm{bo}} = 1$.  In Fig.~\ref{fig:zeta}, we plot $\zeta$ as a function of $\eta_{\rm{bo}}$ for several choices of $\chi$.

A few important relationships are apparent from studying equations~\ref{zetachieta} and~\ref{zeta} along with Fig.~\ref{fig:zeta}.  First, we see that if any two of $\zeta$,  $\chi$, and $\eta_{\rm{bo}}$ are equal to 1, then the third must necessarily also be 1.  Second, we note that if $\chi \le 1$, then $\zeta \ge 1$ always holds.  This captures our earlier conclusion (see Section~\ref{sec:nonequilibrium}) that an INERT breakout spectrum (which requires $\zeta < 1$) is not possible when $t_{\rm{bo}}\ge t_{\rm{lc}}$ (i.e., when $\chi \le 1$).  Finally, we point out that $\zeta$ strictly increases or stays constant with increasing $\eta_{\rm{bo}}$ (i.e., $\mathrm{d}\zeta/\mathrm{d}\eta_{\rm{bo}} \ge 0$), and when $\eta_{\rm{bo}}=1$ and $\chi > 1$, we have $\zeta = \chi^{-1} < 1$.  This means that for any given $\chi >1$, the necessary conditions for an INERT breakout spectrum ($\zeta <1$, $\chi > 1$, $\eta_{\rm{bo}} >1$) are \textit{guaranteed to be met} for $\eta_{\rm{bo}}$ in the range $1<\eta_{\rm{bo}} <\eta_{\rm{bo,}\zeta=1}(\chi)$, where
\begin{equation}
    \label{eta1}
    \eta_{\rm{bo,}\zeta=1}(\chi) = [1+\ln(\chi)]^{p_\eta} \chi^{1/6}.
\end{equation}
We illustrate this feature in Fig.~\ref{fig:zeta}
with a heavier line.  Note that for $\chi >1$, the value of $\eta_{\rm{bo,}\zeta=1}$ is always greater than 1.  The value increases with $\chi$, so that the larger $\chi$ becomes, the easier it is to obtain an INERT spectrum.  This makes sense because increasing $\chi$ requires either increasing $t_{\rm{lc}}$, or decreasing $t_{\rm{bo}}$ (which decreases $t_{\rm{eq,ls}}$ in turn), and either of these options makes the condition $t_{\rm{eq,ls}} < t_{\rm{lc}}$ easier to satisfy.

These relationships between $\chi$, $\zeta$, and $\eta_{\rm{bo}}$ also have a geometric interpretation.  Let the value of $n$ be fixed, and consider the 3-dimensional parameter space $(\rho_{\rm{bo}}, v_{\rm{bo}}, R_{\rm bo})$.  In this space, the equations $\zeta=1$, $\chi=1$, and $\eta_{\rm{bo}} = 1$ each define 2-dimensional surfaces.  We have shown that if two of these equations are satisfied, then the third is also satisfied automatically; this means that all three surfaces meet in a single 1-dimensional curve.  In log-space, the $\chi=1$ and $\eta_{\rm{bo}} = 1$ surfaces are planes, so their curve of intersection is simply a line, which has the equation 
\begin{align}
    \label{line}
    \log_{10} R_{\rm bo,14} &= -\frac{16}{15}\left[\log_{10} \rho_{\rm{bo,}-11} -0.39 -0.5\log_{10}\left(q(1+n)^{7/8}\right)\right] \nonumber\\
    &= -32\left[\log v_{\rm{bo,}-1} + 0.34 +0.25\log_{10}\left(q(1+n)^{-9/8}\right)\right] .
\end{align}  
The $\zeta=1$ surface is curved and more complicated, but it still passes through the same line.  Now, since $\zeta < 1$ is not possible if $\chi \le 1$ (see Fig.~\ref{fig:zeta}), the $\zeta=1$ surface is only defined on one side of the $\chi=1$ plane, where $\chi > 1$.  Furthermore,  for $\chi > 1$, we have shown that when $\zeta = 1$, $\eta_{\rm{bo}} = \eta_{\rm{bo,}\zeta=1} > 1$ via equation~\ref{eta1}.  This means that the $\zeta = 1$ surface also lies entirely on the $\eta_{\rm{bo}} > 1$ side of the $\eta_{\rm{bo}} = 1$ plane.  So, the $\chi=1$ and $\eta_{\rm bo}=1$ planes divide the space into four regions, and the $\zeta=1$ surface further subdivides the region where $\chi > 1$ and $\eta_{\rm bo} >1$ into two subregions where $\zeta < 1$ or $\zeta >1$, creating five regions in total.  The wedge-shaped region between the $\eta_{\rm{bo}} = 1$ plane and the $\zeta = 1$ surface, on the $\chi > 1$ side of the $\chi = 1$ plane, is where an INERT breakout is achieved.

To summarize, the 3-dimensional parameter space can be broken into the following 5 distinct regions with different spectral properties (for more about the spectra in each case, see Section~\ref{sec:model}, as well as Paper II):
\begin{enumerate}[I.]
\item $\chi > 1$, $\eta_{\rm{bo}} < 1$, and $\zeta < 1$:  The breakout is in equilibrium, with the duration set by the light-crossing time, and the spectrum is a blackbody.
\item $\chi > 1$, $\eta_{\rm{bo}} > 1$, and $\zeta < 1$:  The breakout is out of equilibrium, with the duration set by the light-crossing time, and the spectrum simultaneously exhibits both blackbody and free-free emission components.
\item $\chi > 1$, $\eta_{\rm{bo}} > 1$, and $\zeta > 1$:  The breakout is out of equilibrium, with the duration set by the light-crossing time, and the spectrum is a Comptonized free-free spectrum. 
\item $\chi < 1$, $\eta_{\rm{bo}} < 1$, and $\zeta = 1$:  The breakout is in equilibrium, with the duration set by the diffusion time of the breakout layer, and the spectrum is a blackbody.
\item $\chi < 1$, $\eta_{\rm{bo}} > 1$, and $\zeta > 1$:  The breakout is out of equilibrium, with the duration set by the diffusion time of the breakout layer, and the spectrum is a Comptonized free-free spectrum.
\end{enumerate}
All 5 regions meet along the line give by equation~\ref{line}, since ${\zeta=\chi=\eta_{\rm{bo}} = 1}$ there. Regions I and IV (thermal breakout), and III and V (NEST breakout), were realized before and have been discussed by previous authors \citep[e.g., NS10, FS19,][]{katz}. However, region II, which represents the INERT breakout case, has not been explored until now.

\begin{figure*}
    \begin{center}    \includegraphics[width=\textwidth]{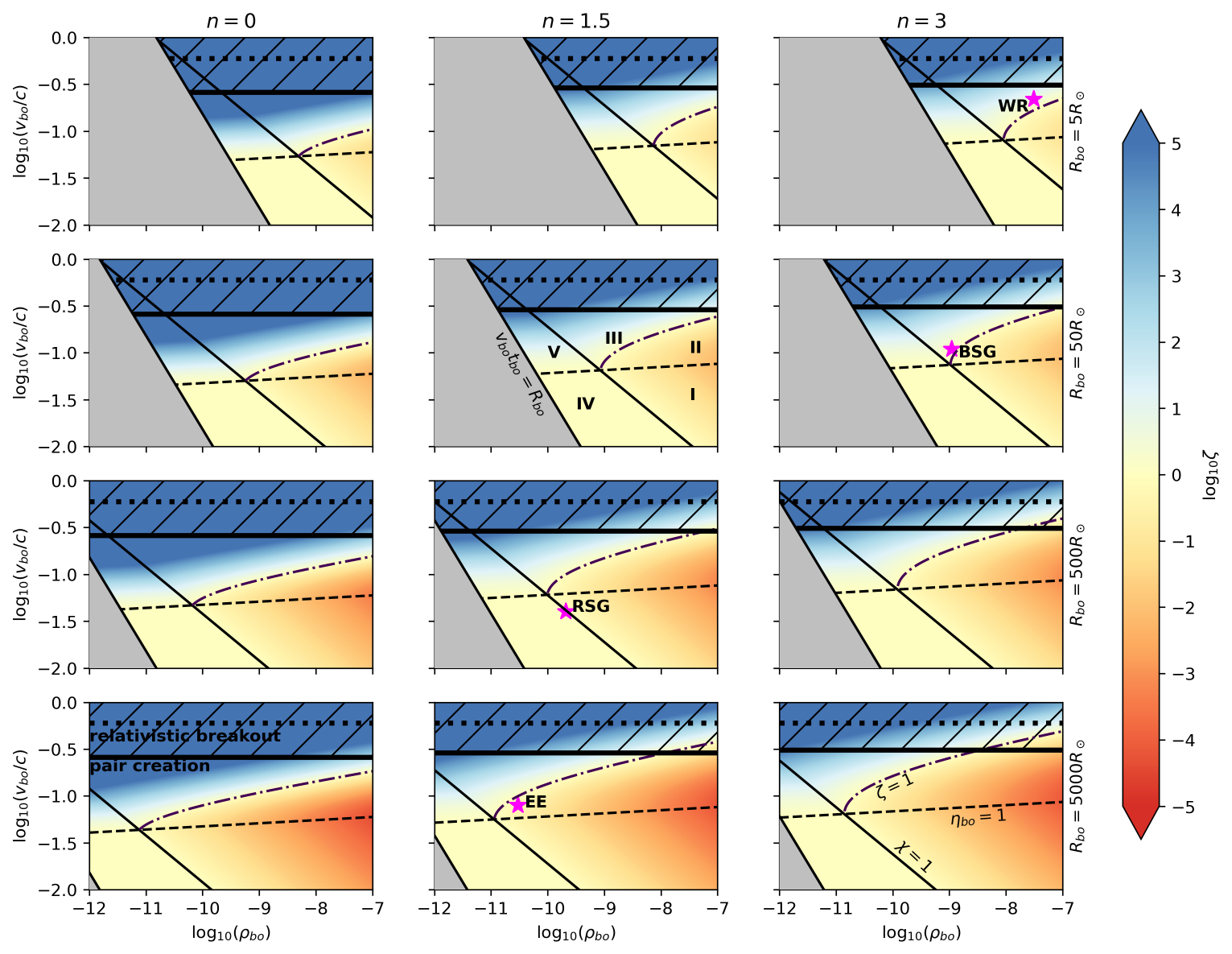}
    \end{center}
    \caption{The $v_{\rm{bo}}$ versus $\rho_{\rm{bo}}$ parameter space for $n = (0,1.5,3)$ (columns, left to right) and $R_{\rm bo} = (5,50,500,5000)\,\mathrm{R}_\odot$ (rows, top to bottom). All panels have the same limits for both axes.  Colour indicates the value of $\zeta$ (red for $\zeta \ll 1$, yellow for $\zeta \sim 1$, and blue for $\zeta \gg 1$; values greater than $10^5$ or less than $10^{-5}$ are solid blue or red, respectively).  Lines indicate the conditions $\chi = 1$ (solid), which separates diffusion-dominated and light-travel dominated breakouts; $\eta_{\rm{bo}} = 1$ (dashed), which separates thermal and non-thermal breakouts; and $\zeta = 1$ (dot-dashed), which determines whether or not thermalization occurs within a light-crossing time. The pair creation threshold $kT_{\rm{obs,bo}} \ga 50\,$keV, as well as the relativistic shock breakout condition $v_{\rm{bo}} \ga 0.6 c$ suggested by \citet{katz} and \citet{fs2}, are also shown as heavy solid and dotted lines, respectively (as labelled in the bottom left panel).  The diagonal hatching indicates where our model breaks down due to pair creation.  In the grey region, the breakout takes place far from the edge of the system and our assumption of $R_{\rm bo} \approx R$ is invalid.  The line $v_{\rm bo} t_{\rm bo} = R_{\rm bo}$, which marks the boundary of the gray region, is labelled in the $n=1.5$ and $R = 50\,\mathrm{R}_\odot$ panel. In that same panel, we have also labelled the regions I--V corresponding to the five types of breakout discussed in the text.  The magenta stars show estimated values for several progenitors: a red supergiant (RSG), a blue supergiant (BSG), a Wolf-Rayet star (WR), and a breakout in an extended convective envelope (EE).  All results assume a value of $q=1$.}
    \label{fig:rhovplot}
\end{figure*}

To give an idea of the general behaviour, we plot several 2D slices of the 4D $(n,R_{\rm bo},\rho_{\rm{bo}},v_{\rm{bo}})$ parameter space in Fig.~\ref{fig:rhovplot}.  In each panel, $\rho_{\rm{bo}}$ varies from $10^{-12}$\,g\,cm$^{-3}$ to $10^{-7}$\,g\,cm$^{-3}$; $v_{\rm{bo}}$ varies from $0.01\,c$ to $0.2\,c$; $n$ is fixed to 0, 1.5, or 3; and $R_{\rm bo}$ is fixed to 5, 50, 500, or 5000$\mathrm{R}_\odot$.  The reason for considering the non-standard $n=0$ case will become clear later, in Section~\ref{sec:n0}.  As noted in Section~\ref{sec:nonequilibrium}, for temperatures $kT_{\rm{obs,bo}} \ga 50\,$keV, pair production becomes important \citep[e.g., NS10, FS19,][]{katz,fs2}, and our model, which neglects pairs, breaks down. For this reason, our model is invalid within the hatched region in the figure, which roughly corresponds to $v_{\rm bo} \ga 0.3\,c$ (the relation between $v_{\rm bo}$ and $T_{\rm obs,bo}$ is derived in Section~\ref{sec:comptonization}). In addition, as discussed in Section~\ref{sec:nonequilibrium}, our model only applies when $v_{\rm bo} t_{\rm bo} < R_{\rm bo}$.  Since ${t_{\rm bo} \propto (\rho_{\rm bo} v_{\rm bo}^2)^{-1}}$ according to equation~\ref{tbo}, this condition becomes relevant for low densities, slow velocities, or small radii, as shown by the grey region in the figure.

We highlight four panels in particular: $R_{\rm bo} = 5 \mathrm{R}_\odot$ and $n=3$, corresponding to a typical Wolf-Rayet star; $R_{\rm bo} = 50 \mathrm{R}_\odot$ and $n=3$, corresponding to a typical blue supergiant; $R_{\rm bo} = 500 \mathrm{R}_\odot$ and $n=1.5$, corresponding to a typical red supergiant; and $R_{\rm bo} = 5000 \mathrm{R}_\odot$ and $n=1.5$, corresponding to a star with an extended envelope supported by convection.  In each of these cases we mark the typical expected values of $\rho_{\rm{bo}}$ and $v_{\rm{bo}}$ with a star (for the extended envelope case, the typical values are less clear, but we adopt $\rho_{\rm{bo}}= 3\times10^{-11}\,\text{g}\,\text{cm}^{-3}$ and $v_{\rm{bo}} = 0.08\,c$ as an illustrative example).  

We see that for red supergiants, $\eta_{\rm{bo}} \ll 1$, $\chi \approx 1$, and $\zeta \approx 1$.  As expected, we find that a red supergiant shock breakout is situated near the boundary of regions I and IV, meaning that the breakout is deep in the blackbody regime, and light-crossing and diffusion effects are comparable. For Wolf-Rayet stars, $\eta_{\rm{bo}} \gg 1$, $\chi \gg 1$ and $\zeta \approx 5$, placing the breakout in region III.  Thus, a Wolf-Rayet shock breakout is deep in the free-free regime, and the duration is set by the light-crossing time rather than the diffusion time.  

So far, these results agree with previous studies (e.g, NS10, FS19).  However,  interestingly, we find that in the blue supergiant case, ${\eta_{\rm{bo}} \approx 5}$, $\chi \approx 2$, and $\zeta\approx 1.4$.  This places a blue supergiant shock breakout marginally within region III, but very close to the boundary of region II, the INERT regime.  We conclude that \textit{shock breakouts from the surface of blue supergiants might sometimes show blackbody and free-free emission simultaneously.}  The shock breakout spectrum and light curve in the blue supergiant case might therefore differ from what was calculated in previous studies like NS10 and FS19, which did not consider the possibility that the equilibrium time could be shorter than the breakout duration.  Updated modelling of blue supergiant shock breakout is beyond the scope of this paper, but will be considered in future work.

In the extended envelope case, we find that the situation is similar to the blue supergiant case: $\eta_{\rm{bo}} \approx 4$, $\chi \approx 5$ and $\zeta \approx 0.5$.  Of course, here we have deliberately tuned the value of $\rho_{\rm{bo}}$ and $v_{\rm{bo}}$ to produce a breakout in region II.  But the chosen values of $\rho_{\rm{bo}}= 3\times10^{-11}\,\text{g}\,\text{cm}^{-3}$ and $v_{\rm{bo}} = 0.08\,c$ are not so different from the typical values expected for $ll$GRB envelopes \citep[see][and Paper III for further discussion]{nakar,ic16}.  This demonstrates that shock breakout from an extended envelope could plausibly satisfy the requirements for an INERT breakout.

To wrap up this section, let us examine the regions in Fig.~\ref{fig:rhovplot} to see what conclusions can be drawn about  favourable conditions for INERT breakout.  First, in order to be in region II, $v_{\rm{bo}}$ and $\rho_{\rm{bo}}$ must be larger than some minimum values.  The minimum value occurs at the point where all five regions meet, which is located along the line described by equation~\ref{line}.  The minimum breakout velocity is $\approx 0.04\,c$, with an extremely weak dependence on the breakout radius ($v_{\rm{bo}} \propto R_{\rm bo}^{-1/32}$ via equation~\ref{line}), and only a slight dependence on $n$.  The minimum breakout density depends mostly on $R_{\rm bo}$, as $\rho_{\rm{bo,}-11} \ga 0.4 R_{\rm bo,14}^{-15/16}$, with a weak dependence on $n$.  The breakout velocity also has a maximum value, since if the shock is too fast the breakout will be be in region III instead of region II.  The maximum velocity has a non-trivial dependence on $\rho_{\rm{bo}}$, $R_{\rm bo}$, and $n$, but in general it lies somewhere in the range of $0.04$--$0.2\,c$.  We point out that a velocity estimate of $0.1\,c$ for $ll$GRBs (as mentioned in Section~\ref{sec:nonequilibrium}, see also Paper III) is consistent with the typical range in which INERT breakout is relevant.

In summary, provided that $0.04\,c \la v_{\rm{bo}} \la 0.2\,c$, the breakout will always be in region II for a sufficiently high $\rho_{\rm{bo}}$, so increasing the density is almost always favourable for obtaining an INERT breakout.  For a fixed breakout density, a larger value of $R_{\rm bo}$ is favourable because it lowers the minimum density required (however, see Section~\ref{sec:envelopebreakout}).  Increasing $n$ also seems favourable, because it increases the maximum allowed value of $v_{\rm{bo}}$ (however, again, see Section~\ref{sec:envelopebreakout}).

\subsection{Shock breakout from an extended envelope}
\label{sec:envelopebreakout}

Based on the discussion in Section~\ref{sec:parameterspace}, it seems that shock breakout in an extended low-mass envelope is a promising scenario for producing an INERT breakout. We will now consider the behaviour in that case.  We suppose that an energy $E_0$ is deposited in an envelope of mass $M_{\rm{env}}$ and radius $R_{\rm{env}}$, with a density profile $\rho \propto (R_{\rm{env}}-r)^n$ in the outer layers of the envelope.  In order to translate the results of Section~\ref{sec:parameterspace} to these parameters, we need to express $\rho_{\rm{bo}}$ and $v_{\rm{bo}}$ in terms of $E_0$, $M_{\rm{env}}$, $R_{\rm{env}}$ and $n$.  To accomplish this, we must first specify the density profile.

We consider here an envelope which extends from an inner radius $R_*$ to an outer radius $R_{\rm{env}}$, with $R_{\rm{env}} \gg R_*$.  We adopt an envelope density profile of the form
\begin{equation}
    \label{densityprofile}
    \rho(r) = \begin{cases}
        \rho_{\rm{env}} \left(\dfrac{r}{R_{\rm{env}}}\right)^{-s}, & R_* < r \ll R_{\rm{env}} \\
        \rho_{\rm{env}} \left(\dfrac{R_{\rm{env}}-r}{\Lambda R_{\rm{env}}}\right)^n, & r-R_{\rm{env}} \ll R_{\rm{env}}
    \end{cases},
\end{equation}
where $\Lambda$ is a dimensionless parameter which controls the thickness of the region where the outer density profile applies.  In general $\Lambda$ could be treated as a free parameter, but here we consider a simple case where it is a function of $n$ and $s$. By requiring that the two functions in equation~\ref{densityprofile} and their derivatives join smoothly at some transition radius $R_t$, we obtain 
\begin{equation}
\label{Lambda}
\Lambda = (n/s)(1+n/s)^{-(1+s/n)}
\end{equation}
and $R_t = R_{\rm{env}}/(1+n/s)$.  In this picture, for fixed $s$, the value of $n$ determines both the steepness and the thickness of the outer density profile.  If $n \ga s$, then the outer density profile is thick, with $\Lambda \sim 1$.  On the other hand, if $n \ll s$, the outer density profile is thin, with $\Lambda \ll 1$.  As $n/s \rightarrow 0$, $\Lambda \rightarrow 0$ and as $n/s \rightarrow \infty$, $\Lambda \rightarrow 1$.  Assuming that $s < 3$, so that most of the envelope mass is at large radii, the density $\rho_{\rm{env}}$ can be estimated by
\begin{equation}
    \label{rhoenv}
    \rho_{\rm{env}} \approx \dfrac{(3-s)M_{\rm{env}}}{4\uppi R_{\rm{env}}^3}
\end{equation}
Here, for simplicity we have approximated the envelope mass by assuming that the $\rho \propto r^{-s}$ behaviour continues out to $r = R_{\rm{env}}$.  For $s=2$ and $n\le 3$, this overestimates the true envelope mass by at most $50$ per cent.

The appropriate values of $s$ and $n$ depend on the structure of the envelope and how it was formed.  If the envelope is in hydrostatic equilibrium, with a polytropic equation of state, then $n$ will be the polytropic index, and the inner region could be approximated by a power-law with $s$ somewhere between 1 and 3.  On the other hand, if the envelope was produced by a steady wind, then $s=2$.  In this scenario, $n$ would depend on the initial ramping-up of the wind velocity and mass-loss rate, before reaching a steady-state with constant mass-loss rate.  If the wind became steady on a short time-scale compared to its age, it could have a sharp edge, corresponding to a small $n$.  Another possibly is that the envelope was produced by a stellar eruption or precursor explosion.  In this case, the expected profile has an inner region of bound material where $\rho \propto r^{-3/2}$ and an outer region of unbound ejecta with $\rho \propto r^{-10}$ or steeper \citep[e.g.,][]{mm,kuriyama,tsuna,ko}.  This situation differs from our assumptions in that the density does not go to zero at a finite radius, and the velocity in the steep region may not be well described by the plane-parallel solution of \citet{sakurai}.  However, for such a steep outer density gradient, the density changes over scales much less than $r$, so in that sense the behaviour might be similar to the $\Lambda \ll 1$ case of our model.  Thus, the eruption case might be roughly described by taking $s=3/2$ and $n \ll 1$.

Based on this discussion, we will adopt a fiducial value of $s=2$, and consider $n=0$, 1.5, and 3 as in Section~\ref{sec:parameterspace}.  The meaning of the $n=0$ case will be discussed in more detail in Sections~\ref{sec:n0} and~\ref{sec:connections}.

Now, if a large energy $E_0$ is deposited at the base of the envelope, either by the outer layers of the supernova as in \citet{ic16} or by a choked jet as in \citet{nakar}, the ensuing explosion drives a strong 
shock through the envelope (assumed here to be spherically symmetric).  In the inner density profile, the shock will behave like a blast wave,\footnote{We assume for simplicity here that the ejecta driving the shock have a density profile $\rho_{\rm ej}\propto r^{-\omega}$ with $\omega \le 5$ so that the evolution is like a Sedov-Taylor blast wave.  If $p > 5$, the velocity will evolve as $v \propto r^{(s-3)/(\omega-3)}$ instead \citep[e.g.,][]{chevalier83}.  This simplification does not significantly influence our results.} with velocity $v\propto r^{(s-3)/2} \propto \rho^{(3-s)/2s}$, while in the outer density profile, the shock will follow the self-similar solution of \citet{sakurai}, with $v \propto \rho^{-\mu}$.  Substituting equation~\ref{densityprofile} for $\rho$ in these relations, we can write the shock velocity as
\begin{equation}
    \label{velocity profile}
    v(r) \approx \begin{cases}
        v_0 \left(\dfrac{r}{R_{\rm{env}}}\right)^{(s-3)/2}, & R_* < r \ll R_{\rm{env}} \\
        v_0 \left(\dfrac{R_{\rm{env}}-r}{\Lambda R_{\rm{env}}}\right)^{-\mu n}, & R_{\rm{env}}-r \ll R_{\rm{env}}
    \end{cases},
\end{equation}
where
\begin{equation}
    \label{v0}
    v_0 \approx \dfrac{2}{5-s}\left(\dfrac{4\uppi E_0}{(3-s)M_{\rm{env}}}\right)^{1/2}
\end{equation}
is the velocity of a Sedov-Taylor blast wave with energy $E_0$ that has swept up a mass of $M_{\rm{env}}$.

Conveniently, we note that only the combinations $(3-s)M_{\rm{env}}/4\uppi$ and $[2/(5-s)]^2 E_0$ appear in equations~\ref{rhoenv} and~\ref{v0}, so if a different value of $s$ is desired, our results can simply be rescaled.  For other values of $s<3$, the implied values of $M_{\rm{env}}$ and $E_0$ can be related to the $s=2$ case via ${M_{\rm{env}} = M_{\mathrm{env,}s=2}/(3-s)}$ and ${E_0 = (5-s)^2 E_{0,s=2}/4}$.

Next, let us derive the properties of the breakout shell.  Provided that the envelope's total optical depth $\tau_*$ is larger than $c/v_*$ initially, where $v_*$ is the shock velocity at $r=R_*$, the shock will be radiation-mediated in the envelope and a shock breakout will occur once ${\tau(r) = c/v(r)}$.  Using $\tau(r) = \kappa \int_{R_{\rm{env}}-r}^{R_{\rm{env}}}\rho(r)\mathrm{d} r$, along with equations~\ref{densityprofile}--\ref{v0}, we can find an expression for the shock velocity $v_{\rm{bo}}$ at the breakout location,
\begin{align}
    \label{vbo}
    v_{\rm{bo}} & \approx v_0 (\Lambda\Psi)^{\mu n/(1+n-\mu n)}\nonumber \\
    & \approx \Lambda^{\frac{0.19n}{1+0.81n}}\left(\frac{\kappa}{c(n+1)}\right)^{\frac{0.19n}{1+0.81n}} \left(\frac{4}{9}E_0\right)^{\frac{1+n}{2(1+0.81n)}} \nonumber\\
    & \quad \times \left(\frac{M_{\rm{env}}}{4\uppi}\right)^{-\frac{1+0.62n}{2(1+0.81n)}} R_{\rm{env}}^{-\frac{0.38n}{1+0.81n}}, 
\end{align} 
where we defined
\begin{equation}
    \label{Psi}
    \Psi = \dfrac{\kappa \rho_{\rm{env}} v_0 R_{\rm{env}}}{c(n+1)}.
\end{equation}
The quantity $\Psi$ is effectively the ratio of the characteristic optical depth scale of the envelope, $\tau_{\rm{env}} = \kappa \rho_{\rm{env}} R_{\rm{env}} (n+1)^{-1}$, to $c/v_0$.  Our assumption that the shock breakout occurs near the edge of the envelope, rather than deep in its interior (i.e., $R_{\rm bo} \approx R_{\rm env}$), corresponds to $\Psi \ga 1$.  This requirement is equivalent to the condition $v_{\rm bo} t_{\rm bo} < R_{\rm bo}$ used in Sections~\ref{sec:nonequilibrium} and~\ref{sec:parameterspace} when $\Lambda \approx 1$.  However, when $n$ is small and $\Lambda \ll 1$, the situation is more subtle, as will be addressed in Section~\ref{sec:n0}.

From equation~\ref{vbo} and the relation $v_0/v_{\rm{bo}} = (\rho_{\rm{env}}/\rho_{\rm{bo}})^{-\mu}$, the breakout density is 
\begin{align}
    \label{rhobo}
    \rho_{\rm{bo}} &\approx \rho_{\rm{env}} (\Lambda \Psi)^{-n/(1+n-\mu n)}\nonumber\\
    & \approx \Lambda^{-\frac{n}{1+0.81n}}\left(\frac{\kappa}{c(n+1)}\right)^{-\frac{n}{1+0.81n}} \left(\frac{4}{9}E_0\right)^{-\frac{n}{2(1+0.81n)}} \nonumber\\
    &\quad\times\left(\frac{M_{\rm{env}}}{4\uppi}\right)^{\frac{2+0.62n}{2(1+0.81n)}} R_{\rm{env}}^{-\frac{3+0.43n}{1+0.81n}}.
\end{align}
Setting $\rho = \rho_{\rm{bo}}$ in equation~\ref{densityprofile} and solving for $R_{\rm{env}}-r$ yields the thickness of the breakout shell, $d_{\rm{bo}} = \Lambda R_{\rm{env}} (\rho_{\rm{bo}}/\rho_{\rm{env}})^{1/n}$.  Combining this with equations~\ref{vbo} and~\ref{Psi}, the dynamical time of the breakout shell $t_{\rm{bo}} = d_{\rm{bo}}/v_{\rm{bo}}$ becomes 
\begin{align}
    \label{tbo1}
    t_{\rm{bo}} & \approx \dfrac{c(n+1)}{\kappa \rho_{\rm{env}} v_0^2} (\Lambda \Psi)^{n(1-2\mu)/(1+n-\mu n)} \nonumber\\
    & \approx \Lambda^{\frac{0.62n}{1+0.81n}}\left(\frac{\kappa}{c(n+1)}\right)^{-\frac{1+0.19n}{1+0.81n}} \left(\frac{4}{9}E_0\right)^{-\frac{2+n}{2(1+0.81n)}}\nonumber\\ 
    &\quad\times\left(\frac{M_{\rm{env}}}{4\uppi}\right)^{\frac{0.62n}{2(1+0.81n)}} R_{\rm{env}}^{\frac{3+1.19n}{1+0.81n}}.
\end{align}
When equations~\ref{vbo} and~\ref{rhobo} are substituted into equation~\ref{tbo1}, we recover equation~\ref{tbo}.  The ratio of the light travel time $t_{\rm{lc}} = R_{\rm{env}}/c$ to the breakout time is then
\begin{align}
    \label{chi1}
    \chi &\approx c^{-1} \Lambda^{-\frac{0.62n}{1+0.81n}}\left(\dfrac{\kappa}{c(n+1)}\right)^{\frac{1+0.19n}{1+0.81n}} \left(\frac{4}{9}E_0\right)^{\frac{2+n}{2(1+0.81n)}}\nonumber\\
    &\quad\times\left(\dfrac{M_{\rm{env}}}{4\uppi}\right)^{-\frac{0.62n}{2(1+0.81n)}} R_{\rm{env}}^{-\frac{2(1+0.19n)}{1+0.81n}}.
\end{align}

In order to compare our results with observations, and with the model parameters used in Section~\ref{sec:model}, it will also be useful to have expressions for the breakout shell's luminosity and observed temperature. The breakout luminosity $L_{\rm{bo}}$ can be estimated by comparing the photon energy density set by diffusion,  $L_{\rm{bo}}\tau_{\rm{bo}}/4\uppi R_{\rm{env}}^2c$ (e.g., NS10), to the energy density behind the shock $\rho_{\rm{bo}}v_{\rm{bo}}^2 \approx aT_{\rm{BB,bo}}^4$. Applying equations~\ref{TBB}, \ref{vbo}, and~\ref{rhobo} leads to 
\begin{align}
    \label{Lbo1}
    L_{\rm{bo}} & \approx 4 \uppi R_{\rm{env}}^2 \rho_{\rm{env}} v_0^3 (\Lambda\Psi)^{-n(1-3\mu)/(1+n-\mu n)} \nonumber\\
    &  \approx 4\uppi \Lambda^{-\frac{0.43n}{1+0.81n}}\left(\dfrac{\kappa}{c(n+1)}\right)^{-\frac{0.43n}{1+0.81n}} \left(\frac{4}{9}E_0\right)^{\frac{3+2n}{2(1+0.81n)}}\nonumber \\
    &\quad\times\left(\dfrac{M_{\rm{env}}}{4\uppi}\right)^{-\frac{1+1.24n}{2(1+0.81n)}} R_{\rm{env}}^{-\frac{1-0.05n}{1+0.81n}} .
\end{align}

To calculate $T_{\rm{obs,bo}}$, we first need to know the blackbody temperature $T_{\rm{BB,bo}}$, which we approximate via equation~\ref{TBB}, and then we need to determine $\eta_{\rm{bo}}$.  Using equation 9 in NS10, with the number density of blackbody photons $n_{\rm{BB,bo}} \approx (a/3k)T_{\rm{BB,bo}}^3$, the time spent by photons in the breakout shell given by equation~\ref{tbo1}, and the free-free photon production rate per unit volume $\dot{n}_{\rm{ff,bo}} \approx \frac{1}{2}A_{\rm{ff}} (7 \rho_{\rm{bo}})^2 T_{\rm{BB,bo}}^{-1/2}$, where $A_{\rm{ff}} = 3.5\times 10^{36}\,\text{g}^{-2}\text{cm}^3\,\text{s}^{-1}\,\text{K}^{1/2}$, the factor of $\frac{1}{2}$ adjusts for a composition of helium instead of hydrogen, and the factor of 7 takes into account shock compression for an adiabatic index of 4/3, we finally arrive at 
\begin{align}
    \label{etabo1}
    \eta_{\rm{bo}} & \approx \dfrac{2 q \kappa a^{1/8}}{147 k A_{\rm{ff}} c(n+1)} \rho_{\rm{env}}^{-1/8} v_{0}^{15/4} (\Lambda \Psi)^{n(1+30\mu )/[8(1+n-\mu n)]} \nonumber\\
    & \approx 4.8 \times 10^{-25}\,q \Lambda^{\frac{6.70n}{8(1+0.81n)}}\left(\dfrac{\kappa}{c(n+1)}\right)^{\frac{8+13.18n}{8(1+0.81n)}} \left(\frac{4}{9}E_0\right)^{\frac{30+31n}{16(1+0.81n)}}\nonumber\\
    &\quad\times\left(\dfrac{M_{\rm{env}}}{4\uppi}\right)^{-\frac{32+19.22n}{16(1+0.81n)}} R_{\rm{env}}^{\frac{3-10.97n}{8(1+0.81n)}} ,
\end{align}
where in the second equality all quantities are expressed in cgs units.  Here, as in Section~\ref{sec:nonequilibrium}, we have introduced the dimensionless parameter $q$ to crudely account for physics not included by NS10.  For the observed temperature, using equation~\ref{TBB} and $T_{\rm{obs,bo}} \xi_{\rm{bo}}^2 \approx T_{\rm{BB,bo}} \eta_{\rm{bo}}^2$ (NS10), and then replacing $v_{\rm{bo}}$, $\rho_{\rm{bo}}$, and $\eta_{\rm{bo}}$ by applying equations~\ref{vbo}, \ref{rhobo}, and~\ref{etabo1}, we obtain
\begin{align}
    \label{Tobsbo}
    T_{\rm{obs,bo}}\xi_{\rm{bo}}^2 & \approx  \left(\dfrac{2 q \kappa}{147 A_{\rm{ff}} k c(n+1)}\right)^2 v_0^8 (\Lambda \Psi)^{8\mu n/(1+n-\mu n)} \nonumber \\
    & \approx 7.9 \times 10^{-46} q^2 \Lambda^{\frac{1.52n}{1+ 0.81n}} \left(\dfrac{\kappa}{c(n+1)}\right)^{\frac{2(1+1.57n)}{1+0.81n}} \nonumber\\
    &\quad\times\left(\frac{4}{9}E_0\right)^{\frac{4(1+n)}{1+0.81n}} \left(\dfrac{M_{\rm{env}}}{4\uppi}\right)^{-\frac{4(1+0.62n)}{1+0.81n}} R_{\rm{env}}^{-\frac{3.04n}{1+0.81n}},
\end{align}
where again all quantities are in cgs units in the second equation. Equation~\ref{Tobsbo} must be used in conjunction with a Comptonization model to solve for $T_{\rm{obs,bo}}$ and $\xi_{\rm{bo}}$ self-consistently (see Section~\ref{sec:comptonization}).  As expected, when inserting equations~\ref{vbo} and~\ref{rhobo} into equations~\ref{etabo1} and~\ref{Tobsbo}, and taking $\kappa \approx \kappa_{\rm H}/2$ for helium, where $\kappa_{\rm H} = 0.34\,\text{cm}^2\,\text{g}^{-1}$ is the electron scattering opacity for ionized hydrogen, equations~\ref{etabo1} and~\ref{Tobsbo} reduce to the simpler forms given in equations~\ref{etabo} and~\ref{Tinequality}, with $g_{\rm B}$ set to 1.  If we additionally set $q=1$ and ignore all factors of $n+1$ in equation~\ref{etabo1}, we obtain equation 10 in NS10.

\begin{figure*}
    \begin{center}    \includegraphics[width=\textwidth]{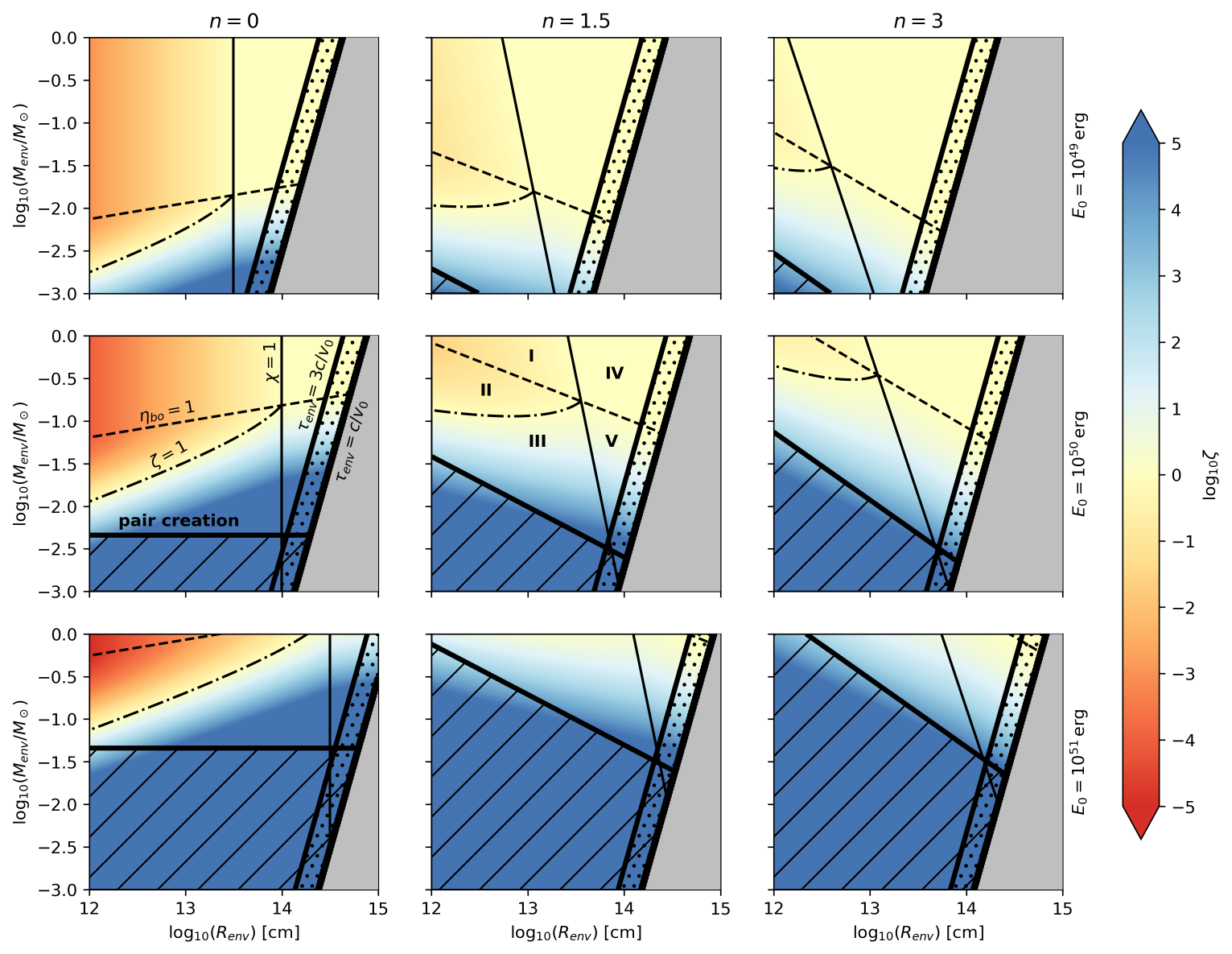}
    \end{center}
    \caption{The $M_{\rm{env}}$ versus $R_{\rm{env}}$ parameter space for $n = (0,1.5,3)$ (columns, left to right) and $E_0 = (10^{49},10^{50},10^{51})\,$erg (rows, top to bottom). All panels have the same limits for both axes.  As in Fig.~\ref{fig:rhovplot}, colour indicates the value of $\zeta$ (using the same scale), and we show the curves $\chi = 1$ (solid), $\eta_{\rm{bo}} = 1$ (dashed), and $\zeta = 1$ (dot-dashed); refer to the labels in the middle left panel.  The breakout regimes I--V are denoted in the central panel. In the gray region, $\tau_{\rm{env}}<c/v_0$ and shock breakout occurs far from the edge of the envelope, contrary to our assumptions.  Hatched regions show where our model assumptions break down, either because pair creation is relevant (diagonal hatching), or because $\tau_{\rm{env}} \sim c/v_0$ (dotted hatching).  The boundaries of these regions are also labelled in the middle left panel. All results assume a value of $q=1$.}
    \label{fig:MRplot}
\end{figure*}

\begin{figure*}
    \begin{center}    \includegraphics[width=\textwidth]{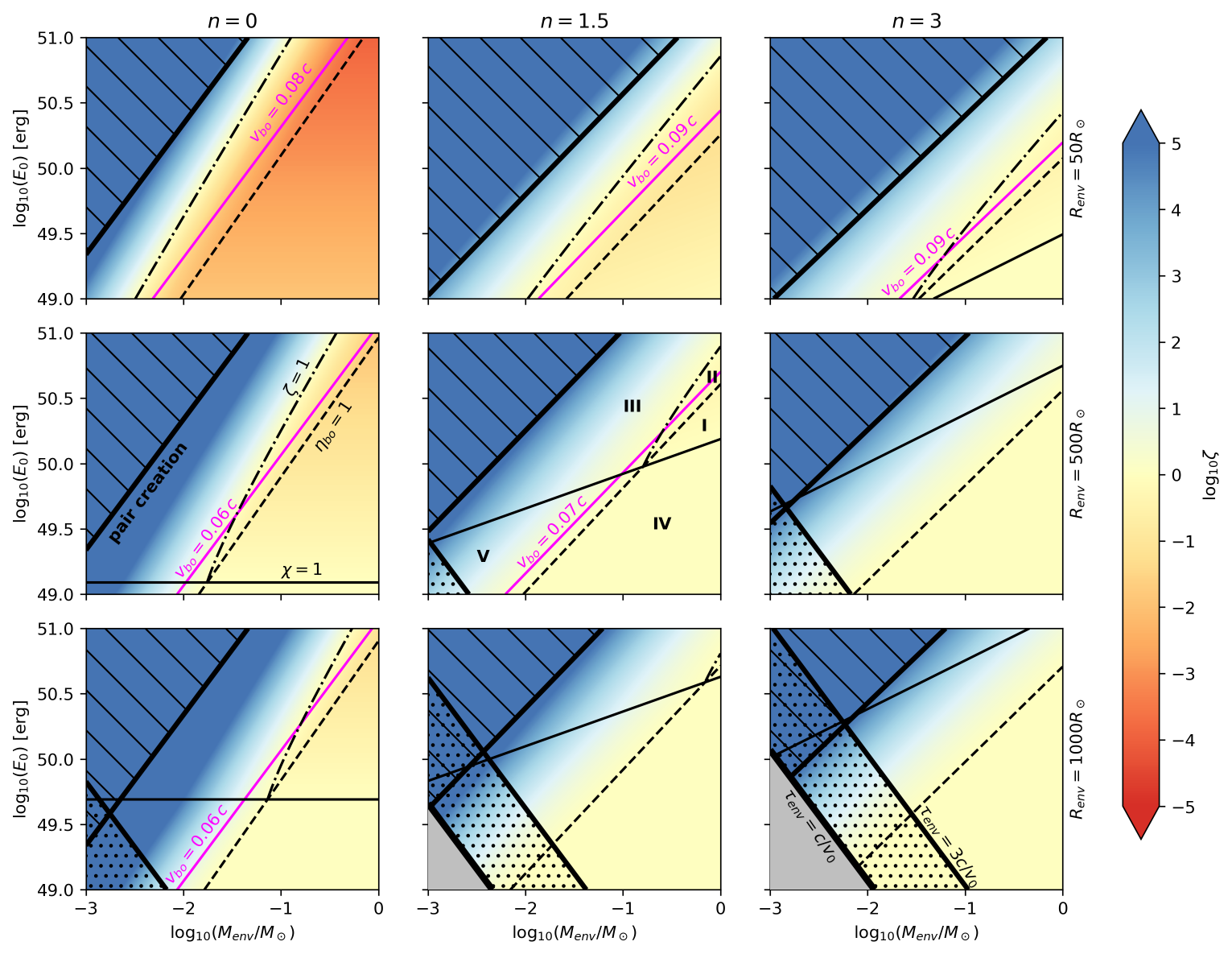}
    \end{center}
    \caption{The $E_0$ versus $M_{\rm{env}}$ parameter space for $n = (0,1.5,3)$ (columns, left to right) and $R_{\rm{env}} = (50,500,1000)\,\mathrm{R}_\odot$ (rows, top to bottom). All panels have the same limits for both axes.  The colouring, hatching, and linestyles are as in Fig.~\ref{fig:MRplot}.  The lines $\chi = 1$ (solid), $\eta_{\rm{bo}} = 1$ (dashed), and $\zeta = 1$ (dot-dashed) are labelled in the middle left panel, along with the boundary where pair creation becomes relevant.  The breakout regimes I--V are denoted in the central panel.  In cases where region II is present, we have plotted a magenta line indicating the approximate velocity for which INERT breakout can occur. The locations where $\tau_{\rm{env}} = c/v_0$ or $\tau_{\rm{env}} = 3c/v_0$ are marked in the lower right panel.  All results assume a value of $q=1$.}
    \label{fig:MEplot}
\end{figure*}

As desired, we have obtained expressions linking the breakout parameters to the envelope properties and the injected energy.  We can now repeat the procedure of Section~\ref{sec:parameterspace}.  First, we calculate $\chi$ and $\eta_{\rm{bo}}$ using equations~\ref{chi1} and~\ref{etabo1}. Then, as before $\zeta$ can be determined using equation~\ref{zeta}.  The results of carrying out this procedure over a large grid of possible parameter values are shown in Figs.~\ref{fig:MRplot} and~\ref{fig:MEplot}.  Fig.~\ref{fig:MRplot} shows the dependence on $M_{\rm{env}}$ and $R_{\rm{env}}$ for fixed $n$ and $E_0$, while Fig.~\ref{fig:MEplot} shows the dependence on $M_{\rm{env}}$ and $E_0$ for fixed $n$ and $R_{\rm{env}}$.  As discussed above, our model is valid only for $\Psi \ga 1$, which holds for large masses, small radii, and large energies.  This condition is violated in the grey region of the figure, where $\Psi < 1$, and it is marginal in the region with dotted hatching, where $1 \le \Psi \le 3$.

As in Fig.~\ref{fig:rhovplot}, we see that the parameter space is divided into five regions, which all meet at a single point.  Another similarity is that there is a preferred range of velocities to achieve an INERT breakout.  This can be seen in Fig.~\ref{fig:MEplot}: since $v_0 \propto E_0^{1/2} M_{\rm{env}}^{-1/2}$, and the scaling of $v_{\rm{bo}}$ is almost the same (ranging from $v_{\rm{bo}} \propto E_0^{1/2} M_{\rm{env}}^{-1/2}$ for $n=0$ to $E_0^{0.58} M_{\rm{env}}^{-0.42}$ for $n=3$), lines of constant $v_{\rm{bo}}$ approximately follow $E_0 \propto M_{\rm{env}}$.  As an example we include several such constant-$v_{\rm{bo}}$ lines in Fig.~\ref{fig:MEplot}.  We see from the figure that if the energy is large enough and the breakout velocity is close to a certain value (typically 5--10 per cent of $c$), then we almost always expect to obtain an INERT breakout.  This feature arises due to the fact that $\eta_{\rm{bo}}$ depends very strongly on $v_{\rm{bo}}$, but only weakly on $\rho_{\rm{bo}}$, as seen in equation~\ref{etabo}.  Consequently, surfaces of constant $v_{\rm{bo}}$ will always be nearly parallel to surfaces of constant $\eta_{\rm{bo}}$ regardless of the chosen parametrization.

At the same time, there are several key differences when comparing Figs.~\ref{fig:MRplot} and~\ref{fig:MEplot} to the results of Section~\ref{sec:parameterspace}. First, whereas the value of $n$ played only a minor role in Fig.~\ref{fig:rhovplot}, in Figs.~\ref{fig:MRplot} and~\ref{fig:MEplot} the results significantly depend on $n$.  For $n=0$, a substantial part of parameter space results in INERT breakout.  For $n=1.5$, INERT breakout is only possible for a small envelope radius $\la 3 \times 10^{13}\,$cm, and an envelope mass of at least $10^{-2}\,\mathrm{M}_\odot$.  For $n=3$, over the range of values we consider, there is hardly any region of parameter space where an INERT breakout can occur.  Moreover, whereas in in Fig.~\ref{fig:rhovplot} the size of region II increased with increasing breakout radius, in Figs.~\ref{fig:MRplot} and~\ref{fig:MEplot} we find that region II is only present for sufficiently small $R_{\rm{env}}$.  The maximum allowable $R_{\rm{env}}$ is decreased for larger $n$ and smaller $E_0$.

The fundamental reason for these differences is that in Section~\ref{sec:parameterspace}, we varied $R_{\rm bo}$ and $n$ while treating $\rho_{\rm{bo}}$ and $v_{\rm{bo}}$ as independent parameters.  But in this section, we change $\rho_{\rm{bo}}$ and $v_{\rm{bo}}$ according to equations~\ref{vbo} and~\ref{rhobo} as $R_{\rm env}$ and $n$ vary.  In this picture, both $v_{\rm{bo}}$ and $\rho_{\rm{bo}}$ depend strongly on $n$, because larger $n$ leads to stronger shock acceleration, which significantly increases $v_{\rm{bo}}$.  When $v_{\rm{bo}}$ is large, the breakout occurs at a smaller optical depth $c/v_{\rm{bo}}$, so the breakout density $\rho_{\rm{bo}}$ is reduced as well.  In fact, when $n$ is increased, the change in $\rho_{\rm{bo}}$ is much more significant than the change in $v_{\rm{bo}}$, as can be seen by comparing the exponents in equations~\ref{vbo} and~\ref{rhobo}.  While $v_{\rm{bo}}$ does not depend much on $R_{\rm{env}}$ ($v_{\rm{bo}} \propto R_{\rm{env}}^0$ for $n=0$ and $v_{\rm{bo}} \propto R_{\rm{env}}^{-0.33}$ for $n=3$), $\rho_{\rm{bo}}$ does depend strongly on $R_{\rm{env}}$ ($\rho_{\rm{bo}} \propto R_{\rm{env}}^{-3}$ for $n=0$ and $\rho_{\rm{bo}} \propto R_{\rm{env}}^{-1.25}$ for $n=3$).

The differences between this and the previous section can in fact be understood purely from this strong dependence of $\rho_{\rm{bo}}$ on $R_{\rm{env}}$ and $n$.  As we saw in Section~\ref{sec:parameterspace}, in order to achieve an INERT breakout, $\rho_{\rm{bo}}$ must be above some minimum value depending on $R_{\rm bo}$. In the envelope breakout case, $R_{\rm bo}\approx R_{\rm{env}}$ so this minimum value scales as $\rho_{\rm{bo,}min} \propto R_{\rm{env}}^{-15/16}$.  Suppose for some value of $R_{\rm{env}}$ and $n$ this condition is met and the breakout is in region II.  What will happen if we then increase $R_{\rm{env}}$ or $n$?  If $R_{\rm{env}}$ is increased, then as discussed above, for $n\le 3$ the value of $\rho_{\rm{bo}}$ will decrease at least as fast as $\propto R_{\rm{env}}^{-1.25}$.  That is to say, $\rho_{\rm{bo}}$ drops off faster than $\rho_{\rm{bo,}min}$, so at large enough $R_{\rm{env}}$ the condition $\rho_{\rm{bo}} > \rho_{\rm{bo,}min}$ will no longer be possible to satisfy.  This is why there is an upper limit on the value of $R_{\rm{env}}$ which can produce an INERT breakout in Fig.~\ref{fig:MRplot}.  Similarly, if $n$ is increased instead of $R_{\rm{env}}$, this will also decrease $\rho_{\rm{bo}}$ and make the minimum density condition harder to satisfy.  To compensate the decrease in $\rho_{\rm{bo}}$ from increasing $n$, we would have to slightly decrease $R_{\rm{env}}$, or greatly increase $M_{\rm{env}}$.  This is why region II moves up and to the left in Fig.~\ref{fig:MRplot} when $n$ is increased.  Finally, if $M_{\rm{env}}$ is increased to offset the change in $n$, we also need to change $E_0$ roughly in proportion to $M_{\rm{env}}$, or else $v_{\rm{bo}}$ will also change and we will no longer be in region II.  This is why region II moves up and to the right, roughly along a line of constant velocity, as $n$ is increased in Fig.~\ref{fig:MEplot}.

In summary, for a realistic extended envelope breakout model, the ideal conditions to achieve an INERT breakout are that $n$ should be as small as possible ($n=0$ is the best case),  $R_{\rm{env}}$ should be less than some maximum value which increases with $E_0$ and decreases with $n$, and $E_0/M_{\rm{env}}$ should lie in a relatively narrow range to ensure $v_{\rm bo} \sim 0.1$.  In light of these findings, we will further investigate the behaviour in the $n\rightarrow 0$ limit in the next section.

\subsection{The $n\rightarrow 0$ limit}
\label{sec:n0}

From Fig.~\ref{fig:MRplot}, we see that when $n=0$, an INERT breakout is expected for $M_{\rm{env}} \sim 10^{-2}$--$10^{-1}\,\mathrm{M}_\odot$ and $R_{\rm{env}} \la 10^{14}$\,cm. Interestingly, these values are similar to the values of $M_{\rm{env}} \sim 10^{-2} \mathrm{M}_\odot$ and ${R_{\rm{env}} \sim 10^{13}\text{--}10^{14}\,\text{cm}}$ inferred from modelling the early optical peak in double-peaked SNe \citep[e.g.,][]{np}.  This connection is especially interesting in the context of $ll$GRBs, which are also thought to have extended envelopes with similar properties \citep[e.g.,][]{nakar,ic16}, and which are known to exhibit both thermal and non-thermal components in their spectra as discussed in Section~\ref{sec:introduction}.  We discuss the potential connection to $ll$GRBs more in Paper III.  Here, we simply point out that the $n=0$ case may be relevant to observations, so it is worthwhile to develop a better physical intuition for the non-standard case with small $n$.
 
Equations~\ref{vbo}--\ref{Tobsbo} greatly simplify in the limit $n\rightarrow 0$.  Although at first glance it appears that the breakout quantities will vanish since $\Lambda \rightarrow 0$ as $n\rightarrow 0$, it turns out that in each equation $\Lambda$ is raised to a power $\propto n$, and $\Lambda^n \rightarrow 1$ as $n \rightarrow 0$ (via equation~\ref{Lambda}), so the equations are well-behaved.  Thus the term involving $\Lambda \Psi$ simply goes to unity as its exponent trends to zero, and we obtain
\begin{equation}
    \label{vbo0}
    v_{\rm{bo}} \approx v_0 \approx 0.056 c\, E_{0,50}^{1/2} M_{\rm{env,}-1}^{-1/2}
\end{equation}
\begin{equation}
    \label{rhobo0}
    \rho_{\rm{bo}} \approx \rho_{\rm{env}} \approx 1.6 \times 10^{-11}\,\text{g}\,\text{cm}^{-3} M_{\rm{env,}-1} R_{\rm{env,}14}^{-3}
\end{equation}
\begin{equation}
    \label{tbo0}
    t_{\rm{bo}} \approx \dfrac{c}{\kappa \rho_{\rm{env}} v_0^2} \approx 3400\,\text{s}\,E_{0,50}^{-1} R_{\rm{env,}14}^3
\end{equation}
\begin{equation}
    \label{chi0} 
    \chi \approx \dfrac{\kappa \rho_{\rm{env}} R_{\rm{env}}v_0^2}{c^2}
    \approx 1.0 E_{0,50} R_{\rm{env,}14}^{-2}
\end{equation}
\begin{equation}
    \label{Lbo0}
    L_{\rm{bo}} \approx 4 \uppi R_{\rm{env}}^2 \rho_{\rm{env}} v_0^3 \approx 9.3\times10^{45}\,\text{erg}\,\text{s}^{-1} E_{0,50}^{3/2} M_{\rm{env,}-1}^{-1/2} R_{\rm{env,}14}^{-1}
\end{equation}
\begin{equation}
    \label{etabo0}
    \eta_{\rm{bo}} \approx 2.3 q E_{0,50}^{15/8} M_{\rm{env,}-1}^{-2} R_{\rm{env,}14}^{3/8} 
\end{equation}
\begin{equation}
    \label{Tobsbo0}
    kT_{\rm{obs,bo}}\xi_{\rm{bo}}^2 \approx 0.1\,\text{keV}\, q^2 E_{0,50}^4 M_{\rm{env,}-1}^{-4}
\end{equation}
for the $n=0$ and $s=2$ case, where $E_0$, $M_{\rm{env}}$, and $R_{\rm{env}}$ are scaled to values of $10^{50}\,$erg, $10^{-1}\mathrm{M}_\odot$, and $10^{14}\,$cm respectively.  These units are chosen to place the breakout in the vicinity of the INERT regime in Figs.~\ref{fig:MRplot} and~\ref{fig:MEplot}.  

As discussed in Section~\ref{sec:envelopebreakout}, in our model the $n \ll s$ case corresponds to a situation in which the steep outer part of the density profile, where $\rho \propto (R_{\rm{env}}-r)^n$, is confined to a geometrically thin layer at the edge of the medium.   In other words, the envelope has a sharp boundary where the density goes from $\rho_{\rm{env}}$ to 0 over a scale much smaller than $R_{\rm{env}}$.  We will refer to this region as the `acceleration layer,' since it is where shock acceleration primarily takes place.  In our model, the thickness of this layer is approximately $\Lambda R_{\rm env}$. In the $n=0$ case where $\Lambda \rightarrow 0$, the density jumps from $\rho_{\rm{env}}$ to 0 discontinuously at $r=R_{\rm{env}}$, i.e. there is no acceleration layer and no shock acceleration.  While this may be unphysical (realistically a transition layer, however thin, should be expected), we will argue that with regards to the shock breakout, for sufficiently small $n$ the behaviour is the same as if $n$ were 0.  This is because as $n$ decreases, the acceleration layer grows thinner and thinner, and the optical depth through it also decreases accordingly.  When $n$ becomes small enough (roughly when $n/s \la \Psi^{-1}$), the optical depth in the acceleration layer no longer satisfies $\tau > c/v_0$, so the breakout must take place \textit{before the shock reaches the acceleration layer}.  In this scenario, the acceleration layer can only influence the breakout through its negligible contribution to the optical depth, so the properties of the breakout should not depend on $n$.  Instead, they depend on the conditions just inside of the acceleration layer, where regardless of $n$ the shock velocity is $\approx v_0$ and the density is $\approx \rho_{\rm{env}}$.  But this situation, where the breakout takes place in a thin shell near the edge of the inner density profile and the shock acceleration is negligible, is exactly the same as what occurs in the $n=0$ case, where there is no acceleration layer at all.  

Another way to put it is that the limit $n\rightarrow0$ corresponds to a situation where the total optical depth of the envelope is $\gg c/v_{\rm{bo}}$, but the optical depth of the steep outer part is $\la c/v_{\rm{bo}}$,  so that the breakout happens inside of the outer density profile, but still near the edge of the inner density profile.  In other words, it is the case where $\Psi \ga 1$, but $\Lambda \Psi \la 1$.  To see this, we can first combine equations~\ref{vbo}, \ref{Psi}, and~\ref{tbo1} to obtain $\Lambda \Psi = (\Lambda R_{\rm env}/v_{\rm bo} t_{\rm bo})^{1+n-\mu n}$.  If $\Psi > \Lambda^{-1}$, this implies that $v_{\rm bo} t_{\rm bo} < \Lambda R_{\rm env}$, meaning that the breakout takes place in the acceleration layer, as in the standard scenario for breakout from a stellar edge.  On the other hand, if $\Psi \approx 1$, the breakout takes place inside of the acceleration layer.  In this case, we have $v_{\rm bo} \approx v_0$ and $\rho_{\rm bo} \approx \rho_{\rm env}$, since acceleration is unimportant. Then we see from equations~\ref{tbo} and~\ref{Psi} that $\Psi \approx 1$ corresponds to $v_{\rm bo} t_{\rm bo} \approx R_{\rm env}$.  Thus, for $1 \la \Psi \la \Lambda^{-1}$, we find that $\Lambda R_{\rm env} \la v_{\rm bo} t_{\rm bo} \la R_{\rm env}$, so that the breakout shell is thicker than the acceleration layer but still thin compared to $R_{\rm env}$.   We stress that this intermediate regime is only relevant if $\Lambda \ll 1$, so that the acceleration layer is confined to a thin shell.  When $\Lambda \sim 1$, the breakout is guaranteed to happen in the outer density profile if $\Psi \gg 1$.

The model presented here can be compared with the recent work of \citet{khatami}, who also considered how the breakout differs when it occurs in a flat inner density profile versus a steep outer density profile.  Their light edge breakout case corresponds to the $n\rightarrow \infty$ limit in our model, and gives the same scalings when $\mu=0.2$ and $\Psi$ is identified as their parameter $\xi$.  On the other hand, the $n\rightarrow 0$ limit in our model is similar to the light interior breakout case of their model, where their parameter $k_0$ is taken to be small so that the breakout occurs at a radius $R_{\rm{bo}} \approx R_{\rm{env}}$.  In that case, our scaling for $L_{\rm{bo}}$ agrees with theirs, but we obtain a slightly different scaling for the breakout duration due to a subtlety in how the  $t_{\rm{bo}}$ is calculated.  The difference is that they approximate $t_{\rm{bo}} \approx R_{\rm{bo}}/v_{\rm{bo}}$ in their interior breakout case, whereas in our model we use $t_{\rm{bo}} \approx (R_{\rm{env}}-R_{\rm{bo}})/v_{\rm{bo}}$.  After accounting for this change, the two models are brought into agreement.  Another difference is that in our model, the steep outer density profile is only finite in extent, and it can be thin compared with $R_{\rm{env}}$.  Because of this, in our model it is possible to achieve the necessary conditions for their interior breakout case even when $\Psi > 1$, as discussed above.

In fact, as we will see in the next section, the $n=0$ case is also deeply connected to previous work on shock breakout in stellar winds, and to the cooling envelope model often used to describe double-peaked SNe.

\subsection{Connections with previous work}
\label{sec:connections}

The envelope breakout scenario discussed in Sections~\ref{sec:envelopebreakout} and~\ref{sec:n0}, which is based on the standard treatment of NS10, can be compared with other models in the literature involving a shock interacting with optically thick material surrounding the star.  Shock breakout from a stellar wind, as has been studied analytically by various authors \citep[e.g.,][]{ci11,hp}, offers one point of comparison.  As shown by \citet{ci11}, there are two possible scenarios, depending on how the radial extent of the wind, $R_{\rm{w}}$ compares with the characteristic diffusion length scale, $R_{\rm{d}}$. If $R_{\rm{w}} > R_{\rm{d}}$, the breakout occurs while the the shock is still deep inside the envelope, while if $R_{\rm{w}} < R_{\rm{d}}$, the breakout occurs when the shock approaches the edge of the wind.  For our purposes, the $R_{\rm{w}} < R_{\rm{d}}$ case is relevant.

Another point of comparison is the cooling envelope model of \citet{np}, which describes the early optical peak in double-peaked Type IIb or Ibc SNe.  Their picture involves an extended stellar envelope heated by the passage of a shock, which expands adiabatically and eventually cools once the radiation trapped within can escape.  

Here, we will show that these two seemingly distinct models can actually be unified.  In fact, both can be viewed as special cases of the envelope shock breakout model presented here, in the limit $n \rightarrow 0$ where, as discussed above, shock acceleration is not very important.  

First, let us consider the shock breakout emission.  From equations~\ref{tbo0} and~\ref{Lbo0}, we have $t_{\rm{bo}} \propto (c/\kappa) R_{\rm{env}}^3 E_0^{-1}$ and ${L_{\rm{bo}} \propto R_{\rm{env}}^{-1} E_0^{3/2} M_{\rm{env}}^{-1/2}}$ for the $n=0$ case.  This gives a total radiated energy of $E_{\rm bo} \propto t_{\rm bo}L_{\rm bo} \propto (c/\kappa) R_{\rm env}^2 E_0^{1/2} M_{\rm env}^{-1/2}$.  In Appendix~\ref{sec:appendixb}, we verify that these scalings for the observables are identical to the ones given by \citet{ci11} and \citet{hp}, although they parametrized the problem differently.  Thus, we see that in terms of the duration, luminosity, and radiated energy of the shock breakout, the model considered here, when taking the limit $n \rightarrow 0$, is effectively equivalent to the $R_{\rm{w}} < R_{\rm{d}}$ case of shock breakout from a stellar wind.

Now, let us consider the cooling emission.  According to NS10, following the breakout, the luminosity first decays as $t^{-4/3}$ for $t < t_{\rm s} \approx R_{\rm{env}}/v_{\rm{bo}}$.  Then, for $t>t_{\rm s}$, the luminosity scales as $L \propto t^{(2-2.28n)/3(1+1.19n)}$ as the luminosity shell recedes into the breakout ejecta.\footnote{In this discussion, we neglect for simplicity the logarithmic correction of FS19, as it has a negligible effect on the bolometric luminosity, and also does not significantly affect the temperature at times $t>t_{\rm s}$.}  Interestingly, we see that when $n \le 0.87$, the resulting light curve actually rises again after the initial breakout emission decays.  Taking $n=0$, we have  $L \propto t^{2/3}$ for $t>t_{\rm s}$, with $t_{\rm s}$ given by  $t_{\rm s} \approx R_{\rm{env}} E_0^{-1/2} M_{\rm{env}}^{1/2}$. 

The light curve will continue to rise until all of the shocked envelope material has been exposed, i.e. until $m_{\rm{ls}} \approx M_{\rm{env}}$ where $m_{\rm{ls}}$ is the mass of the luminosity shell.  From NS10, we have $m_{\rm{ls}} = m_{\rm{bo}} (t/t_{\rm s})^{2(1+n)/(1+1.19n)}$ in the spherical phase.  Setting $n=0$ and $m_{\rm{ls}} \approx M_{\rm{env}}$, noting that $m_{\rm{bo}} \approx 4 \uppi R_{\rm{env}}^2 c/\kappa v_{\rm{bo}} \approx c R_{\rm{env}} t_{\rm s}/\kappa$, and solving for $t$, 
we find that the cooling emission peaks at the time\footnote{Equation~\ref{tp} is equivalent to the familiar expression for the diffusion time, $(\kappa M/4\uppi cv)^{1/2}$, of an expanding medium with mass $M$ and velocity $v$, with $M=M_{\rm{env}}$ and $v \approx (E_0/M_{\rm{env}})^{1/2}$.}
\begin{equation}
    \label{tp}
    t_{\rm{p}} \approx \left(\dfrac{4 \uppi c}{\kappa}\right)^{-1/2} E_0^{-1/4} M_{\rm{env}}^{3/4} \approx 4 \times 10^4\,\text{s}\, E_{0,51}^{-1/4} M_{\rm{env,}-2}^{3/4},
\end{equation}
where $E_{0,51} = E_0/(10^{51}\,\text{erg})$ and $M_{\rm{env,}-2} = M_{\rm{env}}/(10^{-2} \mathrm{M}_\odot)$.  Next, for $n=0$, the luminosity when $t>t_{\rm s}$ is given by ${L = L_{\rm{bo}} (t_{\rm s}/t_{\rm{bo}})^{-4/3} (t/t_{\rm s})^{2/3}}$. Plugging in $t = t_{\rm{p}}$, and using equations~\ref{ts}, \ref{tbo1}, and \ref{tp} we obtain the bolometric luminosity of the cooling peak,\footnote{The meaning of equation~\ref{Lp} becomes clear when rewriting the right hand side as $[E_0(R_{\rm{env}}/v_0 t_{\rm{p}})]/t_{\rm{p}}$ using equation~\ref{tp}. The quantity $E_0 (R_{\rm{env}}/v_0 t_{\rm{p}}$) is the internal energy remaining in the ejecta after it expands adiabatically to a radius $v_0 t_{\rm{p}}$, so equation~\ref{Lp} is simply the internal energy divided by the diffusion time.}
\begin{equation}
    \label{Lp}
    L_{\rm{p}} \approx \left(\dfrac{4 \uppi c}{\kappa}\right) R_{\rm{env}} E_0 M_{\rm{env}}^{-1} \approx 9 \times 10^{45}\,\text{erg}\,\text{s}^{-1} R_{\rm{env,}14} E_{0,51} M_{\rm{env,}-2}^{-1}.
\end{equation}
We can also consider the temperature at the time $t_{\rm{p}}$.  In \citet{np}, the temperature at the peak is approximated by the effective temperature, $T_{\rm{p}} \approx [L_{\rm{p}}/4 \uppi \sigma_{\rm{B}} (v_0 t_{\rm{p}})^2]^{1/4}$, where $\sigma_{\rm{B}}$ is the Stefan-Boltzmann constant.  If we make the same approximation,\footnote{Since the ejecta is scattering-dominated and the surface of last absorption is not at $\tau = 1$, a better approximation for the observed temperature would be $T_{\rm{p}} \approx [\tau_{\rm{cs}} L_{\rm{p}}/4 \uppi \sigma_{\rm{B}} (v_0 t_{\rm{p}})^2]^{1/4}$ (see for instance the discussion in Section 2.1 of NS10), which would give a different scaling. However, we do not expect this to significantly alter the results, since $T_{\rm{p}} \propto \tau_{\rm{cs}}^{1/4}$ and the optical depth of the color shell $\tau_{\rm{cs}}$ should not be too large at the late times considered here.} then from equations~\ref{tp} and~\ref{Lp}, we arrive at 
\begin{equation}
    \label{Tp}
    kT_{\rm{p}} \approx \left(\dfrac{4 \uppi c^2 k^4}{\sigma_{\rm{B}} \kappa^2}\right)^{1/4} R_{\rm{env}}^{1/4} E_0^{1/8} M_{\rm{env}}^{-3/8} \approx 10\,\text{eV}\,R_{\rm{env,}14}^{1/4} E_{0,51}^{1/8} M_{\rm{env,}-2}^{-3/8}.
\end{equation}

Again, we confirm in Appendix~\ref{sec:appendixb} that equations~\ref{tp}--\ref{Tp} are identical to the expressions in \citet{np} up to order-unity factors.  Therefore, at late times we find that our model with $n=0$ indeed reduces to the cooling envelope model of \citet{np}.

To summarize, we have introduced a model which is based on NS10, but generalizes their model to the case where the usual ${\rho \propto (R-r)^n}$ behaviour smoothly connects to a power-law behaviour $\rho \propto r^{-s}$ at smaller radii. In our setup we allow the thickness of the acceleration layer to decrease with $n$ in such a way that the system is well-behaved even for $n=0$.  For typical order-unity values of $n$, we recover the results of NS10.  When taking $n\rightarrow 0$, however, we instead obtain the results for breakout from the edge of a stellar wind, as in \citet{ci11} and \citet{hp}.  Moreover, the late-time behaviour of our model when $n=0$ also captures the cooling envelope model for double-peaked SNe devised by \citet{np}.  The model we present here can therefore be viewed as a generalized model for shock breakout from the edge of a region, which unifies these seemingly unconnected models in the literature into a single framework.

\section{Spectral modelling}
\label{sec:model}

Having determined the physical conditions that lead to INERT breakout, our aim now is to get a better idea of what the spectrum might look like in this case.  To accomplish this, we need to develop a spectral model which takes into account non-equilibrium effects, rapid thermalization, and the smearing of the spectrum due to unequal light arrival times, and that will be the topic of this section. We will first describe the model assumptions in Section~\ref{sec:modelparameters}, and then show how to construct the spectrum.  Our basic approach will be to first consider the evolution of the spectrum when the light-crossing time is neglected (Section~\ref{sec:sourceframe}), and then to carry out an integral over the arrival time to account for the fact that the high-latitude emission arrives with a delay (Section~\ref{sec:observedspectrum}).  For simplicity, we consider here only the case of spherical symmetry, and only the evolution during the planar phase ($t<t_{\rm s}$), where the shock radius is approximately constant.  Additionally, in Sections~\ref{sec:modelparameters}--\ref{sec:observedspectrum}, we adopt $q=1$.

\subsection{Model description}
\label{sec:modelparameters}
  
The spectral behaviour is completely determined by six quantities which can be computed via the breakout model discussed in Sections~\ref{sec:basicidea} and~\ref{sec:conditions}. The first three are the time-scales $t_{\rm{bo}}$, $t_{\rm{lc}}$ and $t_{\rm{eq,ls}}$, given respectively by equations~\ref{tbo}, \ref{tlc}, and~\ref{teq}.  The light-crossing time $t_{\rm{lc}}$ controls the overall duration of the bolometric emission; the breakout time-scale $t_{\rm{bo}}$ is effectively the adiabatic cooling time of the breakout shell, which affects the temperature and luminosity evolution;  and the equilibrium time of the luminosity shell $t_{\rm{eq,ls}}$ determines when thermal emission starts to be observed. The next two quantities are the breakout shell's bolometric luminosity,  $L_{\rm{bo}}$ (equation~\ref{Lbo}), and its observed temperature, $T_{\rm{obs,bo}}$ (equation~\ref{Tinequality}).  These respectively set the normalization of the bolometric light curve and the peak energy.  The last parameter, $\alpha$, is a power-law index given by equation~\ref{Tpowerlaw}, which governs the time dependence of the observed temperature for $t<t_{\rm{eq,ls}}$.  Since our breakout model has only four free parameters ($\rho_{\rm bo}$, $v_{\rm bo}$, $R_{\rm bo}$, and $n$), the six quantities determining the spectrum are not independent, and are subject to two additional closure relations which will be derived below.

In terms of these six quantities, the bolometric luminosity and observed temperature as functions of time are given by\footnote{In this section, the value of $L$ is taken at the luminosity shell, and the values of $T_{\rm{obs}}$, $T_{\rm{BB}}$, $\nu_{\rm{a}}$, $\tau$, and $\xi$ are taken at the colour shell (as labelled in Table~\ref{table1}).  For notational simplicity we drop the subscripts `ls' and `cs'.} 
\begin{equation}
\label{Levolution}
    L(t) = 
    \begin{cases}
        L_{\rm{bo}}, & t < t_{\rm{bo}} \\
        L_{\rm{bo}}\left(\dfrac{t}{t_{\rm{bo}}}\right)^{-4/3}, & t_{\rm{bo}} < t < t_{\rm s}
    \end{cases}
\end{equation}
and
\begin{equation}
\label{Tevolution}
    T_{\rm{obs}}(t) =
    \begin{cases}
        T_{\rm{obs,bo}},  & t < t_{\rm{bo}} \\
        T_{\rm{obs,bo}} \left(\dfrac{t}{t_{\rm{bo}}}\right)^{-\alpha}, & t_{\rm{bo}} < t < \min(t_{\rm{eq,ls}},t_{\rm s}) \\
        T_{\rm{obs,bo}} \left(\dfrac{t_{\rm{eq,ls}}}{t_{\rm{bo}}}\right)^{-\alpha} \left(\dfrac{t}{t_{\rm{eq,ls}}}\right)^{-1/3}, & \min(t_{\rm{eq,ls}},t_{\rm s}) < t < t_{\rm s}
    \end{cases}.
\end{equation}
Here, we have ignored the logarithmic correction of FS19 for the luminosity and for the temperature in the equilibrium phase, instead  using the simple power-laws $L\propto t^{-4/3}$ and $T_{\rm{obs}} \propto t^{-1/3}$, as expected for a planar adiabatically cooling thermal shell (e.g., NS10).  This is justified due to the very weak dependence of these quantities on the logarithmic term (see equations 29 and 58 in FS19).  The power-law dependence of $T_{\rm{obs}}$ for $t>t_{\rm{eq,ls}}$ is in fact slightly steeper than $t^{-1/3}$, due to an additional weak dependence on the slowly evolving optical depth of the colour shell ($T_{\rm{obs}} \propto t^{-0.36}$ for $n=3$ according to FS19), but for simplicity this difference is also ignored here.  

To fully characterize the spectrum in the non-equilibrium case, we also need to know the self-absorption frequency $\nu_{\rm{a}}$.  This can be expressed in terms of the other quantities above. From the discussion in Section~\ref{sec:thermalization}, we know that $\nu_{\rm{a}}$ evolves roughly as $\nu_{\rm{a}} \propto t^{\lambda(\alpha-2/3)}$, and from equation~\ref{etanuaTobs} we know that it approaches $(\uppi^4/45)^{1/2} kT_{\rm{eq,ls}}$ as $t\rightarrow t_{\rm{eq,ls}}$.  From equation~\ref{Tevolution}, we have 
\begin{equation}
    \label{Teq}
    T_{\rm{eq,ls}} = T_{\rm{obs}}(t_{\rm{eq,ls}}) = T_{\rm{obs,bo}}(t_{\rm{bo}}/t_{\rm{eq,ls}})^{\alpha},
\end{equation}
leading to
\begin{equation}
    \label{nuaevolution}
    \nu_{\rm{a}}(t) = 
    \begin{cases} \nu_{\rm{a,bo}}, & t<t_{\rm{bo}} \\ \nu_{\rm{a,bo}} \left(\dfrac{t}{t_{\rm{bo}}}\right)^{\lambda(\alpha-2/3)}, & t_{\rm{bo}} < t < \min(t_{\rm{eq,ls}},t_{\rm s})
    \end{cases},
\end{equation}
where
\begin{equation}
    \label{nuabovalue}
    \nu_{\rm{a,bo}} = \left(\dfrac{\uppi^4}{3.92(15)}\right)^{1/2} \left(\dfrac{kT_{\rm{eq,ls}}}{h}\right) \left(\dfrac{t_{\rm{bo}}}{t_{\rm{eq,ls}}}\right)^{\lambda(\alpha-2/3)},
\end{equation}
and we have slightly altered the numerical prefactor in equation~\ref{nuabovalue} for reasons that will become clear in the next section.  Throughout this section we adopt $\lambda = 1.2$.

While equation~\ref{Levolution} applies in any case, the behaviour of equations~\ref{Tevolution} and~\ref{nuaevolution} depends on the ordering of $t_{\rm{bo}}$, $t_{\rm{eq,ls}}$, and $t_{\rm s}$. As discussed in Section~\ref{sec:nonequilibrium}, our assumption that the shock breaks out near the edge of the system implies $v_{\rm bo} t_{\rm bo} < R_{\rm bo}$, and therefore $t_{\rm s} > t_{\rm bo}.$ Then, recalling from Section~\ref{sec:basicidea} that the spectrum at $t < t_{\rm{eq,ls}}$ is a Comptonized free-free spectrum, and the spectrum at $t > t_{\rm{eq,ls}}$ is a blackbody, there are three possible behaviours (see also Paper II):  
\begin{itemize}
    \item If $t_{\rm{eq,ls}} = t_{\rm{bo}}$, then the spectrum is always a blackbody and the observed temperature follows $T_{\rm{obs}} \propto t^{-1/3}$ from the beginning.
    \item If $t_{\rm{bo}} < t_{\rm{eq,ls}} < t_{\rm s}$, the spectrum is initially a Comptonized free-free spectrum, with $T_{\rm{obs}}$ steeply declining and $\nu_{\rm{a}}$ steeply increasing with time. 
    As $t\rightarrow t_{\rm{eq,ls}}$, $kT_{\rm{obs}}$ and $h\nu_{\rm{a}}$ approach $kT_{\rm{eq,ls}}$, and the spectrum becomes a blackbody. After $t_{\rm{eq,ls}}$, $T_{\rm{obs}}$ breaks to a shallower $t^{-1/3}$ evolution for the rest of the planar phase.
    \item If $t_{\rm{eq,ls}} > t_{\rm s}$, the temperature decays steeply throughout the entire planar phase, and the spectrum is a Comptonized free-free spectrum.  The system remains out of thermal equilibrium up until $t=t_{\rm s}$, with $h\nu_{\rm{a}} < kT_{\rm{obs}}$, and only thermalizes later, during the spherical phase.
\end{itemize}
Regardless of the early evolution, once the flow enters the spherical phase at $t=t_{\rm s}$, the gas and radiation quickly thermalize (if they have not done so already), and within a few $t_{\rm s}$ the spectrum becomes a blackbody with a temperature obeying $T_{\rm{obs}} \propto t^{-\alpha_{\rm sph}}$, where typically $\alpha_{\rm sph} \approx 0.6$ (as described in, e.g., NS10, FS19).  Here we are interested only in the early evolution, so we do not discuss the spherical phase further.  

In what follows, it will also be useful to have an expression for the electron-scattering optical depth, $\tau$, in terms of other known quantities.  Assuming the photons have an average diffusion speed of $\approx c/\tau$, NS10 obtained\footnote{Note that because the ejecta's opacity is dominated by electron scattering, spectrum formation takes places at an optical depth of $\sim c/v_{\rm{bo}} \gg 1$, and therefore the usual Stefan-Boltzmann law does not apply; see NS10 for further discussion.}
\begin{equation}
\label{Stefan}
    L = 4\uppi R^2  \tau^{-1} ac T_{\rm{BB}}^4,
\end{equation}
which is simply a more general version of equation~\ref{Lbo}.  Now, since we know that $T_{\rm{BB}} = T_{\rm{obs}}$ when $t \ge t_{\rm{eq,ls}}$, and  since $T_{\rm{BB}} \propto (\nu_{\rm{a}} T_{\rm{obs}})^{1/2}$ (see equation~\ref{etanuaTobs}), we obtain
\begin{equation}
\label{tauevolution}
    \tau(t) = 
    \begin{cases}
    \tau_{\rm{bo}}, & t<t_{\rm{bo}} \\
    \tau_{\rm{bo}} \left(\dfrac{t}{t_{\rm{bo}}}\right)^{2(\lambda-1)(\alpha-2/3)}, & t_{\rm{bo}} < t < \min(t_{\rm{eq,ls}},t_{\rm s}) \\
    \tau_{\rm{bo}} \left(\dfrac{t_{\rm{eq,ls}}}{t_{\rm{bo}}}\right)^{2(\lambda-1)(\alpha-2/3)}, & \min(t_{\rm{eq,ls}},t_{\rm s}) < t < t_{\rm s} \end{cases}
\end{equation}
from equations~\ref{Tevolution}--\ref{Stefan}, where
\begin{equation}
    \label{taubo}
    \tau_{\rm{bo}} = \dfrac{4\uppi a c^3 t_{\rm{lc}}^2 T_{\rm{obs,bo}}^4}{L_{\rm{bo}}(t_{\rm{eq,ls}}/t_{\rm{bo}})^{2[\lambda(\alpha-2/3)+\alpha]}}.
\end{equation}
The constancy of $\tau$ for $t>t_{\rm{eq,ls}}$ follows from our assumption that $T_{\rm{obs}} \propto t^{-1/3} \propto L^{1/4}$ at that time.

We point out that if $L_{\rm bo}$, $t_{\rm bo}$, and $t_{\rm lc}$ can be determined from observations, constraints can immediately be placed on the radius, density, and shock velocity of the breakout shell.  Rearranging equations~\ref{tbo}, \ref{tlc}, and~\ref{Lbo} yields
\begin{align}
\label{Rboconstraint}
R_{\rm bo} & \approx c t_{\rm lc} \nonumber \\
& \approx 3 \times 10^{13}\,\text{cm} \left(\dfrac{t_{\rm lc}}{1000\,\text{s}}\right),
\end{align}
\begin{align}
\label{rhoboconstraint}
\rho_{\rm bo} &\approx \dfrac{16 \uppi^2 c^7 t_{\rm lc}^4}{\kappa^3 t_{\rm bo}^3 L_{\rm bo}^2} \nonumber \\
& \approx 4 \times 10^{-11}\,\text{g}\,\text{cm}^{-3} \left(\dfrac{t_{\rm lc}}{1000\,\text{s}}\right)^4 \left(\dfrac{t_{\rm bo}}{100\,\text{s}}\right)^{-3} \left(\dfrac{L_{\rm bo}}{10^{47}\,\text{erg}\,\text{s}^{-1}}\right)^{-2},
\end{align}
and
\begin{align}
\label{vboconstraint}
v_{\rm bo} &\approx \dfrac{\kappa L_{\rm bo} t_{\rm bo}}{4 \uppi c^3 t_{\rm lc}^2} \nonumber \\
& \approx 0.2\,c \left(\dfrac{t_{\rm lc}}{1000\,\text{s}}\right)^{-2} \left(\dfrac{t_{\rm bo}}{100\,\text{s}}\right) \left(\dfrac{L_{\rm bo}}{10^{47}\,\text{erg}\,\text{s}^{-1}}\right),
\end{align}
where we have assumed $\kappa=0.2\,\text{cm}^2\,\text{g}^{-1}$.  Equations~\ref{Rboconstraint}--\ref{vboconstraint} are accurate up to an order-unity factor depending on $n$.  The associated values of $M_{\rm env}$ and $E_0$ can be estimated from equations~\ref{vbo} and~\ref{rhobo}, if the density profile is specified.

As mentioned above, in order to be compatible with our shock breakout model, which has four degrees of freedom, the six parameters describing the spectrum ($t_{\rm bo}$, $t_{\rm eq,ls}$, $t_{\rm lc}$, $L_{\rm bo}$, $T_{\rm obs,bo}$, and $\alpha$) must obey two additional closure relations.  The first can be deduced by considering the blackbody temperature of the breakout shell, $kT_{\rm BB,bo}$.  On one hand, we have from equations~\ref{tbo} and~\ref{TBB} that $T_{\rm BB,bo} \approx (c/a\kappa t_{\rm bo})^{1/4}$.  On the other hand, we have ${kT_{\rm BB,bo} \approx (kT_{\rm obs,bo} h\nu_{\rm a,bo})^{1/2}}$ from equation~\ref{etanuaTobs} (we neglect the weak dependence on $\xi_{\rm bo}$ and the numerical prefactor).  Replacing $\nu_{\rm a,bo}$ using equations~\ref{Teq} and~\ref{nuabovalue}, we obtain
\begin{equation}
\label{closure1}
T_{\rm obs,bo} \left(\dfrac{t_{\rm bo}}{t_{\rm eq,ls}}\right)^{[\alpha+\lambda(\alpha-2/3)]/2} \sim \left(\dfrac{c}{a\kappa t_{\rm bo}}\right)^{1/4}.
\end{equation}
The second closure relation can be found by comparing the luminosity of the breakout shell, $L_{\rm bo}$, to the energy per unit time generated by Comptonized free-free emission, which is $(4\uppi R_{\rm bo}^2 v_{\rm bo} t_{\rm bo})[3 kT_{\rm obs,bo} \dot{n}_{\rm ff}(T_{\rm obs,bo}) \xi_{\rm bo}]$, where $4\uppi R_{\rm bo}^2 v_{\rm bo} t_{\rm bo}]$ is the breakout shell's volume and $\dot{n}_{\rm ff}(T_{\rm obs,bo}) \xi_{\rm bo}$ is the number of photons with an energy of $3k T_{\rm obs,bo}$ generated per unit volume per unit time in a Comptonized free-free spectrum (the factor of $\xi_{\rm bo}$ accounts for additional photons brought to the peak by Comptonization).  If we assume that $\xi_{\rm bo} \sim 1$ (as will be justified for our typical parameters in Section~\ref{sec:comptonization}), we have
\begin{equation}
    \label{closure2}
    L_{\rm bo} t_{\rm bo} \sim 4 \uppi \left(\dfrac{147 c^{13} k A_{\rm ff}}{2 \kappa^5}\right)^{1/4} t_{\rm lc}^2 T_{\rm obs,bo}^{1/8},
\end{equation}
where $\dot{n}_{\rm ff}$ and $A_{\rm ff} = 3.5\times 10^{36}$ (in cgs units) were defined above equation~\ref{etabo1}, and we applied equations~\ref{rhoboconstraint} and~\ref{vboconstraint} to replace $\rho_{\rm bo}$ and $v_{\rm bo}$ with observables.  Equations~\ref{closure1} and~\ref{closure2} can be used to check whether the parameters obtained from fitting the spectrum with our model are indeed compatible with a non-equilibrium shock breakout interpretation. If these relations do not hold to within a factor of several, then a spherical\footnote{In the aspherical case, the emission might differ substantially from the symmetric case, because the properties of the shock breakout vary across the stellar surface \citep[e.g.,][]{il}.} shock breakout origin is unlikely.

\subsection{The spectrum produced at the source}
\label{sec:sourceframe}
As discussed in the preceding section, for non-equilibrium breakouts the spectrum is expected to evolve from a Comptonized free-free spectrum to a blackbody spectrum.  Our goal now is to obtain an explicit expression for the spectral luminosity $L_\nu$ as a function of time, which can describe the behaviour in both regimes.

We will start by considering the blackbody regime, in which case the spectrum is a blackbody with a temperature of $T_{\rm{obs}}=T_{\rm{BB}}$ and a luminosity given by equation~\ref{Stefan}.  Noticing that $L = \int_0^\infty L_\nu \mathrm{d} \nu$ and $(ac/4) T_{\rm{BB}}^4 = \uppi \int_0^\infty B_\nu(T_{\rm{BB}}) \mathrm{d} \nu$, we can rearrange equation~\ref{Stefan} and then differentiate it with respect to $\nu$ to arrive at 
\begin{equation}
    \label{LnuBB}
    L_\nu = 16 \uppi^2 R^2 \tau^{-1} B_\nu(T_{\rm{obs}}),
\end{equation}
where $\tau$ is given by equation~\ref{tauevolution} (note that $\tau$ is independent of $\nu$ for electron scattering), and we used $T_{\rm{obs}}=T_{\rm{BB}}$ for a blackbody. Regardless of the initial conditions, if equilibrium is achieved during the planar phase, the solution for $L_\nu$ should asymptote to the form in equation~\ref{LnuBB}.

Now, to handle the Comptonized free-free regime, we will make the following simplifying assumption, which will be justified later, in Section~\ref{sec:comptonization}.  Motivated by the lack of a clear Wien peak in the observed spectrum of candidate shock breakout events such as SN 2008D and GRB 060218, we assume that during the free-free phase, the spectrum is only modestly affected by Comptonization, i.e. the Comptonization parameter of the luminosity shell $\xi(T_{\rm{obs}})$ is not too large.  In particular, provided that the Comptonization parameter is less than a certain critical value (which will be derived in Section~\ref{sec:comptonization}), the effect of Comptonization is to move the peak energy of the free-free spectrum from $kT$ to $4kT$, with the shape of the spectrum remaining mostly unchanged \citep[see, e.g.,][]{margalit}.  In this case, we can crudely approximate the spectrum as a modified free-free spectrum with a cutoff at $\approx 4kT_{\rm{obs}}$ instead of $kT_{\rm{obs}}$.  This approximation is convenient because it ensures that the peak energy (defined as the energy which maximizes $\nu F_\nu$) does not have a  discontinuity at $t = t_{\rm{eq,ls}}$, since $\nu F_\nu$ peaks at $4kT_{\rm{obs}}$ for both the Comptonized free-free and blackbody cases.  

Under this assumption, the spectral luminosity at a given time takes the form
\begin{equation}
    \label{Lnu}
    L_\nu = 16 \uppi^2 R^2 \tau^{-1} C_L(t) I_\nu,
\end{equation}
where 
\begin{equation}
    \label{Inu}
    I_\nu = (1- e^{-\tau \tau_{\nu\rm{,ff}}})B_\nu(T_{\rm{obs}}),
\end{equation}
and $\tau_{\nu\rm{,ff}}$ is the optical depth to free-free absorption, given by
\begin{equation}
    \label{tauff}
    \tau_{\nu\rm{,ff}} = C_\tau(t) \left(\dfrac{h\nu}{kT_{\rm{obs}}}\right)^{-3} \left(e^{h\nu/kT_{\rm{obs}}}-1\right) e^{-h\nu/3.92 kT_{\rm{obs}}}.
\end{equation}
The dimensionless quantities $C_L(t)$ and $C_\tau(t)$ are to be determined below.  The coefficient of $I_\nu$ in equation~\ref{Lnu} was chosen to ensure that $L_\nu$ takes the form of~\ref{LnuBB} in the limit $\tau \tau_{\nu\rm{,ff}} \gg 1$. Note that equation~\ref{tauff} differs from the standard prescription for free-free absorption due to the presence of a factor of 3.92 in the exponent.  This factor is found by maximizing $\nu B_\nu$ and it is introduced to ensure that throughout the whole evolution, $I_\nu$ always peaks at the same value of ${h\nu/kT_{\rm{obs}} \approx 3.92}$.  By including this factor, the peak of the free-free spectrum is shifted to $\approx 4 kT_{\rm{obs}}$, which roughly accounts for the effect of Comptonization as discussed above.  Although this prescription has a non-standard behaviour at $h\nu \gg kT_{\rm{obs}}$ (i.e., $I_\nu \rightarrow B_\nu$ instead of $I_\nu \rightarrow \tau_{\nu\rm{,ff}}B_\nu$ as $\nu \rightarrow \infty$), for our purposes we are not interested in frequencies above the peak, so this difference does not affect our results.  

The quantity $C_\tau$ appearing in equation~\ref{tauff} can be expressed as 
\begin{equation}
    \label{Ctau}
    C_\tau = \tau^{-1}\left(\dfrac{h\nu_{\rm{a}}}{kT_{\rm{obs}}}\right)^{3} \left(e^{h\nu_{\rm{a}}/kT_{\rm{obs}}}-1\right)^{-1} e^{h\nu_{\rm{a}}/3.92 kT_{\rm{obs}}}.
\end{equation}  
The simplest way to see this is to observe that, by definition, $\nu_{\rm{a}}$ is the frequency satisfying $\tau \tau_{\nu\rm{,ff}}(\nu_{\rm{a}}) = 1$.  Therefore $\tau_{\nu\rm{,ff}}(\nu_{\rm{a}}) = \tau^{-1}$ and equation~\ref{tauff} leads immediately to equation~\ref{Ctau}.  It is desirable for $C_\tau$ to quickly blow up to infinity after $t_{\rm{eq,ls}}$, so that the spectrum converges to a blackbody at all frequencies.  The exact way this is implemented does not make much difference; we do it by letting $\nu_{\rm{a}}$ continue to grow according to equation~\ref{nuaevolution} even after $t_{\rm{eq,ls}}$. 

With $C_\tau$ known, $L_\nu$ can be calculated using equations~\ref{Lnu}--\ref{Ctau}.  The remaining constant $C_L$ is then found numerically, by demanding that $\int_0^\infty L_\nu \mathrm{d}\nu = L(t)$ at all times.  By comparing equations~\ref{LnuBB} and~\ref{Lnu}, we see that $C_L \rightarrow 1$ for $t \gg t_{\rm{eq,ls}}$.  Our choice of the coefficient in equation~\ref{nuabovalue} ensures that $C_L \rightarrow 1$ as $t\rightarrow t_{\rm{bo}}$ as well. In general $C_L \sim 1$ and it peaks when $t$ is an appreciable fraction of $t_{\rm{eq,ls}}.$  As a final consistency check, we confirm that when $h\nu_{\rm{a}} \ll kT_{\rm{obs}}$, after substituting equations~\ref{Tevolution}, \ref{nuaevolution}, and~\ref{tauevolution} into equations~\ref{Lnu}--\ref{Ctau}, $L_\nu$ reduces to the form $L_\nu(t) \approx [L(t)/3.92 kT_{\rm{obs}}(t)] e^{-h\nu/3.92 kT_{\rm{obs}}(t)}$, as expected for free-free emission with luminosity $L$ and peak energy $3.92 kT_{\rm{obs}}$.

We stress that equation~\ref{Inu} is only an empirical relation that smoothly interpolates between the correct asymptotic behaviours at $\nu \ll \nu_{\rm{a}}$ and $\nu \gg \nu_{\rm{a}}$, and is not a solution to the radiative transfer equation in a scattering-dominated medium.  Therefore, the actual behaviour of the spectrum in the neighbourhood of $\nu \sim \nu_{\rm{a}}$ may differ from what equation~\ref{Inu} predicts.  In particular, for large values of the electron scattering optical depth $\tau$, it is known that the spectrum is altered due to `modified blackbody' effects, as discussed by \citet{feltenrees}, \citet{illarionov}, and \citet{rybicki}.  In that case, $I_\nu$ remains basically the same at low and high frequencies, but there is an intermediate regime for $\nu_{\rm{a}}/\tau \la \nu \la \nu_{\rm{a}}$ where $I_\nu \propto B_\nu \tau_{\nu\rm{,ff}}^{1/2} \propto \nu$.  We investigated also a more general form of equation~\ref{Inu} which accounts for this modification.  However, we find that for $\tau \sim 10$ as considered in this work, the change in the spectrum is not too significant ($L_\nu$ changes by $\la 50$ per cent at all frequencies for $\tau = 10$).  Significant differences become apparent only at $\tau \ga 30$.  The basic reason for this is that in the modified blackbody regime, for modest values of $\tau$ the spectrum is affected only over a relatively narrow range of frequencies near $\nu_{\rm{a}}$.  But, even in the case described by equation~\ref{Inu}, the behaviour of the spectrum near $\nu_{\rm{a}}$ is  not so different from $I_\nu \propto \nu$ anyways, because at that frequency the spectrum is transitioning from the $I_\nu \propto \nu^2$ Rayleigh-Jeans behaviour to the $I_\nu \propto \nu^0$ free-free behaviour. However, we caution that for especially slow shocks with large $\tau \sim c/v$, equation~\ref{Inu} may not adequately describe the emergent spectrum, because the $I_\nu \propto \nu$ behaviour could extend over a much larger frequency range. 

We also checked that the results are not significantly affected when the correct time dependence, including the logarithmic corrections discussed in Section~\ref{sec:thermalization}, is used for $\nu_{\rm{a}}$ and $T_{\rm{obs}}$, instead of the approximate power laws assumed in Section~\ref{sec:modelparameters}.  We find that the overall qualitative picture remains unchanged.  The main difference is that, compared to the power-law model, when including the logarithmic terms $T_{\rm{obs}}$ and $\nu_{\rm{a}}$ evolve slightly faster at times $t \sim t_{\rm{bo}}$, and slightly slower at times $t \sim t_{\rm{eq,ls}}$, in such a way that the value of $T_{\rm{obs}}$ ($\nu_{\rm{a}}$) is always lower (higher) than in the power-law case.  As a result, the spectrum for $t_{\rm{bo}} < t < t_{\rm{eq,ls}}$ is slightly brighter for $h \nu_{\rm{a}} < h\nu < kT_{\rm{obs}}$, and slightly fainter for $\nu < \nu_{\rm{a}}$ and $h\nu > kT_{\rm{obs}}$.  The behaviour at times $t \ga t_{\rm{eq,ls}}$ is essentially unchanged due to a weak dependence on the logarithmic term.

\begin{figure}
    \centering
    \includegraphics[width=\columnwidth]{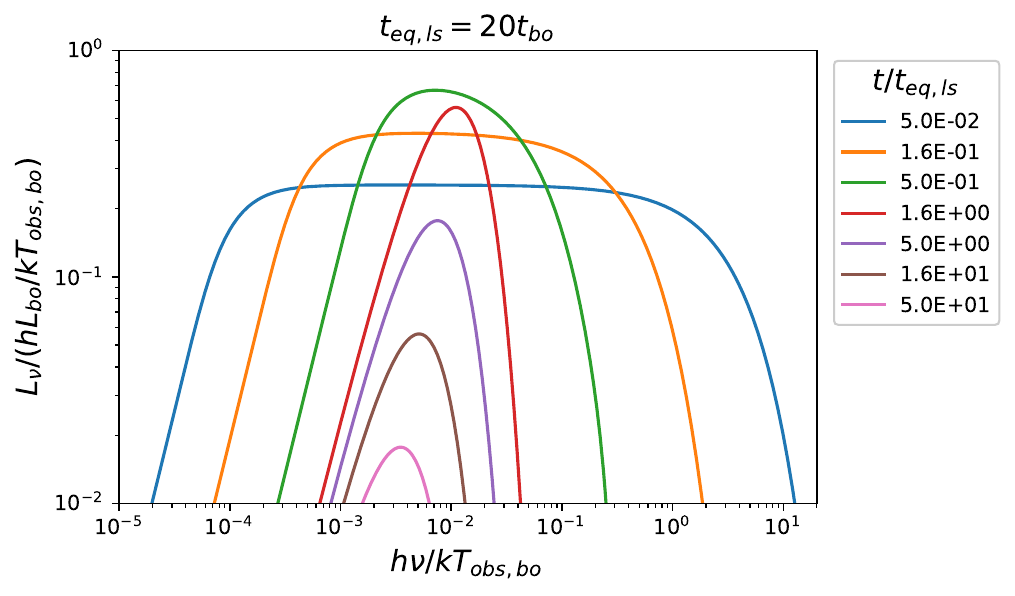}
    \caption{An example of the spectral evolution in the case $t_{\rm{eq,ls}} = 20 t_{\rm{bo}}$ with $\alpha = 1.8$.  $L_{\rm{bo}}$, $T_{\rm{obs,bo}}$, and $t_{\rm{bo}}$ are set to arbitrary values.  The colour indicates time in units of $t_{\rm{eq,ls}}$.}
    \label{fig:sourcespectrumexample}
\end{figure}

To summarize the results of this section, we show an example spectrum computed using our method in Fig.~\ref{fig:sourcespectrumexample}. In the figure, we have taken a large value of $t_{\rm{eq,ls}}/t_{\rm{bo}}$ so that both the free-free and blackbody behaviours show up clearly.  As expected, we see that the spectral is initially a self-absorbed free-free spectrum.  As time goes on, the temperature drops rapidly, while the self-absorption frequency increases.  Since we took $\alpha = 1.8 > 4/3$, the temperature declines faster than the bolometric luminosity, and the peak value of $L_\nu$ increases as $L_\nu \propto L/T_{\rm obs}$ up until $t_{\rm{eq,ls}}$.  At this time, the spectrum becomes a blackbody; the temperature continues to decrease, but much more slowly than before.  At times $t > t_{\rm eq,ls}$, the peak $L_\nu$ gradually fades as $L_\nu \propto L/T_{\rm obs} \propto t^{-1}$.

\subsection{The observed spectrum}
\label{sec:observedspectrum}

The results of the previous section describe the spectrum emitted at the source, which is what an observer would see if the light-crossing time were negligible.  Now we need to account for the delay in arrival due to the light travel time.  Our approach is similar to that of \citet{il}.  As discussed in Section~\ref{sec:blendedspectrum}, emission from material at an angle of $\theta$ from the line of sight arrives with a delay of $t_{\rm{del}}(\theta)$, with $t_{\rm{del}}$ given by equation~\ref{tdel}.  The flux arriving from angles between $\theta$ and $\theta + \mathrm{d}\theta$ is therefore proportional to $L_\nu(t-t_{\rm{del}}(\theta))$.  The flux is also proportional to the projected area contained between $\theta$ and $\theta + \mathrm{d}\theta$, which is $\propto \sin\theta \cos\theta \mathrm{d}\theta$.  The observed luminosity can then be obtained by integrating over $\theta$:
\begin{equation}
    \label{Lnuobs}
    L_{\nu\rm{,obs}}(\nu,t) = 2\int_0^{\uppi/2} L_\nu(\nu, t-t_{\rm{del}}(\theta)) \sin \theta \cos \theta \mathrm{d} \theta,
\end{equation}
where $t_{\rm del}(\theta)$ is given by equation~\ref{tdel}.  The luminosity is taken to be zero when $t-t_{\rm{del}}(\theta)$ is negative.  The normalization factor of 2 comes from the fact that $2\int_0^{\uppi/2} \sin \theta \cos \theta \mathrm{d} \theta = 1$, and ensures that $L_{\nu\rm{,obs}} \approx L_\nu$ when the delay time is negligible, i.e. when $t_{\rm{lc}} \ll t_{\rm{bo}}$ so that the whole surface is revealed in less than the time $t_{\rm{bo}}$.  On the other hand, in the case of significant delay time with $t_{\rm{lc}} \ga t_{\rm{bo}}$, only a small fraction of the surface up to an angle of $\theta_{\rm{bo}} \approx (2t_{\rm{bo}}/t_{\rm{lc}})^{1/2}$ becomes visible within the time $t_{\rm{bo}}$ (as discussed in Section~\ref{sec:blendedspectrum}), so the initial observed luminosity is reduced according to ${L_{\nu\rm{,obs}} \approx L_\nu \theta_{\rm{bo}}^2 \approx 2 L_\nu (t_{\rm{bo}}/t_{\rm{lc}})}$.  This is consistent with the expectation that a non-negligible light-crossing time will not change the total observed energy of the breakout, but will smear out the signal over a time-scale of $t_{\rm{lc}}$, so that a signal with an \textit{intrinsic} duration of $t_{\rm{bo}}$ and luminosity of $L_{\rm{bo}}$ will have an \textit{observed} duration of $t_{\rm{lc}}$ and luminosity of $L_{\rm obs} \approx L_{\rm{bo}} (t_{\rm{bo}}/t_{\rm{lc}})$ after this smearing (see also the discussion surrounding equation~\ref{Lobs}).

\begin{figure}
    \centering
    \includegraphics[width=\columnwidth]{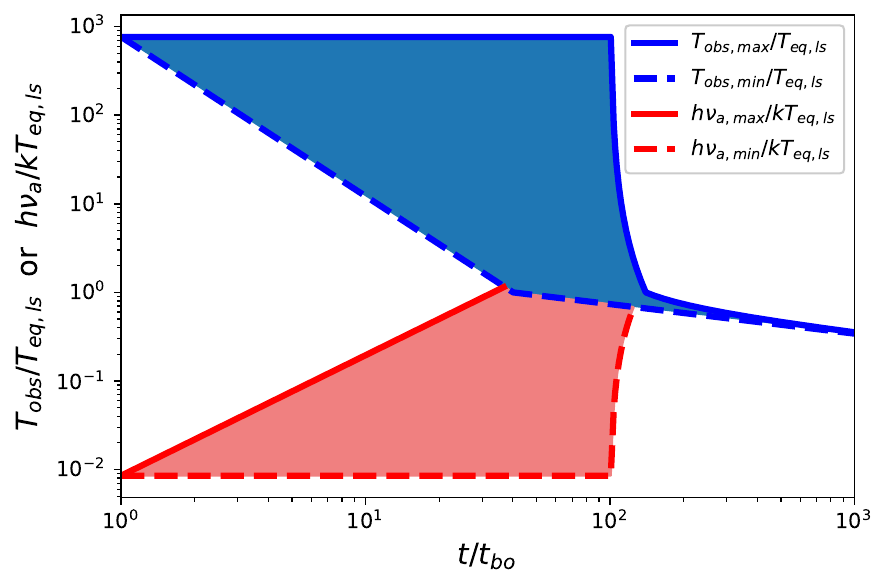}
    \caption{Evolution of the observed temperature (blue) and the self-absorption frequency (red) with time, for $t_{\rm eq,ls}=40 t_{\rm bo}$, $t_{\rm lc} = 100 t_{\rm bo}$, and $\alpha=1.8$. The values of $kT_{\rm obs}$ and $h\nu_{\rm a}$ are scaled to the equilibrium temperature, $kT_{\rm eq,ls}$.  Due to differences in light arrival time, ejecta with a range of temperatures and self-absorption frequencies are observed simultaneously.  We show the observed range with coloured bands, with solid and dashed lines indicating, respectively, the maximum and minimum observed values at a given time.}
    \label{fig:temperatures}
\end{figure}

The delayed arrival of the high-latitude emission not only smears the luminosity over a longer time-scale, it also it also smears the single-temperature spectrum into a multi-temperature spectrum (see also Section~\ref{sec:blendedspectrum}).  In particular, the spectrum is smeared between a minimum temperature $T_{\rm{obs,min}}$ and a maximum temperature $T_{\rm{obs,max}}$, as shown in Fig.~\ref{fig:temperatures}.  The minimum observed temperature $T_{\rm{obs,min}}$ is determined by the material along the line of sight, which has been visible since $t=0$, and it evolves according to equation~\ref{Tevolution} from the beginning, i.e. $T_{\rm{obs,min}} = T_{\rm{obs}}(t)$.  Material at larger latitudes also evolves according to equation~\ref{Tevolution}, but with a delay, so that the emission originating from angle $\theta$ is observed to have a temperature of $T_{\rm{obs}}(\theta) = T_{\rm{obs}}(t-t_{\rm{del}}(\theta))$ at time $t$.  The maximum temperature $T_{\rm{obs,max}}$ is set by the highest-latitude material whose light has had enough time to reach the observer.  If $t<t_{\rm{lc}}$, this material is located at an angle of $\theta \approx (2t/t_{\rm{lc}})^{1/2}$, where the condition $t=t_{\rm{del}}(\theta)$ is satisfied.  If $t>t_{\rm{lc}}$, it is located at the equator, with $\theta = \uppi/2$.  Since the temperature of the breakout is the same across the whole surface in our spherical model, $T_{\rm{obs,max}}$ initially stays constant with a value of $T_{\rm{obs,max}} = T_{\rm{obs}}(0) = T_{\rm{obs,bo}}$ until the whole surface has been revealed at $t = t_{\rm{lc}}$.  For $t > t_{\rm{lc}}$, $T_{\rm{obs,max}}$ tracks the temperature of material in the equatorial plane, which evolves according to $T_{\rm{obs,max}} = T_{\rm{obs}}(t-t_{\rm{lc}})$ to account for its delayed arrival.

At a given frequency $kT_{\rm{obs,min}} < h\nu < kT_{\rm{obs,max}}$, the observed emission is dominated by the emission from the angle satisfying $kT_{\rm{obs}}(\theta) \sim h\nu$.  This emission subtends a projected area $\propto \theta^2$; therefore its luminosity is $\propto L\theta^2$ and its spectral luminosity is $L_\nu \propto L\theta^2/T_{\rm{obs}}(\theta)$.  If $T_{\rm{obs}} \propto t^{-\alpha}$ as in equation~\ref{Tevolution}, then since $L \propto t^{-4/3}$ (equation~\ref{Levolution}) and $\theta \propto t^{1/2}$ (equation~\ref{tdel}), we have $L_\nu \propto [T_{\rm{obs}}(\theta)]^{(1-3\alpha)/(3\alpha)} \propto \nu^{(1-3\alpha)/(3\alpha)}$ for ${kT_{\rm{obs,min}} < h\nu <kT_{\rm{obs,max}}}$.  This power-law scaling agrees with the value obtained by NS10 and FS19, who also considered the behaviour of the spectrum over this frequency range.

However, an effect that was not discussed by NS10 and FS19 is that, in the free-free spectral regime, the self-absorption frequency is also smeared out in a way analogous to the temperature, so that the spectrum at low frequencies is also altered between frequencies $\nu_{\rm{a,min}}$ and $\nu_{\rm{a,max}}$, as shown in red in Fig.~\ref{fig:temperatures}.  Since $\nu_{\rm{a}}$ is increasing instead of decreasing with time, in this case it is $\nu_{\rm{a,max}} = \nu_{\rm{a}}(t)$ which evolves from the beginning, and $\nu_{\rm{a,min}}$ which initially remains steady at a value of $\nu_{\rm{a,max}} = \nu_{\rm{a,bo}}$ for $t<t_{\rm{lc}}$, and then evolves as ${\nu_{\rm{a,max}} = \nu_{\rm{a}}(t-t_{\rm{lc}})}$ for $t > t_{\rm{lc}}$.  By an argument identical to the one used above to determine the spectrum between $kT_{\rm{obs,min}}$ and $kT_{\rm{obs,max}}$, except that we have $\nu_{\rm{a}} \propto t^{\lambda(\alpha-2/3)}$ from equation~\ref{nuaevolution} instead of $T_{\rm{obs}} \propto t^{-\alpha}$, we obtain $L_\nu \propto \nu^{(3\alpha-1)/[\lambda(3\alpha-2)]}$ for ${\nu_{\rm{a,min}} < \nu < \nu_{\rm{a,max}}}$.

\begin{figure}
    \centering
    \includegraphics[width=\columnwidth]{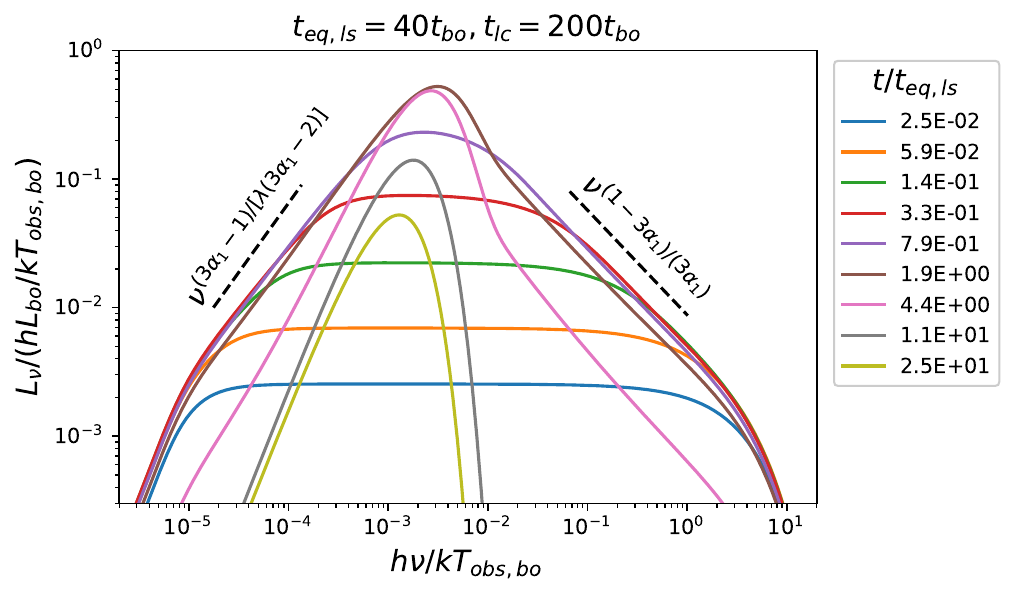}
    \caption{An example of the spectral evolution in the INERT breakout case ${t_{\rm{bo}} < t_{\rm{eq,ls}} < t_{\rm{lc}}}$, where light travel time is not negligible, and blackbody and free-free components coexist in the spectrum.  The chosen parameters are $t_{\rm{eq,ls}} = 40 t_{\rm{bo}}$, $t_{\rm{lc}} = 100t_{\rm{bo}}$, and $\alpha=1.8$.  $L_{\rm{bo}}$, $T_{\rm{obs,bo}}$, and $t_{\rm{bo}}$ are set to arbitrary values.  The colour indicates time in units of $t_{\rm{eq,ls}}$.  The dashed lines show the expected power-laws between $\nu_{\rm{a,min}}$ and $\nu_{\rm{a,max}}$ (left), and between $T_{\rm{obs,min}}$ and $T_{\rm{obs,max}}$ (right).}
    \label{fig:smearedspectrumexample}
\end{figure}

Here, as discussed in Section~\ref{sec:blendedspectrum}, we are mainly interested in the case $t_{\rm{bo}} < t_{\rm{eq,ls}} < t_{\rm{lc}}$ (i.e., the INERT breakout case), where the radiation and gas are initially out of equilibrium, smearing by light travel time is important, and blackbody and free-free components are simultaneously present in the spectrum.  (For example spectra for other orderings of these time-scales, see Paper II.)   Fig.~\ref{fig:smearedspectrumexample} shows an illustrative example of the spectral evolution in this case, for $t_{\rm{eq,ls}} = 40 t_{\rm{bo}}$ and $t_{\rm{lc}} = 200 t_{\rm{bo}}$.  Here, a large value of $t_{\rm{eq,ls}}/t_{\rm bo}$ was adopted so that the spectrum is smeared over a broad frequency range and the expected power-law behaviours are clearly exhibited, and we also took $t_{\rm lc}/t_{\rm eq,ls}$ to be large so that the smearing effect persists for an extended time. For an example with more realistic parameter values, see Paper III.  As expected, we see that the spectrum initially resembles a self-absorbed free-free spectrum.  As time goes on, the high-energy spectrum is smeared into a power-law between $T_{\rm{obs,min}}$ and  $T_{\rm{obs,max}}$, and the low-energy spectrum is smeared into a power-law between $\nu_{\rm{a,min}}$ and $\nu_{\rm{a,max}}$.  We find that the power-law indices are in good agreement with the predictions above, particularly for times $t\ga t_{\rm{eq,ls}}$ when enough time has passed to smear the spectrum over a few decades in frequency. 

Once $t = t_{\rm{eq,ls}}$, the material along the line of sight with temperature $T_{\rm{obs,min}}$ obtains thermal equilibrium, and a blackbody component appears in the spectrum.  As we saw in Section~\ref{sec:sourceframe}, the temperature decay becomes much slower upon entering the blackbody regime.  Therefore, unlike the free-free emission, the blackbody emission is not significantly smeared by the effects of light travel time.  At intermediate times $t_{\rm{eq,ls}} < t < t_{\rm{lc}}$, as expected from Section~\ref{sec:blendedspectrum}, both a blackbody component and a smeared out free-free component are present in the spectrum, as is clearly seen for the brown ($t=1.9t_{\rm{eq,ls}}$) and pink ($t=4.4t_{\rm{eq,ls}})$ lines in Fig.~\ref{fig:smearedspectrumexample}.  As discussed in Section~\ref{sec:blendedspectrum}, the blackbody emission is produced by material with $\theta \la [2(t-t_{\rm{eq,ls}})/t_{\rm{lc}}]^{1/2}$, while the free-free emission comes from material at angles larger than this.  

As $t$ approaches $t_{\rm{lc}}$, the free-free breakout emission arrives from material closer and closer to the equator.  Although this material still emits with a temperature of $T_{\rm{obs,bo}}$, its luminosity is suppressed by projected area effects (i.e., by the factor of $\cos\theta$ appearing in equation~\ref{Lnuobs}).  The hotter material, which comes from larger $\theta$, is suppressed more and as a result, the free-free component of the spectrum starts to become fainter.  The spectrum grows softer at high energies, and harder at low energies, and it no longer follows the expected power-law behaviours described above.  At $t=t_{\rm{lc}}$, the material at $\theta=\uppi/2$ is finally seen, and $T_{\rm{obs,max}}$ starts to drop rapidly.  Within a few $t_{\rm{bo}}$ after $t_{\rm{lc}}$, the equatorial material has cooled significantly and the spectrum becomes very soft at high energies. Ultimately, at $t=t_{\rm{lc}}+t_{\rm{eq,ls}}$, even the material at the equator thermalizes and the free-free component disappears completely. As expected, the late-time spectrum is simply a blackbody.

\subsection{Comptonization}
\label{sec:comptonization}

Finally, let us consider the effect of Comptonization, and show that our assumption 
that it does not strongly influence the spectrum is valid.  Since $\xi_{\rm{ls}}(T_{\rm{obs,ls}})$ is a decreasing function of temperature (e.g., NS10), and the temperatures of all shells inside the breakout shell satisfy $T_{\rm{obs,ls}} \le T_{\rm{obs,bo}}$, the Comptonization parameter of the luminosity shell satisfies $\xi_{\rm{ls}} \le \xi_{\rm{bo}}$ at all times.  Moreover, the rapid evolution of $T_{\rm{obs,ls}}$ (as discussed in Section~\ref{sec:thermalization}) will ensure that $\xi_{\rm ls}$ quickly falls below its initial value.  Therefore, it is sufficient to show that the Comptonization parameter of the breakout shell, $\xi_{\rm{bo}}$, is not too large.  Our discussion here mainly follows FS19 and \citet{margalit} \citep[see also][]{kompaneets,illarionov,rybicki}.

The role of Comptonization in altering the spectrum is basically to bring low-energy photons with energies $h\nu \ll kT$ up to energies of order $4kT$.  This increases the number of photons which are available to share the internal energy, and decreases the temperature.  The importance of Comptonization depends on the number of times a typical photon is scattered before escaping the system.  Photons which are generated with lower energies require a greater number of scatterings to be brought up to $4kT$, and are also more susceptible to self-absorption.  There is a critical frequency, $\nu_{\rm{coh}}$, below which photons are destroyed by self-absorption before they can be Comptonized. If a typical photon undergoes enough scatterings, all photons with $\nu \ga \nu_{\rm{coh}}$ will be brought up to the peak, and the Comptonization will saturate. On the other hand, if not enough scatterings occur, then only photons with frequencies $\nu > \nu_y$ can be brought up to the peak energy, where $h \nu_y \approx kT_{\rm{obs}} e^{-y(c/v \tau)}$ (e.g., FS19), $y \approx (4kT_{\rm obs}/m_{\rm e}c^2)\tau^2$ is the Compton parameter, and $m_{\rm{e}}$ is the electron mass.

For the breakout shell, we have  $T_{\rm{obs}} = T_{\rm{obs,bo}}$ and $\tau = c/v_{\rm{bo}}$, so we find
\begin{align}
    \label{ybo}
    y_{\rm{bo}} &\approx \dfrac{4kT_{\rm{obs,bo}}}{m_{\rm{e}} v_{\rm{bo}}^2} \approx 7.8 \left(\dfrac{kT_{\rm{obs,bo}}}{10\,\text{keV}}\right) v_{\rm{bo,}-1}^{-2} \nonumber\\
    &\approx 10 q^2 (n+1)^{-2} \xi_{\rm{bo}}^{-2} v_{\rm{bo,}-1}^6 
\end{align}
and
\begin{equation}
    \label{nuy}
    h\nu_{y\rm{,bo}} \approx kT_{\rm{obs,bo}} e^{-y_{\rm{bo}}}.
\end{equation}
The value of $y_{\rm bo}$ is very sensitive to $v_{\rm bo}$, and therefore $\nu_{y\rm{,bo}}$ is poorly constrained in general.  For velocities $\la 0.1\,c$, 
equation~\ref{ybo} predicts $y_{\rm{bo}} \la 10$.  In this case $\nu_{y\rm{,bo}}$ falls in the range ${1\text{--}10^4\,}$eV.  Comparing with equation~\ref{nuabovalue} then suggests that $\nu_{y\rm{,bo}} > \nu_{\rm{a,bo}}$, indicating that most photons generated in the breakout shell can diffuse out of the system before they can be Comptonized, although care must be taken here because $\nu_{\rm a,bo}$ also depends sensitively on $v_{\rm bo}$.

It is useful to define two critical values of $y$: $y_{\rm a}$ for which $\nu_{y} = \nu_{\rm a}$, and $y_{\rm sat}$ for which $\nu_y = \nu_{\rm coh}$.  For $y \ga y_{\rm a}$ ($\nu_y < \nu_{\rm{a}}$), the value of $\xi$ saturates because the supply of photons available to Comptonize has effectively been exhausted (e.g., FS19).  Since the photon number spectrum peaks at $\nu_{\rm{a}}$ for self-absorbed free-free emission, the number of photons with ${\nu < \nu_{\rm a}}$ is negligible.  So, even though additional photons from below $\nu_{\rm a}$ can be brought to the peak for $y\ga y_{\rm a}$, their presence does not appreciably change the photon number density at $kT_{\rm obs}$ or the value of $\xi$.  For the breakout shell, a value of ${y_{\rm a,bo} \approx \ln(kT_{\rm obs,bo}/h\nu_{\rm a,bo})}$ is found by setting $\nu_{y\rm{,bo}} = \nu_{\rm a,bo}$ in equation~\ref{nuy}. Note that $y_{\rm a}$ is only defined for out-of-equilibrium systems with $kT_{\rm obs,bo} > h \nu_{\rm a,bo}$, and otherwise we set it to 0.

If $y \ga y_{\rm sat}$ ($\nu_y < \nu_{\rm coh}$), then the $y$-parameter is sufficiently large that the spectrum's shape is significantly affected.  The spectrum saturates to a Wien spectrum for $y \gg y_{\rm sat}$. \citet{margalit}, who studied Comptonization of the X-ray emission in interacting supernovae, found that for $y \la y_{\rm sat}$, the shape of the free-free seed spectrum is not significantly altered, other than the peak energy shifting to $\approx 4kT$ (see their Figure 8).
To estimate $y_{\rm sat}$ for the breakout shell, we first note that for $h \nu_{\rm{coh}} \ll kT_{\rm{obs}}$, the value of $\nu_{\rm{coh}}$ is approximated by $\nu_{\rm{coh}} \approx \nu_{\rm{a}} /y^{1/2}$ \citep[e.g.,][]{rybicki}.  Plugging $\nu_y = \nu_{\rm{coh}}$ into equation~\ref{nuy} then yields ${y_{\rm sat,bo} \approx y_{\rm a,bo} + \frac{1}{2}\ln y_{\rm bo}}$. Note that $y_{\rm sat,bo}$ is only defined when $y_{\rm bo} \ge 1$ and it always satisfies $y_{\rm sat,bo} > y_{\rm a,bo}$. For the values of $kT_{\rm{obs,bo}}$,  $h\nu_{\rm{a,bo}}$, and $y_{\rm bo}$ in equations~\ref{Tinequality}, ~\ref{nuabo}, and~\ref{ybo}, we find $y_{\rm sat,bo} \sim 10$.  Thus, marginally $y_{\rm{bo}} \la y_{\rm sat,bo}$ in our typical case, and as desired the shape of the spectrum should not be greatly affected. 

Now, to improve on these basic estimates and obtain the Comptonization parameter $\xi_{\rm{bo}}$, we first combine equation~\ref{nuy} with ${\xi = 1+ 0.75y + 0.5y^2}$ (FS19)\footnote{Note that FS19 defined $y$ differently than us.  Their quantity $\ln y$ equals our quantity $y$.}, resulting in
\begin{equation}
    \label{xibo}
    \xi_{\rm{bo}} = 1+ 3.0\left(\dfrac{k T_{\rm{obs,bo}}}{m_{\rm{e}} v_{\rm{bo}}^2}\right) + 8.0 \left(\dfrac{k T_{\rm{obs,bo}}}{m_{\rm{e}} v_{\rm{bo}}^2}\right)^2
\end{equation}
for the breakout shell. Note that equation~\ref{xibo} is only valid when ${y < y_{\rm a}}$. In the $y \ga y_{\rm a}$ case, $\xi$ is given by ${\xi_{\rm a} \equiv {\xi(y_{\rm{a}}) \approx 1 + 0.75 y_{\rm{a}} + 0.5 y_{\rm{a}}^2}}$ instead (FS19).  However, as we will confirm later, for our problem $y < y_{\rm a}$ and equation~\ref{xibo} is appropriate.

We now have three equations (\ref{Tinequality}, \ref{ybo}, and~\ref{xibo}) relating the four quantities $v_{\rm bo}$, $T_{\rm obs,bo}$, $y_{\rm bo}$, and $\xi_{\rm bo}$.  We can therefore express any three of them in terms of the fourth.  It is instructive to choose either $v_{\rm bo}$ or $T_{\rm obs,bo}$ as the independent quantity; the former is useful for theory as it has the clearest relation to the properties of the progenitor and explosion, while the latter is best when comparing to observations since $T_{\rm obs,bo}$ is directly observable.  Here we will study $T_{\rm obs,bo}$, $y_{\rm bo}$, and $\xi_{\rm bo}$ as functions of $v_{\rm bo}$.

\begin{figure}
\centering
\includegraphics[width=0.9\columnwidth]{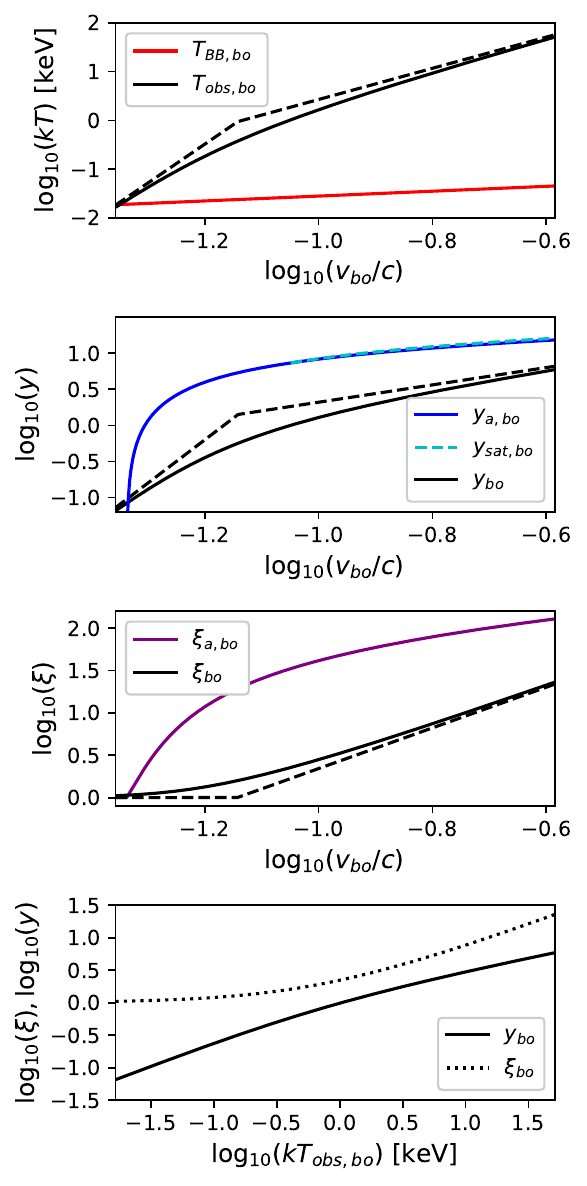}
\caption{Comptonization properties for shock velocities in the range $0.04 \la v_{\rm bo}\la 0.25$, where the breakout is out of equilibrium, but not hot enough to produce pairs, assuming $q=1$ and $n=0$.  Coloured lines indicate quantities which depend on density; in computing those we adopt $\rho_{\rm bo,-11}=1$. Solid and black dashed lines show, respectively, the implicit and approximate forms of equations~\ref{Timplicit}--\ref{xiapprox}. \textit{Top panel:} The observed breakout temperature, $T_{\rm obs,bo}$ and blackbody temperature, $T_{\rm BB,bo}$, versus $v_{\rm bo}$. \textit{Second panel:} The Compton $y$-parameter versus $v_{\rm bo}$ compared to the critical values $y_{\rm a,bo}$ and $y_{\rm sat,bo}.$ \textit{Third panel:} The Comptonization parameter $\xi_{\rm bo}$ compared to its saturation value $\xi_{\rm a,bo}$. \textit{Bottom panel:} $y_{\rm bo}$ (solid) and $\xi_{\rm bo}$ (dotted) as functions of $T_{\rm obs,bo}$ }
\label{fig:compton}
\end{figure}

Let us first consider the behaviour of $T_{\rm obs,bo}$.  By inserting equation~\ref{Tinequality} into equation~\ref{xibo}, we can eliminate $\xi_{\rm{bo}}$ to get the implicit equation
\begin{equation}
    \label{Timplicit}
    kT_{10} \left[1+ 5.9 v_{\rm{bo,}-1}^{-2}kT_{10} + 30.6 v_{\rm{bo,}-1}^{-4}(kT_{10})^2 \right]^{2} = {1.3 \left(\dfrac{q}{n+1}\right)^2 v_{\rm{bo,}-1}^8},
\end{equation}
where $kT_{10} = {kT_{\rm{obs,bo}}}/(10\,\text{keV})$.  Keeping only the first or last term in the brackets leads to the approximation
\begin{equation}
    \label{Tapprox}
    kT_{\rm{obs,bo}} \approx \begin{cases}
        13\,\text{keV}\,\left(\dfrac{n+1}{q}\right)^{-2} v_{\rm{bo,}-1}^8, & v_{\rm{bo,}-1} \la 0.7 \left(\dfrac{n+1}{q}\right)^{1/3} \\
        2.7\,\text{keV}\, \left(\dfrac{n+1}{q}\right)^{-2/5} v_{\rm{bo,}-1}^{16/5}, & v_{\rm{bo,}-1} \ga 0.7 \left(\dfrac{n+1}{q}\right)^{1/3},
    \end{cases}
\end{equation}
which is accurate to within a factor of $\sim 3$.  The first expression in equation~\ref{Tapprox} is the same as equation~\ref{Tinequality} in the limit $\xi_{\rm{bo}} = 1$, so we see that neglecting Comptonization is somewhat acceptable up to about $v_{\rm{bo}} \sim 0.07\,c$.  A comparison of the value of $T_{\rm obs,bo}$ versus $v_{\rm bo}$ obtained from equations~\ref{Timplicit} and~\ref{Tapprox} is given in the top panel of Fig.~\ref{fig:compton}, up to temperatures of $\approx 50\,$keV where pair creation becomes non-negligible.  For comparison we also show the observed temperature that the breakout shell would have it if were in thermal equilibrium, $T_{\rm BB,bo}$, assuming a density of $\rho_{\rm bo,-11}=1$.

With $T_{\rm obs,bo}$ now known, it is straightforward to compute $y_{\rm bo}$ as well. Using equation~\ref{ybo} in conjunction with equations~\ref{Timplicit} and~\ref{Tapprox}, we have
\begin{equation}
\label{yimplicit}
    y_{\rm bo} (1 + 0.75 y_{\rm bo} + 0.5 y_{\rm bo}^2)^2 = 10 \left(\dfrac{q}{n+1}\right)^2 v_{\rm bo,-1}^6
\end{equation}
or approximately
\begin{equation}
\label{yapprox}
    y_{\rm bo} \approx 
    \begin{cases}
        10 \left(\dfrac{n+1}{q}\right)^{-2} v_{\rm bo,-1}^6, &  v_{\rm{bo,}-1} \la 0.7 \left(\dfrac{n+1}{q}\right)^{1/3}\\
        2.1 \left(\dfrac{n+1}{q}\right)^{-2/5} v_{\rm bo,-1}^{6/5}, &  v_{\rm{bo,}-1} \ga 0.7 \left(\dfrac{n+1}{q}\right)^{1/3}.
    \end{cases}
\end{equation}
If values for $\rho_{\rm bo}$, $n$, and $q$ are adopted, we can also calculate $\nu_{\rm a,bo}$ from equation~\ref{nuabo}, and thus determine $y_{\rm a,bo}$ and $y_{\rm sat,bo}$. The results are shown in the second panel of Fig.~\ref{fig:compton} for $\rho_{\rm bo,-11} = 1$, $q=1$, and $n=0$.  We find that $y_{\rm bo} < y_{\rm a,bo} < y_{\rm{sat,bo}}$ holds over the entire range of velocities we consider (excluding a small region where Comptonization is irrelevant since $y_{\rm a,bo} = 0$ and $\xi_{\rm bo}=1$).  Furthermore, $y_{\rm sat,bo}\approx y_{\rm a,bo}$ due to the modest value of $y_{\rm bo}$.  This result is not affected much by our choice of parameters because $y_{\rm sat,bo}$ and $y_{\rm a,bo}$ only depend logarithmically on $\nu_{\rm a,bo}$.  We therefore conclude that our assumption of unsaturated Comptonization is valid, and equation~\ref{xibo} is applicable.

Finally, repeating the steps that led to equation~\ref{Timplicit}, but eliminating $T_{\rm obs,bo}$ from equation~\ref{xibo} instead, we find an implicit equation relating $\xi_{\rm bo}$ and $v_{\rm bo}$, 
\begin{equation}
    \label{xiimplicit}
    \xi_{\rm bo} = 1+ 7.7 \xi_{\rm bo}^{-2} v_{\rm bo,-1}^6 \left(\dfrac{q}{n+1}\right)^2 + 52.2 \xi_{\rm bo}^{-4} v_{\rm bo,-1}^{12} \left(\dfrac{q}{n+1}\right)^4,
\end{equation}
which we can roughly estimate by
\begin{equation}
    \label{xiapprox}
    \xi_{\rm{bo}} \approx \begin{cases}
        1, & v_{\rm{bo,}-1} \la 0.7 \left(\dfrac{n+1}{q}\right)^{1/3} \\
        2.2 \left(\dfrac{n+1}{q}\right)^{-4/5} v_{\rm{bo,}-1}^{12/5}, & v_{\rm{bo,}-1} \ga 0.7 \left(\dfrac{n+1}{q}\right)^{1/3}.
    \end{cases}
\end{equation}
Thus, we see that $\xi_{\rm bo} \sim 1$ up to about $0.1\,c$, although it grows quickly beyond that.  A comparison of equations~\ref{xiimplicit} and~\ref{xiapprox} is shown in the third panel of Fig.~\ref{fig:compton}, where we also indicate the value of $\xi_{\rm a,bo}=\xi(y_{\rm a,bo})$ that would be obtained if all available photons were Comptonized. The fact that $\xi_{\rm bo} < \xi(y_{\rm a,bo})$ virtually everywhere demonstrates again that we are in the regime of unsaturated Comptonization.  

Lastly, to facilitate comparison with observations, we plot $\xi_{\rm bo}$ and $y_{\rm bo}$ versus $T_{\rm bo}$ in the bottom panel of Fig.~\ref{fig:compton}.  Comptonization effects start to become relevant only above temperatures of a few keV.  We therefore expect shock breakout-powered transients peaking in the \textit{Swift} XRT band or the \textit{Einstein Probe} bands to be only modestly influenced by Comptonization.

The results of this section justify our assumption of order-unity $\xi_{\rm bo}$ in Sections~\ref{sec:modelparameters}--\ref{sec:observedspectrum}.  Even if $\xi_{\rm bo}$ is a few, it is unlikely that our results will be significantly affected.  The main effect would be to slightly alter the power-law indices for the time evolution of $T_{\rm obs}$ and $\nu_{\rm a}$, as discussed at the end of  Section~\ref{sec:thermalization}, which would in turn slightly modify the spectral indices appearing in Fig.~\ref{fig:smearedspectrumexample}. 

As a further check that our assumptions about the spectrum are reasonable, our results can be compared to the spectra seen in Monte-Carlo radiative transfer simulations, for example those of \citet{ito1} and \citet{ito2}.  In particular, the wind breakout model presented in \citet{ito2} is relevant.  The top panel of their Figure 7 shows spectra for a case with a shock velocity of $0.1\,$c.  Although they considered a somewhat higher number density than us, $n=10^{15}\,\text{cm}^{-3}$, they also pointed out that another model with $n=10^{12}\,\text{cm}^{-3}$ was not significantly different in terms of spectral shape or peak energy, except for a lower self-absorption frequency.  Since our simple treatment does not consider the effect that the escape of radiation has on the shock structure, our model should be compared with their models with $f_{\rm esc} \ll 1$ \citep[for further discussion of the escape fraction, see, e.g.,][and references therein]{ioka}.  In particular, the $f_{\rm esc}=0.09$ case in \citet{ito2} exhibits a rather flat spectrum, with only modest Comptonization, and it is self-absorbed below $\sim 1\,$eV even in spite of their higher assumed density.  All of these features are consistent with our model assumptions.  The peak energy in their simulation is $3kT_{\rm{obs}}\approx 3\,$keV for a shock velocity of $0.1\,c$, which can be compared with our model to estimate an appropriate value of $q/(n+1)$ (in their $0.25\,c$ and $0.5\,c$ models, pairs are likely to be important so our model cannot be reliably applied).  
Using equation~\ref{Timplicit}, for $v_{\rm{bo}}=0.1\,c$ we find $kT_{\rm{obs,bo}} = (0.6,1.0,1.7,2.6)\,$keV for $q/(n+1)=(0.3,0.5,1,2)$.  
A value of $q/(n+1)=0.5$  therefore seems to give a good agreement with the Monte-Carlo simulations of \citet{ito2}, and for typical values of $n$ this is consistent with $q\sim1$.  We note that if photon escape effects are important, the low-frequency spectrum could be somewhat steeper than what we assume.  However, at least in the case of GRB 060218, a flat spectrum with a photon index of $-1.4$ to $-1.1$ is observationally indicated \citep[e.g.,][]{toma}.

To wrap up this discussion, let us consider what happens after the breakout, as deeper and colder layers are revealed. As discussed above, for a given $\nu_{\rm{a}}$ and $T_{\rm{obs}}$, the allowed values of $\xi$ are bounded between $1\le\xi\le\xi_{\rm a}$.  Now, once ejecta near thermal equilibrium become visible, $h\nu_{\rm{a}} \rightarrow kT_{\rm{obs}}$, so $y_{\rm{a}} \rightarrow 0$ and $\xi_{\rm a}\rightarrow 1$.  Thus, as expected, we find that $\xi \rightarrow 1$ when the system attains thermal equilibrium, and Comptonization eventually becomes unimportant.

\section{Discussion and conclusions}
\label{sec:conclusions}

Utilizing recent advances in shock breakout theory which predict that the observed temperature falls off more steeply than previously thought at early times (FS19), we have shown that a new type of shock breakout can occur under certain conditions (see Section~\ref{sec:basicidea}). In this previously unexplored scenario, the emission initially originates from outer regions of the flow where the gas and radiation are out of equilibrium, but deeper regions in thermal equilibrium are able to be probed on a time-scale $t_{\rm{eq,ls}}$ which is shorter than the light-crossing time of the system $t_{\rm{lc}}$.  The observed spectrum in this regime is significantly different from the standard shock breakout case: it contains both a blackbody component, and a Comptonized free-free component which is smeared out due to the non-negligible light-crossing time.  We call this new shock breakout regime the INERT (`initially non-equilibrium rapid thermalization') regime. 

We studied the necessary conditions to obtain this novel type of breakout in Section~\ref{sec:conditions}, finding that it depends strongly on the breakout velocity, $v_{\rm{bo}}$, and preferentially occurs for  $v_{\rm{bo}} \approx 0.1\,c$ (to within a factor of $\sim 2$ depending on other parameters).  A sufficiently high breakout density, such that $\rho_{\rm{bo,}-11} \ga 0.4 R_{\rm bo,14}^{-15/16}$ where $R_{\rm bo}$ is the breakout radius, is also required.  Motivated by observations of $ll$GRBs which suggest potentially extended progenitors, we also considered the case where the breakout takes place in an extended low-mass medium in Section~\ref{sec:envelopebreakout}, and investigated the dependence on the outer density profile of the medium, i.e. on the parameter $n$ in $\rho(r) \propto (R-r)^n$.  We constructed a model that is sufficiently general to encapsulate both the classical case of breakout from the edge of a polytropic medium (for order-unity $n$), and the case of breakout from the edge of a compact stellar wind or CSM (for ${n \ll 1}$), as discussed in Sections~\ref{sec:n0} and~\ref{sec:connections}.  The small $n$ limit was determined to be advantageous for obtaining an INERT breakout.  We also demonstrated that in this limit, an early optical peak due to cooling emission is expected, as suggested by \citet{np}.  The inferred properties of the peak are similar to those obtained in other studies of $ll$GRBs \citep[e.g.,][]{nakar,ic16}.  

In Section~\ref{sec:model}, we constructed a simple spectral model for a shock breakout in this unexplored regime.  The model depends on six parameters -- the diffusion time of the breakout shell $t_{\rm bo}$, the light-crossing time $t_{\rm lc}$, the equilibrium time $t_{\rm eq,ls}$, the luminosity $L_{\rm bo}$, the observed temperature $T_{\rm obs,bo}$, and the temporal power-law index $\alpha$ describing the temperature evolution -- all of which can be directly constrained by observations (for an example, see Paper III).  In the INERT regime, for which $t_{\rm bo} < t_{\rm eq,ls} < t_{\rm lc}$, the emergent spectrum features both a free-free component, smeared into a multi-colour power-law by the effects of light travel time, and a less-smeared blackbody component which is initially subdominant, but eventually comes to dominate (see Fig.~\ref{fig:smearedspectrumexample}).  The free-free component is initially very broad, with a self-absorption frequency of $h\nu_{\rm a}\sim 0.1\,$eV and a temperature of $kT_{\rm obs}\sim 10\,$keV, resulting in strong emission across the electromagnetic spectrum at early times.  The blackbody component peaks in the soft X-ray range, typically around $\sim 0.1\,$keV.  If the observed temperature decreases as $T_{\rm obs} \propto t^{-\alpha}$, then due to differences in light arrival time, the free-free emission is smeared to form a broken power-law spectrum, with characteristic spectral indices of $(3\alpha-1)/[\lambda(3\alpha-2)]$ at low frequencies, and $(1-3\alpha)/(3\alpha)$ at high frequencies, as shown in Fig.~\ref{fig:smearedspectrumexample} (for further discussion, refer to Paper II).  The high energy index is in agreement with previous studies (e.g., NS10, FS19).  We provide estimates for the radius, density, and velocity of the breakout shell in terms of observables, as well as two closure relations which can be used to test whether shock breakout can plausibly explain a given event.

In terms of applications, the most interesting seems to be $ll$GRBs, as they exemplify the subrelativistic velocities ($\approx 0.1\,c$) and $\sim 10\,$keV temperatures for which the physics studied here are most relevant.  Indeed, the qualitative similarities between the model proposed here and the most well-studied $ll$GRB, GRB 060218, are striking.  Many key features of our model, including strong early optical emission, fast early temperature evolution, an optical peak on a time-scale of half a day, and a spectrum with a blackbody hump and a power-law tail, were also exhibited by this peculiar GRB.  The rich connection between this work and $ll$GRBs is studied more quantitatively in a subsequent paper (Paper III). 

As discussed in Section~\ref{sec:connections}, explosions occurring in extended envelopes with the requisite properties for INERT breakout are also expected to result in a peak in the optical light curve on a time-scale of hours to days.  We therefore predict that events exhibiting INERT breakout features in X-rays will often be associated with double-peaked SNe.  Interestingly, two $ll$GRBs, GRB 060218 \citep{campana} and GRB 171205A \citep{izzo}, both of which had distinct thermal and non-thermal components in their spectra, have been conclusively linked to double-peaked SNe.  In a third event, GRB 100316D, the existence of an early optical peak could not be established, but excess blue emission at early times hinted that such a peak may have been missed \citep{cano}.  Yet another double-peaked event, the broad-lined Type Ic SN 2020bvc, had no detected prompt gamma-ray counterpart, but exhibited a radio and X-ray afterglow markedly similar to GRB 060218 \citep{ho}.  The recently discovered soft X-ray transient EP240414a was also associated with a double-peaked Type Ic-bl SN \citep{sun,vandalen}.  Based on these events, the connection between peculiar X-ray transients and double-peaked SNe seems promising, but so far there are not enough well-studied $ll$GRBs to form a proper statistical sample.  

Thankfully, \textit{Einstein Probe} should dramatically increase the number of known $ll$GRBs in the coming years.  However, although it will be an excellent tool for discovering X-ray transients of interest, \textit{Einstein Probe} alone will not suffice to test our spectral model due to its narrow wavelength coverage. A joint detection of a nearby $ll$GRB by \textit{Einstein Probe} and \textit{Swift}, \textit{SVOM}, or \textit{Fermi} would be ideal, although these satellites could only detect relatively nearby $ll$GRBs.  Rapid UV/optical/IR follow-up is also  essential, and upcoming high-cadence UV missions like \textit{ULTRASAT} and \textit{UVEX} will further contribute by providing crucial constraints in the far UV.  All of these missions will help us to piece together a complete picture of the early spectrum of stellar transients, providing at long last a chance to rigorously test shock breakout theory.  We look forward to these and other missions advancing our understanding of shock breakout physics in the near future.

\section*{Acknowledgements}
We thank T. Faran, E. Nakar, and R. Sari for helpful discussions, and K. Murase for valuable comments.  This work is supported by the JST FOREST Program (JPMJFR2136) and the JSPS Grant-in-Aid for Scientific Research (20H05639, 20H00158, 23H01169, 23H04900). 

\section*{Data Availability} 

The data underlying this article will be shared on reasonable request
to the corresponding authors.






\appendix

\section{Glossary of symbols}
\label{sec:appendixa}

For reference, we list important quantities appearing in the paper, the equations defining them (where applicable), and a brief description of their meanings in Table~\ref{table1}.

\begin{table*}
\centering
\begin{threeparttable}
\begin{tabular}{l c l}
Symbol & Equation & Meaning \\
\hline
\multicolumn{3}{l}{Notation} \\
\hline
bo & & Subscript denoting the breakout shell, where the optical depth is $c/v_{\rm bo}$ \\
ls & & Subscript denoting the luminosity shell, where the observed luminosity is generated \\
cs & & Subscript denoting the colour shell, where spectrum formation takes place\tnote{a} \\
\hline 
\multicolumn{3}{l}{Important time scales} \\
\hline 
$t_{\rm bo}$ & \ref{tbo} & The diffusion time, and dynamical time, of the breakout shell \\
$t_{\rm lc}$ & \ref{tlc} & The light-crossing time of the system \\
$t_{\rm eq,ls}$ & \ref{teq} & The time when thermal radiation is first observed \\
$t_{\rm s}$ & \ref{ts} & The time it takes for the outflow's radius to double \\
\hline
\multicolumn{3}{l}{Initial properties of the breakout} \\
\hline
$\rho_{\rm bo}$ & & The initial density of the breakout shell, before being shocked \\
$v_{\rm bo}$ & & The shock velocity encountered by the breakout shell \\
$R_{\rm bo}$ & & The radius of the shock at breakout \\
$L_{\rm bo}$ & \ref{Lbo} & The luminosity produced by the breakout shell \\
$T_{\rm obs,bo}$ & \ref{Tinequality} & The observed temperature of the breakout shell \\
$T_{\rm BB,bo}$ & \ref{TBB} & The temperature that the breakout shell would have, if in thermal equilibrium \\
$\nu_{\rm a,bo}$ & \ref{nuabo} & The self-absorption frequency of the breakout shell \\
$y_{\rm bo}$ & \ref{ybo} & The Compton $y$-parameter of the breakout shell \\
$\eta_{\rm bo}$ & \ref{etabo} & A  parameter controlling whether the breakout shell is in thermal equilibrium ($\eta_{\rm bo} > 1$) or not ($\eta_{\rm bo} < 1$) \\
$\xi_{\rm bo}$ & \ref{xibo} & A  parameter controlling whether the spectrum is initially Comptonized ($\xi_{\rm bo} > 1$) or not ($\xi_{\rm bo} = 1$)\\
$\chi$ & \ref{chi0} & A parameter controlling whether the breakout duration is set by diffusion ($\chi < 1$) or light-crossing ($\chi > 1$) \\
\hline
\multicolumn{3}{l}{Quantities describing the post-breakout planar evolution\tnote{b}} \\
\hline
$L_{\nu}(\nu,t)$ & \ref{Lnu}--\ref{tauff} & The time-dependent spectral luminosity produced by the source \\
$L_{\nu\rm,obs}(\nu,t)$ & \ref{Lnuobs} & The observed spectral luminosity, accounting for projected area and light arrival time effects \\
$L_{\rm ls}(t)$ & \ref{Levolution} & The power generated by the luminosity shell  \\
$T_{\rm obs,cs}(t)$ & \ref{Tevolution} & The observed temperature at the colour shell \\
$\nu_{\rm a,cs}(t)$ & \ref{nuaevolution} & The self-absorption frequency at the colour shell \\
$T_{\rm eq,ls}$ & \ref{Teqls} & The observed temperature when thermalized ejecta are first revealed, at $t=t_{\rm eq,ls}$ \\
$\alpha$ & \ref{Tpowerlaw} & A power-law index describing the early temperature evolution, $T_{\rm obs,cs} \propto t^{-\alpha}$ \\
$\beta$ & \ref{nuapowerlaw}--\ref{betavsalpha} & A power-law index describing the early evolution of the self-absorption frequency, $\nu_{\rm a,cs} \propto t^{\beta}$ \\
$\zeta$ & \ref{zetadef}--\ref{zetachieta} & A parameter controlling whether equilibrium is achieved during the initial breakout flash ($\zeta < 1$), or after it ($\zeta > 1$) \\
\hline 
\multicolumn{3}{l}{Progenitor properties} \\
\hline
$R_*$ & & The radius of the progenitor star \\
$M_*$ & & The mass of the progenitor star \\
$R_{\rm env}$ & & The radius of the extended envelope \\
$M_{\rm env}$ & & The mass contained in the extended envelope \\
$\rho_{\rm env}$ & \ref{rhoenv} & The typical density in the extended envelope \\
$s$ & & A power-law index describing the inner part of the envelope, $\rho \propto r^{-s}$ \\
$n$ & & A power-law index describing the outer part of the envelope, $\rho \propto (R_{\rm env}-r)^n$ \\
$\Lambda$ & \ref{Lambda} & A function of $n$ and $s$ which controls the scale height of the outer density profile \\
$E_0$ & & The total energy deposited in the envelope \\
$v_0$ & \ref{v0} & The typical shock velocity in the bulk of the envelope \\
$\Psi$ & \ref{Psi} & A parameter controlling whether the breakout occurs at the edge of the envelope $(\Psi > 1)$, or deeper inside it $(\Psi < 1)$ \\
\hline
\end{tabular}
\begin{tablenotes}
    \item[a] In a non-equilibrium breakout, the colour and luminosity shells are equivalent until thermal equilibrium is achieved.
    \item[b] The subscripts `ls' and `cs', which were dropped in Section~\ref{sec:model}, are included here for clarity.
\end{tablenotes}
\end{threeparttable}
\caption{Glossary of symbols.}
\label{table1}
\end{table*}

\section{Comparison of our scaling relations with past work}
\label{sec:appendixb}

We parametrize our envelope breakout model for $n=0$ in terms of the envelope mass, $M_{\rm{env}}$, the envelope radius, $R_{\rm{env}}$, and the energy deposited in the envelope, $E_0$.  In Section~\ref{sec:connections}, we gave scalings for the observables $t_{\rm{bo}}$, $L_{\rm{bo}}$, and $E_{\rm{bo}}$ in terms of these parameters.  Here we confirm that our scalings are identical to those of \citet{ci11} and \citet{hp}.

\citet{hp} instead parametrize the observables in terms of $R_{\rm{env}}$, the shock velocity $v_{\rm s}$, and the wind parameter $D$ defined by $\rho(r) = D r^{-2}$.  With the replacements ${v_{\rm s} \propto E_0^{1/2} M_{\rm{env}}^{-1/2}}$ and ${D \propto M_{\rm{env}} R_{\rm{env}}^{-1}}$, we obtain ${t_{\rm{bo}} \propto (c/\kappa) R_{\rm{env}}^2 D^{-1} v_{\rm s}^{-2}}$, ${{E_{\rm{bo}} \propto (c/\kappa) R_{\rm{env}}^2 v_{\rm s}}}$, and ${L_{\rm{bo}} \propto D v_{\rm s}^3}$, in agreement with the scalings presented in \citet{hp}.\footnote{In \citet{hp}, $R_{\rm{env}}$, $t_{\rm{bo}}$, $E_{\rm{bo}}$, and $L_{\rm{bo}}$ are called $R_{\rm{w}}$, $t_{\rm{rise}}$, $E_{\rm{sbo}}$, and $L_{\rm{sbo}}$ respectively.}  

\citet{ci11} also used $D$ and $R_{\rm{env}}$ as free parameters, but replaced $v_{\rm s}$ with the supernova energy $E_{\rm{SN}}$ and the supernova ejecta mass $M_{\rm{e}}$ via the relation $v_{\rm s} \propto (E_{\rm{SN}}/M_{\rm{e}})^{1/2} (M_{\rm{e}}/M_{\rm{env}})^{1/4}$, which comes from the \citet{chevalier83} self-similar solution for an ejecta with an outer density profile $\rho \propto r^{-7}$ colliding with a stellar wind with $\rho \propto r^{-2}$.  Note that the energy $E_{\rm{SN}}$, which is the total supernova energy, is different from the energy $E_0$ in our model, which represents only the energy deposited in the low-mass envelope; the two are related by ${E_0 \approx E_{\rm{SN}} (M_{\rm{env}}/M_{\rm{e}})^{1/2}}$, assuming the same ejecta density profile as above.  When this relation and ${D\propto M_{\rm{env}}/R_{\rm env}}$ are substituted into our scaling relations, we obtain ${t_{\rm{bo}} \propto (c/\kappa) R_{\rm{env}}^{5/2} E_{\rm{SN}}^{-1} M_{\rm{e}}^{1/2} D^{-1/2}}$ and ${E_{\rm{bo}} \propto (c/\kappa) R_{\rm{env}}^{7/4} E_{\rm{SN}}^{1/2} M_{\rm{e}}^{-1/4} D^{-1/4}}$, which combine to give ${L_{\rm{bo}} \propto R_{\rm{env}}^{-3/4} E_{\rm{SN}}^{3/2} M_{\rm{e}}^{-3/4} D^{1/4}}$, in agreement with the results of \citet{ci11} in the $R_{\rm{w}} < R_{\rm{d}}$ case (see their erratum, as there was an error in these scalings in their original paper).\footnote{In \citet{ci11}, $E_{\rm{SN}}$, $R_{\rm{env}}$, $t_{\rm{bo}}$, $E_{\rm{bo}}$, and $L_{\rm{bo}}$ are called $E$, $R_{\rm{w}}$, $t_{\rm{r}}$, $E_{\rm{rad}}$, and $L$ respectively.}

As discussed in Section~\ref{sec:basicidea}, the values of $L_{\rm{bo}}$ and $t_{\rm{bo}}$ reflect the intrinsic luminosity and duration of the breakout, which is only equal to the observed luminosity $L_{\rm bo}$ when $t_{\rm{bo}} < t_{\rm{lc}}$ so that light travel effects can be neglected.  We may also consider the situation where $t_{\rm{lc}} > t_{\rm{bo}}$, in which case the observed luminosity is given by $L_{\rm{obs}} \approx L_{\rm{bo}}t_{\rm{bo}}/t_{\rm{lc}}$.  We then have ${L_{\rm{obs}} \propto (c/\kappa) R_{\rm{env}} E_0^{1/2} M_{\rm{env}}^{-1/2}}$, ${L_{\rm obs} \propto (c/\kappa) R_{\rm{env}} v_{\rm s}}$, or ${L_{\rm obs} \propto (c/\kappa) R_{\rm{env}}^{3/4} E_{\rm{SN}}^{1/2} M_{\rm{e}}^{-1/4} D^{-1/4}}$, respectively for the parametrizations of this paper, \citet{hp}, and \citet{ci11}. 

It remains to check that our scalings for $t_{\rm{p}}$, $L_{\rm{p}}$, and $T_{\rm{p}}$ given by equations~\ref{tp}--\ref{Tp} reproduce those of \citet{np}. Like \citet{ci11}, \citet{np} parametrize the problem in terms of the total supernova energy $E_{\rm{SN}}$, instead of the energy $E_0$ which is deposited in the extended envelope.\footnote{In \citet{np}, $E_{\rm{SN}}$, $v_0$, $R_{\rm{env}}$, $M_{\rm{env}}$, and $E_0$ are called $E$, $v_{\rm{ext}}$, $R_{\rm{ext}}$, $M_{\rm{ext}}$, and $E_{\rm{ext}}$, respectively.}  In their model, there is a massive core inside of the envelope, and the core mass $M_{\rm c} \gg M_{\rm{env}}$ is also a free parameter.  They assume that the transition of the shock from the core to the envelope is described by the self-similar solution of \citep{sakurai}, which results in the relationship ${E_0 \propto E_{\rm{SN}} (M_{\rm{env}}/M_{\rm c})^{0.7}}$.  We then obtain $t_{\rm{p}} \propto \kappa^{0.50} E_{\rm{SN}}^{-0.25} M_{\rm c}^{0.17} M_{\rm{env}}^{0.57}$ from equation~\ref{tp}, $L_{\rm{p}} \propto \kappa^{-1} R_{\rm{env}} E_{\rm{SN}} M_{\rm c}^{-0.7} M_{\rm{env}}^{-0.3}$ from equation~\ref{Lp}, and ${T_{\rm{p}} \propto \kappa^{-0.25} R_{\rm{env}}^{0.25} E_{\rm{SN}}^{-0.13} M_{\rm c}^{0.09} M_{\rm{env}}^{-0.46}}$ from equation~\ref{Tp}.  After some rearranging, we find ${M_{\rm{env}} \propto E_{\rm{SN}}^{0.43} \kappa^{-0.87} M_{\rm c}^{-0.30} t_{\rm{p}}^{1.75}}$, ${R_{\rm{env}} \propto E_{\rm{SN}}^{-0.87} \kappa^{0.74} M_{\rm c}^{0.61} L_{\rm{p}} t_{\rm{p}}^{0.51}}$, and ${T_{\rm{p}} \propto \kappa^{-1/4} R_{\rm{env}}^{1/4} t_{\rm{p}}^{-1/2}}$, which agrees with equations 10, 12, and 13 in \citet{np}.  


\bsp	
\label{lastpage}
\end{document}